\documentclass[11pt]{article}
\pdfoutput = 1

\usepackage[utf8]{inputenc}
\usepackage{color,graphicx}
\usepackage{epsfig}
\usepackage{amsmath}
\usepackage{verbatim}
\usepackage{amssymb}
\usepackage{mathabx}
\usepackage[mathcal]{eucal}
\usepackage{stmaryrd}
\usepackage{empheq}
\usepackage{esint}
\usepackage{physics}
\usepackage{amsfonts}
\usepackage{cite}
\usepackage{array}
\usepackage{setspace}
\usepackage{braket}
\usepackage{float}
\usepackage{url}
\usepackage{mathtools}
\usepackage{tensor}
\usepackage{bbold}
\usepackage{slashed}
\usepackage{tikz}
\usepackage{pgfplots}
\usepackage{graphicx}
\usepackage{epstopdf}
\usepackage{subcaption}
\usepackage[labelfont=bf,font={sf}]{caption}
\usepackage{pgfplots}
\usepackage{mathrsfs}
\usepackage{fancybox}
\usepackage{eurosym}
\usepackage{tcolorbox}
\usepackage{tensor}
\usepackage[normalem]{ulem}
\allowdisplaybreaks
\usepackage{changepage}

\usetikzlibrary{arrows.meta, decorations.markings}

\usepackage[margin = 2.2cm]{geometry}
\usepackage[ragged]{footmisc}
\usepackage[bookmarks=true,bookmarksnumbered=false,
hyperindex=true,bookmarksopen=true,hyperfigures=true,
colorlinks=true,linkcolor=webblue,citecolor=webgreen,urlcolor=webblue,breaklinks]{hyperref}
\definecolor{webgreen}{rgb}{0, 0.5, 0}
\definecolor{webblue}{rgb}{0, 0, 0.5}
\definecolor{webred}{rgb}{0.5, 0, 0}
\definecolor{darkgray}{rgb}{0,0.5,0}

\newcommand{\slt}{$\TT{SL}(2,\mathbb{C})$}
\newcommand{\slz}{$\TT{SL}(2,\mathbb{Z})$}

\newcommand{\cls}{$\mathbb{C}$LS}

\newcommand{\de}{\partial}
\newcommand{\ca}{\mathcal}
\newcommand{\LL}{\left}
\newcommand{\RR}{\right}
\newcommand{\TT}{\text}

\newcommand{\ishiket}[1]{\vert{#1}\rangle \hspace{-.07cm}\rangle \hspace{-.0cm}}
\newcommand{\ishibra}[1]{\hspace{-.0cm} \langle \hspace{-.07cm} \langle{#1}\vert}

\def\is{\!&\! = \! & \!}

\def\sfb{{\mathsf b}}
\def\sfq{{\mathsf q}}

\def\spc{\hspace{1pt}}

\def\ben{\begin{equation}}
\def\een{\end{equation}}

   \let\d=\delta 
   
     \let\r=v

\def\be{\begin{equation}}
\def\ee{\end{equation}}
\def\ba{\begin{array}}
\def\ea{\end{array}}

\def\dalemb#1#2{{\vbox{\hrule height .#2pt
       \hbox{\vrule width.#2pt height#1pt \kern#1pt
               \vrule width.#2pt}
       \hrule height.#2pt}}}

\newcommand{\bea}{\begin{eqnarray}}
\newcommand{\eea}{\end{eqnarray}}

\let\tilde=\widetilde

\def\nspc{\hspace{-1pt}}
\renewcommand{\d}{\mathrm{d}}
\renewcommand{\i}{\mathrm{i}}

\numberwithin{equation}{section}

\begin{document}

\thispagestyle{empty}
    ~\vspace{5mm}
\begin{adjustwidth}{-1cm}{-1cm}
\begin{center}
     {\LARGE \bf 
An observer's quantization of 3d de Sitter
}
   \vspace{0.4in}

     {\bf Andreas Blommaert$^{1}$, Damiano Tietto$^2$, Herman Verlinde$^2$}
     \end{center}
    \end{adjustwidth}
\begin{center}
    \vspace{0.4in}
    {$^1$School of Natural Sciences, Institute for Advanced Study, Princeton, NJ 08540, USA\\
    $^2$Department of Physics, Princeton University, Princeton, NJ 08544, USA}
    \vspace{0.1in}
    
    {\tt blommaert@ias.edu, dtietto@princeton.edu, verlinde@princeton.edu}
\end{center}

\vspace{0.4in}

\begin{abstract}
\noindent What is the density of states of the de Sitter static patch? We propose a definition and calculation of such a density in 3d dS. Our proposal involves a sum over an SL(2,$\mathbb{Z}$) set of Euclidean no-boundary Kerr-lens spacetimes sourced by a line-defect with given energy and spin - which in Lorentzian time represents an observer's worldline at the center of the dS static patch. We develop an exact quantum computation of the spectral density using a holographic duality between dS$_3$ gravity and two copies of $\mathbb{C}$LS, the complex Liouville string. The SL(2,$\mathbb{Z}$) Kerr-lens spacetimes map under the duality to an SL(2,$\mathbb{Z}$) family of generalized crosscap geometries. We compute the $\mathbb{C}$LS $\otimes$ $\mathbb{C}$LS crosscap amplitudes  and show that they match the semi-classical gravity prediction. For the simplest non-trivial Kerr-lens space, the $\mathbb{C}$LS $\otimes$ $\mathbb{C}$LS description is, in turn, dual to two copies of the $G\Sigma$ effective field theory of the double scaled SYK model. The $G\Sigma\otimes G\Sigma$ theory lives on an observer's worldline in the static patch, setting the stage for developing a microscopic worldline hologram of 3d de Sitter.
\end{abstract}

\pagebreak
\setcounter{page}{1}
\setcounter{tocdepth}{2}

\addtolength{\baselineskip}{-.6mm}
\tableofcontents
\addtolength{\baselineskip}{.6mm}
\section{{Introduction}}

One of the ways in which general relativity hints at a more fundamental description of quantum gravity is via Euclidean partition functions. Gibbons and Hawking \cite{gibbons1977cosmological} proposed that the Euclidean gravitational path integral on the sphere, describing the analytic continuation of that static patch of de Sitter space, computes the number of states of the dS static patch. Specializing to three dimensions the GH procedure identifies the entropy of 3d dS spacetime with  \cite{gibbons1977cosmological}:
\bea 
\label{gh-entropy}
S_{\rm dS} = \frac{A}{4 G}\,.
\eea
Here $A = 2\pi R_{\rm dS}$ is the area of the cosmological horizon, surrounding the static patch. The GH entropy formula \eqref{gh-entropy} can be naturally interpreted as counting the spectral density of the microscopic quantum system associated to the static patch. A better understanding of the gravitational path integral beyond the semiclassical approximation is then desirable to help uncover a microscopic theory of de Sitter quantum gravity. 

The GH proposal that the Euclidean sphere computes an entropy was given strong supporting evidence in the  $G\to 0$ limit in\cite{Chandrasekaran:2022cip}, using  Lorentzian algebraic QFT techniques.\footnote{Interesting studies about a state counting interpretation of the sphere path integral at one loop were presented in \cite{Anninos:2020hfj,Anninos:2021ene,Anninos:2026hia,Anninos:2023exn}.} Their crucial insight was that $A/4 G$ counts the entropy from the perspective of an observer in the static patch, and that this observer should be considered an integral part of the gravitational theory \cite{Anninos:2011af} and thus backreacts on the area of the cosmological horizon.\footnote{Other recent work about cosmological implications of describing gravity from an observer's perspective includes \cite{Abdalla:2025gzn,Harlow:2025pvj,Akers:2025ahe,Blommaert:2025bgd,Kudler-Flam:2024psh,Cotler:2025gui,Maldacena:2024spf,Ivo:2025yek,Chen:2025jqm,Shi:2025amq,Chen:2024rpx,Blommaert:2025rgw,Goto:2026ipq,Tietto:2025oxn}.}  CLPW \cite{Chandrasekaran:2022cip} consider observers that introduce tiny perturbations of empty dS space, as required by the $G\to 0$ limit. But which entropy (or spectral density) would the Euclidean gravitational path integral predict for an observer with general energy $E$ and spin $J$ to all orders in $G$?

We shall propose an exact quantization of the static patch of 3d dS from an observer's perspective, and leverage this to give an exact prediction for an observer's spectral density. The logical generalization of the GH proposal in this situation, is to consider the analytic continuation to Euclidean signature of the static patch metric of the Kerr-de Sitter spacetime (from hereon we set $R_{\rm dS} = 1$):
\bea
\label{kds-metric}
    \d s^2=-N^2\d t^2+\frac{\d r^2}{N^2}+r^2\bigg(\d \varphi-\frac{J}{ 2 r^2}\,\d t\bigg)^2\,,\quad N^2=E-r^2+\frac{J^2}{4 r^2}\,,\quad 0\leq E\leq 1\,.
\eea
The KdS metric \eqref{kds-metric} includes the backreaction due to the $(E,J)$ observer worldline. Pure dS spacetime corresponds to setting $(E,J) = (1,0)$. We will study the physical properties of the space of KdS metrics in more detail in section \ref{sect:2.1solutions}.

We aim to compute the Euclidean no-boundary \cite{hartle1983wave} path integral ending on a Wick rotated neighborhood of the observer's worldline. In 3d, diff invariance kills any local metric fluctuations, simplifying the computation of the gravitational path integral. Nevertheless, the gravitational theory is non-trivial because of the presence of global degrees of freedom, and of topologically distinct saddles. We propose that the no-boundary path integral of Euclidean Kerr-de Sitter (KdS) space computes the spectrum of the static patch surrounding the observer, following GH:
\bea
\label{kds-exact-spectrum}
    \boxed{\ \mathcal{Z}_\text{KdS}(E,J)=\rho(E,J)\,\strut }
\eea
The leading semiclassical computation for the spectral density gives that \cite{gibbons1977cosmological} 
\bea
\label{kds-gh-spectrum}
\rho(E,J)\, \to \, e^{\frac{A(E,J)}{4G}}\,,
\eea
with $A(E,J)$ the area of KdS cosmological horizon
\bea
\label{kds-area}
 {A(E,J)} = {2\pi} \sqrt{\frac{E+ \sqrt{E^2+J^2}}{2}}
\eea
The arrow $\to$ in equation \eqref{kds-gh-spectrum} indicates the semiclassical approximation. Our main goal in this paper is to go beyond this leading approximation, and obtain the exact quantum gravity result for the left-hand-side of \eqref{kds-exact-spectrum}, to all orders in $G$. For this computation we will leverage recent lessons learned from studies of BTZ black holes in AdS$_3$, including the usefulness of Virasoro TQFT \cite{Collier:2023fwi} and the role of the mapping class group, generalizing the approach of \cite{Maloney:2007ud,Keller:2014xba} to computing the BTZ partition function in pure AdS$_3$ gravity  and the early insightful work \cite{Castro:2011xb} applying these ideas to pure 3d dS.\footnote{The analysis in \cite{Castro:2011xb} did not include a heavy object along an observer's worldline and computed amplitudes via SU$(2)$ CS gauge theory \cite{Castro:2011xb,Hikida:2021ese}. We will include the observer worldline and use the modular matrices of complex Virasoro TFT.}

There are several key differences between 3d KdS and 3d BTZ. First and foremost, unlike AdS$_3$, the KdS static patch does not have some asymptotic boundary where gravity decouples. So, our amplitude \eqref{kds-exact-spectrum} is not the partition function of a 2d CFT with a real central charge. A similar structure does arise in dS$_3$ when considering global wavefunctions which end at future infinity, where gravity also decouples. This is the context of dS/CFT \cite{Strominger:2001pn,Witten:2001kn,Anninos:2011ui}. In this case, the 3d wavefunction is a 2d CFT partition function with complex central charge. See for instance \cite{Chakravarty:2025sbg,Castro:2012gc,Collier:2025lux,Collier:2024kmo,Godet:2025bju,Cotler:2019nbi}. 

We are interested in describing physics from the perspective of an observer, in the dS static patch, with a hope of ultimately developing a static patch worldline hologram \cite{Anninos:2011af,Witten:2023xze}. In this case, the boundary at future infinity is not accessible. Gravity is fundamentally dynamical, on every 2d slice. The natural hope for a 2d holographic dual is hence not a 2d CFT with complex central charge on a fixed 2d metric, but a 2d CFT coupled to dynamical 2d gravity. This defines a string worldsheet with complex central charge. An example is the Complex Liouville string ($\mathbb{C}$LS) \cite{Collier:2024kmo}, which shall feature prominently in our observer's quantization of 3d dS.

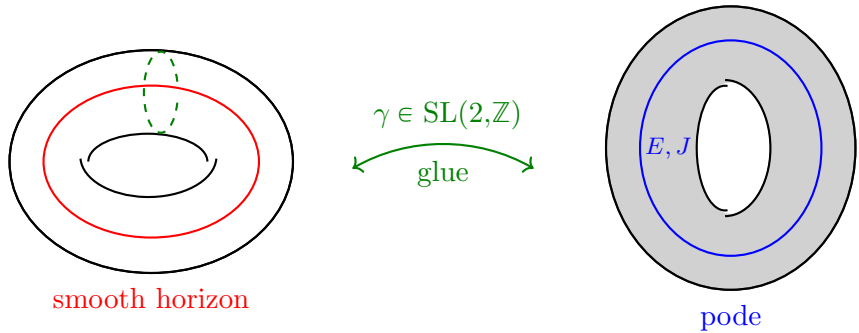
\begin{figure}[t]
\centering
\raisebox{3mm}{\begin{tikzpicture}[rotate=0,xscale=.75,yscale=-.67] 
\tikzset{
    partial ellipse/.style args={#1:#2:#3}{
        insert path={+ (#1:#3) arc (#1:#2:#3)}
    }
}   
\draw[black,thick,fill=white] (0,0) [partial ellipse=-180:180:2.5cm and 2.2cm];
\draw[black,thick] (0,0) [partial ellipse=-180:180:2.5cm and 2.2cm];
\draw[white,fill=white] (-.05,0) [partial ellipse=-180:180:1cm and .6cm];
\draw[black,thick] (-.06,0) [partial ellipse=-0:-180:1.05cm and .55cm];
\draw[black,thick] (-.05,-.1) [partial ellipse=03:177:1.2cm and .8cm];
 \draw[red,thick] [partial ellipse=0:360:1.9cm and -1.5cm];
\draw[darkgray,thick,dashed](0.17,-1.37) [partial ellipse=15:345:.3cm and .8cm];
\draw[blue] (0,2.7) node {\mbox{\textcolor{red}{smooth horizon}}};
\end{tikzpicture}}
\hspace{5mm}
 \begin{tikzpicture}[scale=.4,baseline={([yshift=-2.6cm]current bounding box.center)}]
        \draw[<->, thick,darkgray] (60:1) arc (60:120:6) node[midway, below] {glue\large\strut} node[midway, above] {\mbox{ $\gamma \in$ SL(2,$\mathbb{Z})$\large$\strut$}} ;
\end{tikzpicture}\hspace{8mm}
\begin{tikzpicture}[rotate=90,xscale=.75,yscale=-.75] 
\tikzset{
    partial ellipse/.style args={#1:#2:#3}{
        insert path={+ (#1:#3) arc (#1:#2:#3)}
    }
}   
\draw[black,thick,fill=lightgray,opacity=.71] (0,0) [partial ellipse=-180:180:2.5cm and 2.2cm];
\draw[white,fill=white] (-0.0,0) [partial ellipse=-180:180:1.15cm and .6cm];
\draw[fill=white] (0,-.1) [partial ellipse=10:170:1.2cm and .8cm];
\draw[black,thick] (0,0) [partial ellipse=-180:180:2.5cm and 2.2cm];
\draw[black,thick] (0,-.1) [partial ellipse=-0:180:1.2cm and .8cm];
\draw[black,thick] (0,-.1) [partial ellipse=05:-185:1.1cm and .5cm];
 \draw[blue, thick] [partial ellipse=0:360:1.9cm and -1.6cm];
\draw[blue] (-3,0) node {\mbox{\textcolor{blue}{pode}}};
 \draw[blue] (0,-1.1) node {\mbox{\footnotesize $E,J$}};
\draw[white,fill=white] (0,2.5) [partial ellipse=-180:180:.1cm and .05cm];
\end{tikzpicture}
\caption{The observer's density of states computed as the overlap of two 2d torus universe wavefunctions, one containing the worldline of the observer at the pode and one that represents the smooth Kerr-de Sitter horizon. One sums over possible ways of gluing these two tori together, labeled by the elements $\gamma \in$ SL(2,$\mathbb{Z}$).  \label{fig1intro}}
\end{figure}

The static patch metric near the observer does not fix the Euclidean topology. As shown in figure \ref{fig1intro} the map between the smooth Euclidean horizon region and the neighborhood of an observer worldline allows for the choice of an element $\gamma\in$ SL(2,$\mathbb{Z}$). This is analogous to the case of the Euclidean 3d BTZ black holes and underlies the MWK \cite{Maloney:2007ud,Keller:2014xba} proposal for the pure AdS$_3$ gravitational partition function, or in a more microscopic setting, the black hole Farey tail \cite{Dijkgraaf:2000fq}. Our computation is the direct de Sitter generalization of that of MWK, except that for given $(E,J)$ our gravitational path integral will just be the number $\rho(E,J)$, rather than some CFT (or gravitational) partition function.

In order to exactly quantize 3d KdS we consider two complementary proposals which reproduce the same final formula $\rho(E,J)$. These proposals are discussed in more detail in the summary \textbf{section \ref{sect:summary}}. 

\begin{figure}[t]
\centering
\raisebox{10mm}{\input{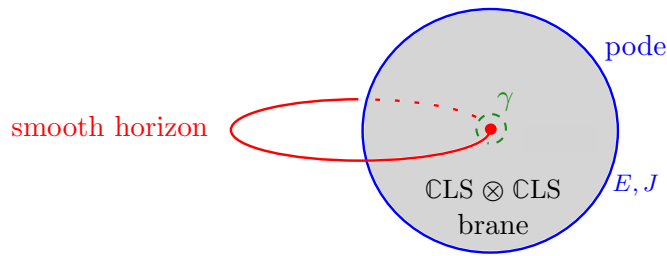}}
\caption{The observer's density of states computed as the $\mathbb{C}\text{LS} \otimes \mathbb{C}\text{LS}$ amplitude on a 2d ``brane'' anchored onto the observer's worldline. The puncture imposes generalized crosscap conditions, labeled by $\gamma \in \text{SL}(2,\mathbb{Z})$.  \label{fig2intro crosscap}}
\vspace{-.5cm}
\end{figure}

In the \textbf{first approach}, we will treat the radial direction as a Euclidean time direction, along which one applies radial quantization.\footnote{Radial quantization of torus universes was discussed in \cite{Moncrief:1989dx, Hosoya:1989sy, Carlip:1994ap,Carlip:2004ba}.} The constant $r$ slices have the topology of a torus, which we can think of as the thickened Euclidean observer's worldline. Wave functions can then be obtained by performing the gravitational path integral over the doughnut-shaped interior and exterior regions of the torus, as shown in figure \ref{fig1intro}, and are functionals (for instance) of the volume $e^\phi$ and the modulus $m$ of the torus. Physical states are labeled by complex Virasoro primaries. The modular properties of complex Virasoro blocks are known, and can in principle be used to build a dS analogue of Virasoro TQFT. We will find that the associated modular matrices (describing the transformation of torus characters), compute an exact answer for the observer's spectral density \eqref{kds-exact-spectrum}. 

In the \textbf{second approach}, we propose a holographic duality between the static patch and two copies of $\mathbb{C}$LS living on a 2d ``brane'', anchored on the Euclidean observer's worldline. In this description, the Euclidean horizon (which is topologically a circle) maps to a closed string operator insertion on the 2d brane. Different saddles, labeled by $\gamma\in$ SL(2,$\mathbb{Z}$), map to generalized crosscap closed string states $\ket{\text{C}_\gamma}$. Wavefunctions of these states and the eventual $\mathbb{C}$LS $\otimes$ $\mathbb{C}$LS amplitude follow from open-closed duality \cite{blumenhagen2009boundary}, and reproduce the aforementioned modular matrices. In turn, this $\mathbb{C}$LS $\otimes$ $\mathbb{C}$LS theory is holographically related with $G\Sigma \otimes G\Sigma$: two copies of SYK's collective field theory in the double scaling limit \cite{Blommaert:2025eps}, living on the observer's worldline. Standard SYK is related with one particular choice for $\gamma$. We thus pinpoint the precise embedding of DSSYK in 3d quantum cosmology. For earlier work on this see \cite{Narovlansky:2023lfz,Verlinde:2024znh,Susskind:2021esx,Narovlansky:2025tpb,Tietto:2025oxn,Blommaert:2025eps, Marini:2026zjk,Gaiotto:2024kze}.\footnote{Other interesting work on DSSYK includes \cite{Berkooz:2018qkz,Blommaert:2024ydx,Blommaert:2025avl,Blommaert:2023wad,berkooz2019towards,Lin:2022nss,Lin:2022rbf,Lin:2023trc,Belaey:2025ijg, vanderHeijden:2025zkr,Schouten:2025tvn,jafferis2022jt,Okuyama:2023kdo,Heller:2024ldz,Blommaert:2024whf,Blommaert:2023opb,Aguilar-Gutierrez:2025mxf,Aguilar-Gutierrez:2026nmd,Cui:2025sgy,Almheiri:2024xtw}.} We illustrate this holographic duality in figure \ref{fig2intro crosscap}.

Prior to summarizing our results, we clarify one further element of our setup. What is the definition of the observer that we consider, in terms of the gravitational path integral? To implement an observer with data $(E,J)$ (and following for instance \cite{Abdalla:2025gzn}), we introduce a dynamical single particle path integral inside the gravitational path integral. For a spinless particle one inserts the path integral:
\begin{equation}
    \text{observer}\quad \leftrightarrow\quad \oint\mathcal{D}x\,\exp\bigg(-m(E)\oint \d \tau \sqrt{g_{\mu\nu}\frac{\d x^\mu}{\d\tau}\frac{\d x^\mu}{\d\tau}}\, \bigg)
\end{equation}
with the classical relation $m(E)=(1-E)/8G$. So, we are not imposing any boundary conditions along the observer's worldline. Rather, we are fixing parameters $(E,J)$ in the dynamical action. This means we do not have to worry about various issues that arise when imposing Dirichlet or Neumann boundary conditions along a timelike tube \cite{Anninos:2022ujl,Anninos:2024wpy,An:2021fcq,anderson2008boundary}. These ``dynamical observer boundary conditions'' are closely related to inserting Wilson lines in 3d dS \cite{Castro:2020smu} via SL(2,$\mathbb{C}$) CS theory. Details are found in \textbf{section \ref{sect2.6boundaryconditions}}.

\subsection{Extended summary}\label{sect:summary}

Which geometries should we include in the gravitational path integral? We shall not try to answer this question here in full generality. However, one minimal reasonable assumption is to impose a bulk version of modular invariance.\footnote{This argument can be made more precise by defining the torus wave functions by means of a path-integral, say with fixed boundary conditions $(e^\phi,m)$, labeling respectively the volume and the modulus of the boundary torus. Tori with modulus $m$ related by SL(2,$\mathbb{Z}$) modular transformations are identical, hence the resulting wavefunction may only depend on $m$ modulo modular transformations. Whether this restriction is a fundamental part of the Lorentzian theory is less obvious, as Euclidean and Lorentzian path integrals need not per se agree. Our working assumption is that the Euclidean path-integral is the appropriate formalism for computing the spectral density. } As explained in \textbf{section \ref{sect:2.1solutions}} this is realized by summing over \textbf{Dehn fillings} of the observer's worldline's complement. These are labeled by the mapping class group elements:
 \bea
 \gamma = \biggl( \begin{matrix} a\! & b \\[-.5mm] c \! &  d \, \end{matrix}\biggr) \; \in \; \text{SL(2,}\mathbb{Z})
\eea
This leads us to decompose the observer's density \eqref{kds-exact-spectrum} as a sum over geometries pictures in figure \ref{fig1intro}:
\begin{equation}
    \boxed{\rho(E,J)
     =\sum_{\gamma \, \in\, \mathbb{Z}\backslash \TT{PSL}(2,\mathbb{Z})}\, \rho_{\gamma}(E, J)\,}\quad \label{zkds-sum}
\end{equation}
As we explain in \textbf{section \ref{sect:3torusquant}}, in radial quantization, each term in this sum is computed as the overlap between a state $\ket{E,J}$ produced by the gravitational path integral over the neighborhood of the observer worldline and a state $\ket{\text{smooth}_\gamma}$ associated with the smooth Dehn filling:\footnote{We will show that the $\gamma$ dependence of \eqref{rho-overlap} for general values of $(E,J)$ depends only on co-prime $(c,d)$. The dependence on the matrix element $a$ drops out and the value of $b$ is determined by the condition $\det(\gamma) = 1$.}
\bea
\label{rho-overlap}
\quad \boxed{\ \rho_{\gamma}(E,J) = \bra{E,J}\text{smooth}_{\gamma}\rangle \ \strut } 
\eea
For an empty static patch $(E,J)=(1,0)$, computing this overlap reduces to evaluating the gravitational no-boundary path integral of a lens spaces L$(d,c)$. More generally, we are interested in the lens space partition function in the  presence of a heavy object labeled by $(E,J)$. We call these \textbf{Kerr-lens spaces}. For standard dS $(c,d)=(0,1)$. We picture these contributions by two circles embedded in $S^3$:
\begin{equation}
    \begin{tikzpicture}[baseline={([yshift=-.5ex]current bounding box.center)}, scale=0.7]
 \pgftext{\includegraphics[scale=1]{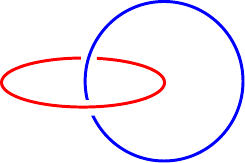}} at (0,0);
    \draw (2.65,-1.1) node {\color{blue}$(E,J)$};
    \draw (-3.8,0) node {\color{red}smooth $(c,d)$};
    \draw (0.7,1.7) node {\color{blue}pode};
  \end{tikzpicture}\hspace{-4mm}
  =\; \rho_{(c,d)}(E,J)\ \to\ e^{\frac{1}{d}\frac{A(E,J)}{4 G}+2\pi \i \frac{c}{d}\frac{J}{8G}}\,.\label{2.54rhoJ}
\end{equation}
The $\to$ on the right-hand side indicates the semiclassical approximation. We aim to compute $\rho_{(c,d)}(E,J)$ exactly in $G$. We present two different perspectives. Both are proposals. The first perspective proposes a generalization of Virasoro TFT \cite{Collier:2023fwi}. The second perspective proposes a holographic duality with two copies of $\mathbb{C}$LS \cite{Collier:2024kmo} or (equivalently) with two copies of double-scaled SYK (DSSYK) \cite{Blommaert:2024ydx,Blommaert:2025eps}.

\subsubsection*{Perspective 1. $\mathbb{V}$irasoro TFT}
Pure 3d dS quantum gravity is exactly solvable \cite{Witten:1988hc}. We derive the phase space of torus-shaped universes (relevant for our radial quantization) and the associated symplectic form in \textbf{section \ref{sect:3torusquant}}. The phase space consists of variables $(\alpha_\pm,\beta_\pm)$ with symplectic form $\Omega = ({\i}/{\hbar})(\d\alpha_+ \wedge \d\beta_+ - \d\alpha_- \wedge \d\beta_-)$, and $\alpha_+^* = \alpha_-$, $\beta_+^* = \beta_-$. The variables $\alpha_\pm$ parametrize the energy and angular momentum of the observer as follows:
\bea
\label{alphapm-def}
\boxed{E = \frac{1}{8\pi^2}\bigl(\alpha_+^2 + \alpha_-^2\bigr)\,,\quad J = - \frac{\i}{8\pi^2}\bigl( \alpha_+^2 - \alpha_-^2 \bigr)\,}
\eea
The conjugate variables $\beta_\pm$ specify the geometry at the $(c,d)$ cosmological horizon. Imposing that the Dehn filling is smooth fixes classically $\beta_\pm(\alpha_\pm,c,d)$, see equation \eqref{2.29smooth}. Notice that this condition does not mix $\pm$ variables. We find that this remains true quantum mechanically, the states $\ket{E,J}$ and $\ket{\text{smooth}_\gamma}$ factorize:
\bea
\ket{E,J}= \ket{\alpha_+}\otimes \ket{\alpha_-} \,,\quad \ket{\text{smooth}_\gamma}=\ket{\mathbb{1}_\gamma}_+\otimes \ket{\mathbb{1}_\gamma}_-\,. 
\eea
Introducing chiral wavefunctions $\mathsf{M}^\gamma_{\mathbb{1}\alpha_\pm}$ this leads to the following formula for the $(c,d)$ spectral density:
\begin{equation}
    \label{rhocd-decom}
    \rho_{(c,d)}(E,J) 
= \mathsf{M}^\gamma_{\mathbb{1}\alpha_+} \mathsf{M}^{\gamma}_{\mathbb{1}\alpha_-}\,,\quad \mathsf{M}^\gamma_{\mathbb{1}\alpha_\pm}\,=\, \langle\alpha_\pm |\mathbb{1}_\gamma\rangle_\pm\,.
\end{equation}
We propose that the exact answer for the chiral wavefunctions is found as follows. Consider the modular matrices for the transformations of the Virasoro torus characters with central charge $c_+=13+\i 3/2 G$:
\begin{equation}
    \boxed{\chi_{\alpha_+} \LL(\frac{a \tau + b}{c \tau + d} \RR) = \int_\mathcal{C} \d \beta_+ \, \mathsf{K}^\gamma_{\alpha_+\beta_+}  \,\chi_{\beta_+}(\tau)\,}\label{1.14char}
\end{equation}
The explicit expression for $\mathsf{K}^\gamma_{\alpha_+\beta_+}$ is given in equation \eqref{4.16K}. The chiral wavefunction $\mathsf{M}^\gamma_{\mathbb{1}\alpha_+}$ is obtained as the modular $S$ transform of this modular matrix, evaluated at the Virasoro identity representation: 
\begin{equation}
    \mathsf{M}^\gamma_{\mathbb{1}\alpha_+} = \int_\mathcal{C}\d \beta_+\; \mathsf{K}^{\gamma}_{\mathbb{1} \beta_+}\,\mathsf{S}_{\beta_+\alpha_+}
 \, = \, \mathsf{K}^{\gamma\cdot S\tiny\strut}_{\mathbb{1} \alpha_+}\,.
\end{equation}
The wavefunction $\mathsf{M}^\gamma_{\mathbb{1}\alpha_-}$ is obtained using a similar procedure, with details discussed in equation \eqref{g tilde definition}. In the semi-classical limit, the exact formula for $\mathsf{M}^\gamma_{\mathbb{1}\alpha_+}\mathsf{M}^\gamma_{\mathbb{1}\alpha_-}$ reduces to the GH entropy formula given in \eqref{2.54rhoJ}.

We now argue why this proposal for the exact wavefunctions is correct, by placing this discussion in a broader framework. For details, see \textbf{section \ref{sect4.1CS}}. For pure $\Lambda>0$ gravity, the only nontrivial data for a given 3d topology are SL(2,$\mathbb{C})$ holonomies (the isometry group of 3d dS) around non-contractible cycles \cite{Witten:1989ip}. The phase space associated to a Heegaard splitting of the 3 manifold on a 2d surface $\Sigma$ is related (as in AdS$_3$) with two copies of Teichmuller space, modulo the mapping class group \eqref{4.10}. The symplectic form is the sum of two copies of the (Weil-Peterson) symplectic form on Teichmuller space, except that for dS$_3$, the $1/\hbar$ prefactor in front of the symplectic form comes with an extra factor of $\pm\i$, and the phase space variables have to satisfy appropriate reality conditions.

As was shown in \cite{Cotler:2019nbi,Verlinde:2024zrh, Collier:2025lux}, correspondingly, the  Hilbert space obtained upon quantizing is the space of conformal blocks on $\Sigma$ of complex Virasoro CFT with a complex central charge $c_\pm = 13\pm \i\spc {3}/{2G}$. In this language, $\ket{\alpha_+}\otimes \ket{\alpha_-}$ would define torus blocks (characters) of complex Virasoro CFT. Equation \eqref{alphapm-def} now takes the familiar form of the parametrization of CFT scale dimension and spin, in terms of (complex) Liouville momenta, see \textbf{section \ref{sec4.1:representations}}. An extension of the familiar rules of Virasoro TQFT \cite{Collier:2023fwi} then indeed associates modular matrices \eqref{rhocd-decom} with our amplitudes of interest.\footnote{In this paper, we will not attempt to give an explicit definition of the inner product between torus characters in terms of a modulus integral. We will instead follow a more pragmatic approach by assuming that the 3D topological gluing rules apply to complex Virasoro TFT and that the Verlinde line operators are hermitian operators with respect to the inner product. In our setting, this implies delta function orthogonality for the Virasoro characters and that modular matrices act unitarily on the space of wavefunctions \cite{witten1989quantum}. We thank Lorenz Eberhardt for helpful discussion on this issue.} This is an extension to the usual surgery rules in CS theories with compact gauge groups \cite{witten1989quantum}, to the non-compact case. We may think of the Kerr-lens amplitudes \eqref{rhocd-decom} as being associated with another TFT, which one might call Complex Virasoro TFT or $\mathbb{V}$TFT \cite{Cotler:2019nbi,Verlinde:2024zrh,Collier:2025lux, Blommaert:2025eps,Witten:1989ip}. The fact that a Virasoro identity representation $\mathbb{1}$ implements smoothness of the $(c,d)$ Dehn filling is obvious from this perspective.

For the standard KdS geometry $(c,d) = (0,1)$, the state $|\mathbb{1}_{(0,1)}\rangle$ represents the block with the identity along the B-cycle, while states $\ket{\alpha_\pm}$ implements a Virasoro primary labeled by the $\mathbb{C}$LS momentum $\alpha_\pm$ along the A-cycle. So the overlap \eqref{rhocd-decom} in this case reduces the modular S-matrix element $\mathsf{S}_{\mathbb{1}\alpha_\pm}$. More generally, the wavefunction $\mathsf{M}^\gamma_{\mathbb{1}\alpha_\pm}$ is the identity matrix element of the modular matrix which maps the A-cycle to  $\gamma$ applied to the $B$-cycle.

Combining equations \eqref{zkds-sum} and \eqref{rhocd-decom}, we find our proposed answer for the exact observer's spectral density: 
\bea
\label{rho-sum}
\boxed{\rho(E,J) = \sum_{\gamma \, \in\, \mathbb{Z}\backslash \TT{PSL}(2,\mathbb{Z})}\mathsf{M}^\gamma_{\mathbb{1}\alpha_+}\mathsf{M}^\gamma_{\mathbb{1}\alpha_-}}
\eea
with $\alpha_\pm$ and $(E,J)$ are related via \eqref{alphapm-def}. This formula is the natural generalization of the MWK \cite{Maloney:2007ud,Keller:2014xba} spectral density of the BTZ black hole, in AdS$_3$. The answer for this sum is discussed in equation \eqref{1.26final} below, with more details in \textbf{section \ref{sect5.5entropy}}

\subsubsection*{Perspective 2. $\mathbb{C}$LS $\otimes$ $\mathbb{C}$LS and SYK}

The second complementary perspective is motivated by the relation between $\mathbb{C}$LS and the $G\Sigma$ collective field theory of DSSYK \cite{Blommaert:2025eps}. We make two basic observations. Firstly, the exact SYK spectral density \cite{Berkooz:2018jqr} equals exactly the (complex) Virasoro modular matrix in \eqref{1.14char} for the case $(c,d)=(-2,1)$, augmented by an additional summation over periodic images of $\alpha_+$:
\begin{equation}
    \boxed{\, \rho_\text{SYK}(\alpha_+)=\sum_{n=-\infty}^\infty\, \mathsf{M}^{\gamma(-2,1)}_{\mathbb{1},\alpha_++4\pi n} =\big(e^{\pm \i\alpha_+};\sfq \big)_\infty\,}
\end{equation}
Secondly, this SYK spectral density is computed exactly in $\mathbb{C}$LS as the string amplitude on a disk with a crosscap inserted \cite{Blommaert:2025eps}:
\bea
    \rho_{\mathbb{C}\rm LS\tiny\strut}^{(-2,1)}(\alpha_+)\, =\quad \begin{tikzpicture}[baseline={([yshift=-.95ex]current bounding box.center)}, xscale=-0.67,yscale=.67]
 \pgftext{\includegraphics[scale=1]{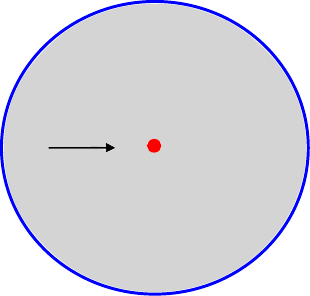}} at (0,0);
    \draw (-2.6,-2) node {\color{blue}$\alpha_+$};
    \draw (-0.6,0.6) node {\color{red}\small $\ket{\text{C}_{(-2,1)}}$};
    \draw (-3.4,1.7) node {\color{blue}\small worldline};
    \draw (-3.4,2.4) node {\color{blue}$G\spc \Sigma$};
    \draw (0,-1.8) node {\color{black}\small crosscap};
    \draw (0,-1.1) node {\color{black}\small $\mathbb{C}$LS};
    \draw (-1.1,-0.3) node {\color{black}$\tau$};
  \end{tikzpicture} 
\eea
In the closed string picture, the crosscap corresponds with a closed string state denoted $\ket{\text{C}_{(-2,1)}}$. This state satisfies crosscap Ishibashi boundary conditions. Open-closed duality fixes its wavefunction to be the identity element of the relevant complex Virasoro modular matrix \cite{blumenhagen2009boundary,Blommaert:2025eps}.

In \textbf{section \ref{sect:5cls}}, we consider the CLS disk amplitude with a closed string state inserted that we denote the \textbf{generalized crosscap} state $\ket{\text{C}_{(c,d)}}$. This state satisfies generalized Ishibashi boundary conditions:
\begin{equation}
    \boxed{\, \big(L_n - e^{-2 \pi \i \frac{d}{c}n} \bar{L}_{-n}\big)\ket{\text{C}_{(c,d)}}=0\,}
\end{equation}
Its wavefunction is determined by imposing open-closed duality \cite{blumenhagen2009boundary}, so the identity representation runs in the open channel:
\begin{equation}
    \bra{\TT{ZZ}} e^{\i \pi \tau H} \ket{\TT{C}_{(c,d)}} = \chi_{\mathbb{1}}\Bigl(- \frac{1}{c^2 \tau} + \frac{a}{c}\, \Bigr)=\int_\mathcal{C}\d \beta_+\,\,\mathsf{K}^{(c,d)}_{\; \mathbb{1}\beta_+}\, \,  \chi_{\beta_+} \Bigl(\tau-\frac{d}{c}\spc\Bigr)\,.
\end{equation}
In \textbf{section \ref{subsec: slz crosscap amplitudes}} we show that this ultimately leads to the following $\mathbb{C}$LS generalized crosscap amplitude:
\begin{equation}
    \boxed{\, \rho_{\mathbb{C}\rm LS}^{(c,d)}(\alpha_+)=\sum_{n=-\infty}^\infty\, \mathsf{M}^{\gamma(c,d)}_{\mathbb{1},\alpha_++4\pi n}\,}\label{1.21clsamp}
\end{equation}
The parameter $\alpha_+$ is implemented in $\mathbb{C}$LS by an FZZT boundary condition \cite{Fateev:2000ik,Teschner:2000md} which distinguishes $\alpha_+$ only modulo $4\pi$.

This reasoning leads us to propose an exact holographic duality between a Kerr-lens space labeled by $(c,d)$ and $\mathbb{C}$LS $\otimes$ $\mathbb{C}$LS on a 2d disk bounded by the observer's worldline, with the Dehn filling implemented by the generalized crosscap state $\ket{\text{C}_{(c,d)}}\otimes \ket{\text{C}_{(-c,d)}}$. In pictures, the $(c,d)$ Kerr-lens amplitude is represented as
\bea\label{1.22pic}
    \begin{tikzpicture}[baseline={([yshift=-.5ex]current bounding box.center)}, xscale=-0.65,yscale=.65]
 \pgftext{\includegraphics[scale=1]{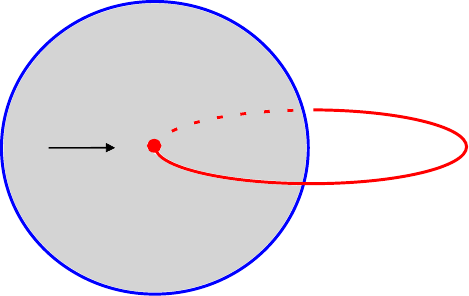}} at (0,0);
    \draw (-4.5,-2) node {\color{blue}$(\alpha_+,\alpha_-)$};
    \draw (5.8,0) node {\color{red}horizon $(c,d)$};
    \draw (-2,0.6) node {\color{red}$\ket{\text{C}_{(c,d)}}$};
    \draw (-4.8,1.7) node {\color{blue}worldline};
    \draw (-1.4,-1.9) node {\color{black}brane};
    \draw (-1.4,-1.2) node {\small \color{black}$\mathbb{C}$LS $\otimes$ $\mathbb{C}$LS};
    \draw (-2.5,-0.3) node {\color{black}$r$};
  \end{tikzpicture}\qquad \qquad \\[-2mm] \notag
\eea
where it is understood that the $\mathbb{C}$LS $\otimes$ $\mathbb{C}$LS amplitude involves the sum over gravitational saddles obtained by independently shifting $(\alpha_+,\alpha_-)$ by multiples of $4\pi$. Our proposed exact gravitational spectral density $\rho_{(c,d)}(E,J)$ for given $E$ and $J$ is of course only one term in this sum.

Asides from the observation that this proposal reproduces perspective 1, and that the semiclassical limit matches with semiclassical gravity \eqref{2.54rhoJ}, this proposal may be intuitively understood as follows. Suppose we cut the 3d KdS path integral on the 2d brane. This prepares a density matrix on the disk. By the usual intuition of CS, one expects the density matrix to have a representation as 2d CFT with complex central charge $c_\pm = 13\pm \i\spc {3}/{2G}$. Since gravity is dynamical on this 2d brane, the CFT has to appear in a generally covariant combination with total central charge $c_++c_-=26$, just like a string world sheet theory. In the case where one slices open the computation of the norm-squared of a dS$_3$ big-bang state at $\mathscr{I}_+$ the resulting string world sheet theory is $\mathbb{C}$LS \cite{Collier:2025lux}. We propose that when we slice open the Kerr-lens space as shown above in equation \eqref{1.22pic}, the theory is ``doubled'' to $\mathbb{C}$LS $\otimes$ $\mathbb{C}$LS, which each $\mathbb{C}$LS representing one chiral half of the total amplitude. We comment further on this heuristic motivation in \textbf{section \ref{sect4.2CLS}}.

In \textbf{section \ref{sect5.3gsigma}} we discuss the potential further 1d holographic interpretation of this proposal. In the case $(c,d)=(-2,1)$, the duality \cite{Blommaert:2025eps} between $\mathbb{C}$LS on a disk with a crosscap and the $G\Sigma$ collective field theory of DSSYK living on the boundary of the crosscap gets doubled. One obtains $G\Sigma \otimes G\Sigma$ as the \textbf{worldline hologram} for the Kerr-lens space KL$(-2,1)$. Indeed, the $G\Sigma \otimes G\Sigma$ theory lives on the observer's worldline, consistently with earlier proposals \cite{Anninos:2011af,Narovlansky:2023lfz,Narovlansky:2025tpb,Goto:2026ipq}. At the level of the spectral density, our proposal clarifies the exact quantum mechanical embedding of DSSYK in 3d quantum cosmology. One of the broader ambitions of our work is indeed to uncover direct hints of a microscopic realization of worldline holography, in which KdS cosmology arises as the dual of a 1d quantum many body system. A relation between SYK and KL$(-2,1)$ was already suggested in \cite{Blommaert:2025eps}. We speculate that general $(c,d)$ have SYK-like microscopic descriptions involving parafermions.

\subsubsection*{Exact spectrum}

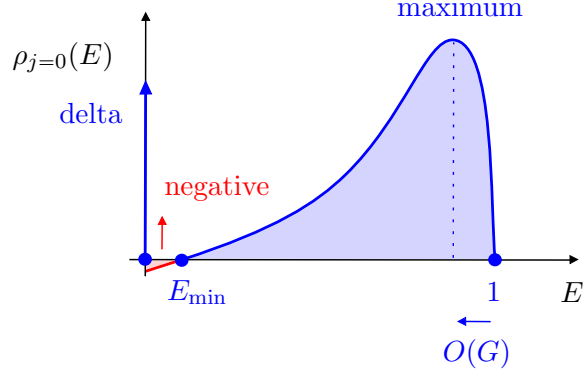
\begin{figure}[t]
    \centering
\begin{tikzpicture}[baseline={([yshift=-.5ex]current bounding box.center)}, xscale=0.74,yscale=.74]
 \pgftext{\includegraphics[scale=1]{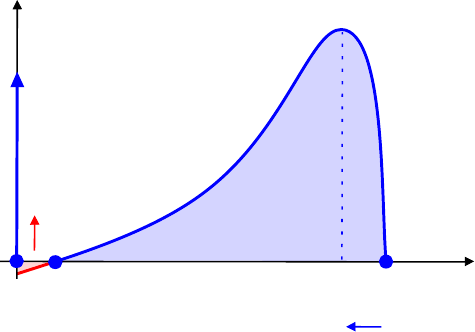}} at (0,0);
    \draw (2.2,-3.3) node {\color{blue}$O(G)$};
    \draw (1.9,2.9) node {\color{blue}maximum};
    \draw (3.9,-2.2) node {$E$};
    \draw (2.5,-2.2) node {\color{blue}$1$};
    \draw (-2.8,-2.2) node {\color{blue}$E_\text{min}$};
    \draw (-4.7,1) node {\color{blue}delta};
    \draw (-5.2,2) node {$\rho_{j=0}(E)$};
    \draw (-2.45,-0.35) node {\color{red}negative};
  \end{tikzpicture}
    \caption{Sketch of our final result for the scalar spectrum $\rho_{j=0}(E)$ where $A(E) = 2 \pi \sqrt{E}$. Empty dS space is defined as the Virasoro identity representation, which leads to a quantum correction in the energy of the maximum entropy state. The spectrum goes negative for $A(E)\to 0$. In this regime, subleading saddles compete and off-shell geometries (as well as potential other saddles) may have to be taken into account.}
    \label{fig: spectral density sketch}
\end{figure}

Motivated by these dual perspectives, we develop an exact quantization of 3d de Sitter cosmology that takes full advantage of the rigid framework of the known modular structure of the complex Liouville string worldsheet theory. Using this insight, we will find that equation \eqref{rho-sum} evaluates to the following exact expression for the Kerr-less spectral density 
\bea
   & & \qquad \qquad  \rho(E,J)\, =\, \sum_{d =1}^\infty \sum_{\substack{c = -\infty \\ (c,d) = 1}}^{+\infty}\, \rho_{(c,d)}(E,J)\,,\notag \\[-3mm]
\label{rho-final-answer}\\[0mm]\notag
    \rho_{(c,d)}(E,J) \is\!\!\! \sum_{\sigma_+,\sigma_-=\pm 1}    \frac{1}{d}\sin(\frac{\sigma_+\alpha_+ - 2 \pi c^*}{2d}) \sin(\frac{\sigma_-\alpha_- + 2 \pi c^*}{2d})\,e^{\frac{1}{d}\frac{A(E,J)}{4 G}+2\pi \i \frac{c}{d}\frac{J}{8G}} \,,
\eea
and $c^* c =1\,\text{mod}\,d$. The above proposed exact formula replaces the semiclassical expression \eqref{2.54rhoJ}.
The details and derivation of this result are found in \textbf{section \ref{sect5.5entropy}}. The sum over $c\to c+nd$ implies that spin is quantized \cite{Maxfield:2020ale}:
\begin{equation}
    j \, \equiv \, \frac{J}{8G}\, \; \in \; \mathbb{Z}.\label{spinquantization}
\end{equation}
The summation over $c$ for ever integer $j$ results in Kloosterman sums, and produces the generalization of the MWK AdS$_3$ spectral density \cite{Benjamin:2020mfz} (or Farey tail sum) to the 3d static patch observer's spectrum:
\begin{equation}
    \boxed{\ \rho_j(E) = - \!\!\! \sum_{\sigma_+,\sigma_- = \pm 1}\sigma_+\sigma_- \sum_{d=1}^\infty 
     \frac{1}{d}K\biggl(j,\mbox{\Large $\frac{\sigma_+-\sigma_-}{2}$},d\biggr)
     \cosh\biggl(\frac{\alpha_+}{d \hbar}\Bigl(2\pi\nspc +\nspc \mbox{\Large{$\frac {\i \hbar\sigma_+}{2}$}}
     \Bigr)\biggr)
     \cosh\biggl(\frac{\alpha_-}{d \hbar}\Bigl(2\pi\nspc +\nspc \mbox{\Large{$\frac{\i \hbar\sigma_-}{2}$}}\Bigr)\biggr)\ }\label{1.26final}
\end{equation}
The scalar spectrum ($j=0$) is sketched in figure \ref{fig: spectral density sketch}. The maximum is close to $E=1$ and matches (to leading order) the GH entropy of empty de Sitter, which is indeed generally expected to be the state with maximal entropy \cite{Bousso:2000nf,Chandrasekaran:2022cip}. Quantum corrections shift the maximum to $1-E\sim G$. 

A more surprising feature is that the spectrum goes \textbf{negative} at $\sqrt{E}\sim G\log(G)$. This is the regime where an observer backreacts heavily and $A(E)$ becomes small. Subleading saddles can then compete. This includes off-shell geometries (which we do not discuss in this work) as well as potential other saddles. The situation is analogous to what happens in AdS$_3$ black holes near extremality \cite{Benjamin:2019stq,Maxfield:2020ale}, where off-shell geometries have been argued to cure such negativity. For now, we interpret spectral negativity as indicating that there are in fact quantum mechanically no dS spacetimes with such small areas, similar to the conclusion found for near-BPS black holes in the spectral gap \cite{Heydeman:2020hhw}. We make some further speculative comments on the role of off-shell geometries in the concluding \textbf{section \ref{sect:6xx}}.

\subsection{Structure}

The remainder of this work is organized as follows. 

In \textbf{section \ref{sect:2.1solutions}}, we discuss the semiclassical KdS spacetime with energy $E$ and spin $J$. We describe the Euclidean solutions from the perspective of a 2d torus universe evolving from the worldline to the Euclidean horizon. We find an SL(2,$\mathbb{Z}$) family of such smooth horizons labeled by $\gamma(c,d)$.  

In \textbf{section \ref{sect:3torusquant}}, we discuss canonical quantization of the aforementioned 2d torus universes and give our no-boundary proposal as summing over SL(2,$\mathbb{Z}$) horizons. When quantization phase space, operator ordering choices have to be made. We show that the most naive torus quantization scheme reproduces the expected thermodynamics for KdS geometries with SL(2,$\mathbb{Z}$) Euclidean horizons. Later, we will use the rigid framework of 2d CFT and string theory to pinpoint the quantization scheme. We also discuss our dynamical observer boundary conditions.

In \textbf{section \ref{sect4guidance}}, we present three dual perspectives on the exact quantization of KdS: $\mathbb{V}$irasoro TFT, $\mathbb{C}$LS $\otimes$ $\mathbb{C}$LS, and DSSYK. We also find the interpretation of $(\alpha_+,\alpha_-)$ in SL(2,$\mathbb{C}$) representation theory. The summary is that we propose and develop a \textbf{holographic triality}. Schematically, the triality reads:
\begin{equation}
    \boxed{\ \text{KdS$_3$}\quad\leftrightarrow\quad \mathbb{C}\text{LS} \otimes \mathbb{C}\text{LS}  \quad \leftrightarrow\quad \text{DSSYK} \otimes \text{DSSYK}\strut\,}
\end{equation}

In \textbf{section \ref{sect:5cls}}, we give the main calculation of this work: an exact computation of the $\mathbb{C}$LS amplitude \eqref{1.21clsamp} and its anti-holomorphic counterpart, predicting the Kerr-lens spectral density $\rho_{(c,d)}(E,J)$ according to the duality \eqref{1.22pic}. We furthermore discuss the sum over $(c,d)$ resulting in the final equation \eqref{1.26final}. More details about this calculation are exiled to \textbf{appendix \ref{app:stringdetails}}.

Finally, in the discussion \textbf{section \ref{sect:concl}}, we will set the stage for establishing a microscopic hologram of 3d quantum cosmology, by presenting aspects of the duality between $\mathbb{C}$LS $\otimes$ $\mathbb{C}$LS generalized crosscap amplitudes and two SYK-like collective field theories $G\Sigma \otimes G\Sigma$ living on an observer's worldline. Details are left for future investigations.

\section{An observer's perspective on KdS} \label{sect:2.1solutions}

In this section we will describe the geometry and classical phase of 3d KdS. Our discussion will at times be somewhat detailed. We hope this will be helpful for subsequent (and possible future) calculations. In \textbf{section \ref{sect2xx}} we discuss the phase space, with emphasis on the perspective where we consider the radial Schwarzschild coordinate as a Euclidean ``time'' coordinate describing the evolution of torus-shaped 2d universes. In \textbf{section \ref{sect2KL}} we discuss the SL$(2,\mathbb{Z})$ family of smooth Euclidean horizons, shown in figure \ref{fig1intro}, from the perspective of this torus phase space. We call these spacetimes Kerr-lens spaces: KL$(c,d)$.

\subsection{KdS as torus cosmology}\label{sect2xx}
Consider the Lorentzian Kerr-de Sitter spacetime that describes an object with an energy $E$ and spin $J$ in three dimensional de Sitter gravity \cite{gibbons1977cosmological,deser1984three,Park:1998qk}:
\bea
    \d s^2=-N^2\d t^2+\frac{\d r^2}{N^2}+r^2\biggl(\d \varphi-\frac{J \d t}{ 2 r^2}\biggr)^2\,,\quad N^2=E-r^2+\frac{J^2}{4 r^2}\,.\label{2.1metric}
\eea
The empty dS static patch corresponds with $(E,J)=(1,0)$. In the case without spin ($J=0$) and with positive energy (above vacuum) $0<E<1$, this indeed describes a massive object in the center of the static patch, with the following Penrose diagram (with the periodic $\varphi$ direction suppressed):
\begin{equation}
    \begin{tikzpicture}[baseline={([yshift=-.5ex]current bounding box.center)}, scale=0.95]
    \pgftext{\includegraphics[scale=1]{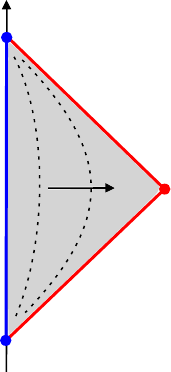}} at (0,0);
    \draw (-1.9,1.5) node {\color{blue}$E$};
    \draw (-2.5,0.7) node {\color{blue}worldline};
    \draw (-2.2,-0) node {\color{blue}pode};
    \draw (-1.9,-1) node {\color{blue}$r_-$};
    \draw (2.4,0) node {\color{red}horizon};
    \draw (1.1,-1) node {\color{red}$r_+$};
    \draw (1,1.8) node {static patch};
    \draw (-0.3,-0.3) node {$r$};
    \end{tikzpicture}
\end{equation}
The periodic identifications of the Euclidean time coordinate $\i t$ and the angular coordinate $\varphi$ determine the inverse temperature $\beta$ and the angular potential $\theta$ of the spacetime:
\begin{equation}
    (\i t,\varphi)\sim (\i t,\varphi+2\pi)\sim (\i t+\beta,\varphi+\theta)\,.\label{2.4periods}
\end{equation}
The variables $E$ and $J$ uniquely specify the spatial metric of the KdS spacetime and, as we will discuss later, they can be completed into two pairs of conjugate phase space variables $(E, T)$ and $(J,\theta)$. Indeed, starting  from equation \eqref{2.1metric}, one may ask why we call $E$ energy and why we call $J$ spin? They deserve these names, because they are conically conjugate to Lorentzian time $T$ and angular potential $\theta$, which appear in the identifications of the metric \eqref{2.4periods}. Evaluation of the on-shell action in section \ref{sect:3.1kdstoruscosmology} indeed reveals the following symplectic form \cite{Park:1998qk,Brown:1994gs}:
\bea
    \omega = \frac{1}{8 G}\d E\wedge \d T+\frac{1}{8G}\d J\wedge \d \theta\,,\quad \beta=\i T\,.\label{2.7omega}
\eea
Phase space is the space of real solutions, which here means that $(E,J,T,\theta)$ are real.\footnote{That Lorentzian time $T$ is to be understood as a real variable upon canonically quantizing gravity was for instance also used in the quantization of JT gravity \cite{Harlow:2018tqv} and sine dilaton gravity \cite{Blommaert:2024ydx}.} One can make the conjugate phase space coordinates $(T,\theta)$ explicit in the metric, by introducing torus time-and space coordinates $(x,y)$ with unit periodicity: 
\begin{equation}
\varphi=2\pi x+\theta y\,, \quad t=T y\,, \quad     (x,y)\sim (x+1,y)\sim (x,y+1)\,.
\end{equation}

The lapse function $N$ can be conveniently rewritten as a function of two radial coordinates $(r_+,r_-)$: 
\bea
    N^2=\frac{(r^2-r_-^2)(r^2_+-r^2)}{r^2}\,,\ & & \  r^2_\pm=\, \frac{E\pm \sqrt{E^2+J^2}}{2}\,.\label{2.4r}
\eea
Note that $r_-$ is imaginary. Often we study this spacetime for $0<r<r_+$, with $r_+$ the radial location of the cosmological horizon. This contour is subtle as the spacetimes are fundamentally singular at $r=0$. We find it more natural and convenient to consider these spacetimes with the complex radial contour:
\bea
    r^2=\frac{r_+^2+r_-^2}{2}-\cos(2\tau)\frac{r_+^2-r_-^2}{2}\,,\quad 0<\tau<\frac{\pi}{2}\,,\quad r(0)=r_-\,,\quad r(\pi/2)=r_+\,. \label{2.5contour}
\eea
The points $r_\pm$ are similar to the inner-and outer horizons for the BTZ black hole. This contour avoids the singular point at $r=0$. The line $r=r_-$ is the observer's classical worldline; we call this the pode.\footnote{The volume of the slice $r\to r_-$ vanishes, meaning this actually describes a one-dimensional worldline, unlike the $r\to 0$ slice which has nonzero area $\sqrt{h}=J\pi\beta$.} Notice:
\bea
\tau\rvert_\text{pode}=0\,,\quad  \tau\rvert_\text{horizon}={\pi}/{2}\,,\quad \sqrt{h}\rvert_\text{pode}=\sqrt{h}\rvert_\text{horizon}=0\,.\label{2.9rooth}
\eea
The metric along the contour \eqref{2.5contour} takes the following form:
\bea
    \d s^2\is \d \tau^2+\frac{E-\sqrt{E^2+J^2}\cos(2\tau)}{2}\d\varphi^2-\frac{E+\sqrt{E^2+J^2}\cos(2\tau)}{2}\d t^2-J\d t\,\d\varphi\,,\quad 0<\tau<\frac{\pi}{2}\,.\ \ \ \label{2.6metric}
\eea
A Euclidean time identification \eqref{2.4periods} makes the pode a circle. The horizon is also topologically a circle as $\varphi\sim \varphi+2\pi$. Slightly thickening either the pode or the horizon shows that fixed $\tau$ slices have a torus topology. We are interested in describing the evolution of the torus shaped universe, as function of $\tau$.\footnote{Evolution of torus universes in dS$_3$ was also discussed for instance by \cite{Godet:2025bju}. However, in that case the torus was homologous to future infinity rather than (the Euclidean continuation of) an observer's worldline as we consider here.}

To build some intuition, consider the non-rotating classical Schwarzschild-de Sitter spacetimes with $J=0$. The metric \eqref{2.6metric} becomes:
\bea
    \d s^2=\d \tau^2+E\sin(\tau)^2\d \varphi^2-E\cos(\tau)^2\d t^2\,,\quad \sqrt{h}=\frac{E\beta}{2}\sin(2\tau)\,,\quad 0<E<1\,.
\eea
In Euclidean time ($\i t$) this describes an $S^3$ with two conical defects. The Euclidean configuration and the Lorentzian Penrose diagram (including the $\varphi$ angular direction) look as follows:
\begin{equation}
    \raisebox{1mm}{\begin{tikzpicture}[baseline={([yshift=-.5ex]current bounding box.center)}, scale=0.69]
    \pgftext{\includegraphics[scale=1]{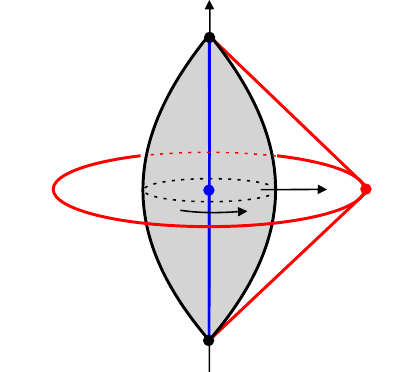}} at (0,0);
    \draw (-0.3,1) node {\color{blue}$E$};
    \draw (-1.5,2) node {$r<r_\text{c}$};
    \draw (4.2,0) node {\color{red}horizon};
    \draw (-2.9,-0.1) node {\color{red}$r_+$};
    \draw (0.7,-1.2) node {\color{blue}$r_-$};
    \draw (1.8,-0.3) node {$r$};
    \end{tikzpicture}}\quad \to \quad \raisebox{-2mm}{\begin{tikzpicture}[baseline={([yshift=-.5ex]current bounding box.center)}, scale=0.62]
\pgftext{\includegraphics[scale=.96]{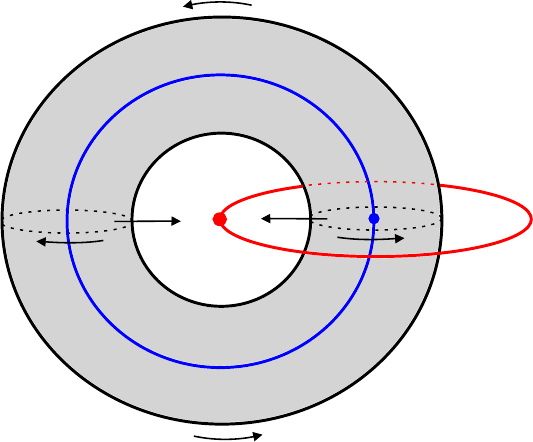}} at (0,0);
    \draw (-0.8,2) node {\color{blue}$E$};
    \draw (-0.8,2.9) node {\color{blue}pode};
    \draw (5.8,0.1) node {\color{red}horizon};
    \draw (4.7,2.5) node {$r<r_\text{c}$ doughnut};
    \draw (5.8,-0.6) node {\color{red} torus};
    \draw (-0.7,-4) node {$\i t$};
    \draw (0.4,-0.2) node {$r$};
    \draw (-2.95,-0.65) node {$\varphi$};
\end{tikzpicture}}\label{2.11torusuniverseV1}
\end{equation}
At the pode $\tau=0$ one finds a defect with opening angle $\gamma_\text{pode}=2\pi\sqrt{E}$. Near the horizon we find an opening angle $\gamma_\text{horizon}=\sqrt{E}\beta$. The metric at the pode and horizon degenerates to lines $ \d s\rvert_\text{pode}=\sqrt{E}\,\i\d t$ and $\d s\rvert_\text{horizon}=\sqrt{E}\,\d \varphi$. Fixed $\tau$ slices have a torus shape, degenerating to lines at the pode and horizon. The torus cycle $S^1$ that degenerates to a point (reducing the torus to a circle) is different at both ends. At the pode, the $\varphi$ direction degenerates; whereas at the horizon the Euclidean time $\i t$ direction does.

\subsubsection*{Parameterization 1. Holonomy variables}

As a preparation for the quantum description of the torus spacetimes, it is useful to introduce another set of phase space variables $(\alpha_\pm,\beta_\pm)$ related to the original coordinates as follows:
\begin{equation}
    \boxed{E = \frac{1}{8\pi^2}\bigl(\alpha_+^2 + \alpha_-^2\bigr)\,,\quad J = - \frac{\i}{8\pi^2}\bigl( \alpha_+^2 - \alpha_-^2 \bigr)\, }\label{2.8EJ}
\end{equation}
These equations may look familiar from the point of view of SL(2,$\mathbb{C}$) representation theory and dS/CFT \cite{Anninos:2011af,Collier:2025lux,Collier:2024kmo,Strominger:2001pn}. See section \ref{sec4.1:representations}. Furthermore, angular velocity and real time are related with $(\alpha_\pm,\beta_\pm)$ as:
\begin{equation}
    \frac{\theta}{\pi}=\frac{\beta_-}{\alpha_-}+\frac{\beta_+}{\alpha_+}\,,\quad \frac{T}{\pi}=\i \frac{\beta_-}{\alpha_-}-\i\frac{\beta_+}{\alpha_+}\,.\label{2.15thetabeta}
\end{equation}
One checks that the transformation \eqref{2.8EJ}/\eqref{2.15thetabeta} to $(\alpha_\pm,\beta_\pm)$ is a canonical transformation:
    \begin{equation}
        \omega=-\frac{\i}{16\pi G}\d \beta_+\wedge \d \alpha_++\frac{\i}{16\pi G}\d \beta_-\wedge \d \alpha_-\,.
    \end{equation}
The symplectic form and reality of $(E,J,T,\theta)$ imply a reality conditions on the new variables $(\alpha_\pm,\beta_\pm)$:
\begin{equation}
    \beta_+^\star=\beta_-\,,\quad \alpha_+^\star=\alpha_-\,.\label{2.9reality}
\end{equation}
In terms of these new variables, the KdS metric \eqref{2.6metric} then takes the following form (familiar from \cite{Carlip:2004ba}):
\bea\label{2.14classical metric}
    \d s^2\is \d \tau^2 + \frac{1}{4}(\alpha_+^2 + \alpha_-^2 - 2 \alpha_+ \alpha_- \cos(2 \tau))\, \d x^2 + \frac{1}{4}(\beta_+^2 + \beta_-^2 - 2 \beta_+ \beta_- \cos(2 \tau) )\, \d y^2\qquad \qquad \\
    & & \qquad\qquad \qquad\qquad \qquad \qquad  \quad + \frac{1}{2}(\alpha_+ \beta_+ + \alpha_- \beta_- - (\alpha_+ \beta_- + \alpha_- \beta_+) \cos(2 \tau) )\, \d x\, \d y\,.\nonumber
\eea
The locations $r_+$ and $r_-$ of the horizon and the observer as well as the horizon area are expressed as:
\bea
    r_\pm = \frac{\alpha_+ \pm \alpha_-}{4 \pi}\,,\quad A(E,J)=2\pi r_+\,.\label{2.16Ar}
\eea
The holonomy variables $(\alpha_+,\alpha_-)$ uniquely determine the spatial geometry. 

\subsubsection*{Parameterization 2. Modular variables}

Our torus perspective on KdS motivates one extra parametrization using the modular parameter $\tau_{\rm torus}$ and the volume $e^\phi$ of constant $\tau$ slices. We rewrite equation \eqref{2.14classical metric} in a way which makes manifest that we are describing the $\tau$ evolution of a torus-topology universe:
\bea
    \boxed{\ \d s^2=\d\tau^2+e^\phi\,\frac{ \abs{\d x + (m_1 + \i m_2) \d y}^2}{m_2} \small{}_{\strut}^{\strut} \ } \quad x \sim x+1 \,, y \sim y +1\,.\label{2.13torus}
\eea
This describes the evolution of some torus $(x,y)$, with the following modular parameter and volume:\footnote{The expressions for the individual moduli $m_1$ and $m_2$  are less instructive:
\bea
\label{2.xxm1m2}
    m_1    = \frac{\beta_+ \alpha_+ + \beta_- \alpha_- - \cos(2\tau)(\beta_+ \alpha_- + \beta_- \alpha_+)}{\alpha_+^2 + \alpha_-^2 - 2 \alpha_+ \alpha_- \cos(2\tau)}  \,, \quad 
    m_2   = \frac{\sin(2 \tau)(\beta_+ \alpha_- - \beta_- \alpha_+)}{\alpha_+^2 + \alpha_-^2 - 2 \alpha_+ \alpha_- \cos(2\tau)} 
    \,.
\eea
Notice that $m_2$ is imaginary according to the reality conditions \eqref{2.9reality}.}
\begin{equation}
    \tau_\text{torus}=m_1+\i m_2=\frac{\beta_+e^{-\i\tau}-\beta_-e^{\i\tau}}{\alpha_+e^{-\i\tau}-\alpha_-e^{\i\tau}}\,,\quad e^{\phi} =  (\alpha_- \beta_+ - \alpha_+ \beta_-) \sin(2 \tau)/4\,,\quad e^\phi\rvert_\text{pode}= e^\phi\rvert_\text{horizon}=0\,.\label{2.21tau}
\end{equation}
As advertised in equation \eqref{2.9rooth} the volume vanishes at the pode and the horizon, but remains finite for other radial times. Notice for future reference
\begin{equation}
    \tau_\text{torus}\rvert_\text{pode}=\frac{\beta_+-\beta_-}{\alpha_+-\alpha_-}\,,\quad \tau_\text{torus}\rvert_\text{horizon}=\frac{\beta_++\beta_-}{\alpha_++\alpha_-}\,.\label{2.24taubdy}
\end{equation}
In section \ref{sect:3.1kdstoruscosmology} we will find it convenient to view fixed $\tau$ slices as Cauchy slices in canonical quantization. In this radial quantization, the ``torus universe'' \eqref{2.13torus} starts at the pode $\tau=0$ as a line (where $e^\phi=0$), then grows and contracts again, until we reach the horizon where again it degenerates to a line ($e^\phi=0$). Is this horizon smooth? We discuss this next.

\subsection{SL(2,$\mathbb{Z}$) horizons and Kerr-lens spacetimes}\label{sect2KL}

In the embedding 3d space, a torus can shrink to a line at the horizon in a smooth way, or in a singular way. If this were an AdS$_3$ black hole, we would impose that the Euclidean horizon is smooth. We first discuss what such a smooth horizon means for the case $J=\theta=0$. According to equation \eqref{2.8EJ}/\eqref{2.15thetabeta}, such spinless configurations have $\alpha_+=\alpha_-$ and $\beta_+=-\beta_-$, such that the metric \eqref{2.14classical metric} reads:
\begin{equation}
   \d s^2=\d \tau^2+\alpha_+^2\sin(\tau)^2\d x^2+\beta_+^2 \cos(\tau)^2\d y^2\,.\label{2.26smoothmetric}
\end{equation}
The horizon $\tau\to \pi/2$ in this metric is described by a conical defect \cite{Louko:1995jw} in the $(y,\tau)$ plane with opening angle $\beta_+$. Imposing that the Euclidean horizon is smooth (that the defect has opening angle $2\pi$) fixes: 
\begin{equation}
    \beta_{+\,\text{smooth}}=2\pi\,,\quad \beta_{-\,\text{smooth}}=-2\pi\,.\label{2.27smooth}
\end{equation}
Notice that fixing $(\beta_+,\beta_-)$ completely determines the geometry in the $(\tau,y)$ plane, according to equation \eqref{2.14classical metric}. Hence, equation \eqref{2.27smooth} is sufficient to impose smoothness at the horizon for general mass and spin $(E,J)$. Equation \eqref{2.27smooth}, combined with \eqref{2.8EJ}/\eqref{2.15thetabeta} reproduces known thermodynamic relations for KdS black holes \cite{Bousso:2001mw}, see equation \eqref{2.89}. Notice that the smoothness condition \eqref{2.27smooth} fixes
\bea
\tau_\text{torus}\rvert_\text{horizon}=0\,.\label{2.28KdSsmooth}
\eea
Enforcing smoothness at the pode would select $(E,J)=(1,0)$ such that $\alpha_+=\alpha_-=2\pi$ and $\tau\rvert_\text{pode}=0$. We will henceforth however generally consider $(E,J)\neq (1,0)$.

Crucially, following the BTZ discussion, we will explain in detail that there is in fact an SL(2,$\mathbb{Z}$) family of manners of imposing Euclidean smoothness away from the pode. We claim that the generalization of the smoothness condition \eqref{2.27smooth} reads
\bea
    \boxed{\ \beta_+= \frac{1}{d}\, 2\pi+\frac{c}{d}\, \alpha_+ \,,\quad \beta_-=-\frac{1}{d}\, 2\pi+\frac{c}{d}\, \alpha_- \, }\label{2.29smooth}
\eea
Here $(c,d)$ are integers. With these conditions one obtains in equation \eqref{2.24taubdy}\footnote{Conversely, we can imagine demanding $\tau_\text{torus}\rvert_\text{horizon}=\frac{c}{d}$. This indeed fixes the horizon metric to \eqref{2.34dshor}. Solutions allow general $\beta_+=-\beta_-$ in equation \eqref{2.41contract}. When $\beta_+\neq 2\pi$, it describes a defect inserted on the horizon. This is visible already for $(c,d)=(0,1)$ in equation \eqref{2.26smoothmetric}.}
\begin{equation}
    \tau_\text{torus}\rvert_\text{horizon}=\frac{c}{d}\,.\label{2.30tausmooth}
\end{equation}
The $J=0$ metrics \eqref{2.14classical metric} subject to the smoothness condition \eqref{2.29smooth} may be rewritten as quotients of the smooth SdS spacetime \eqref{2.26smoothmetric}, for which $(c,d)=(0,1)$:
\bea \label{periodicity kerr-L(p,q)}
    \d s^2 =\d \tau^2 +\alpha_+^2 \sin(\tau)^2 \d x^2 +  4 \pi^2 \cos(\tau)^2 \d y ^2\,,\quad (x, y) \sim (x+1,y) \sim (x+c/d,y+1/d)\,.
\eea 
The second identification implements a $\mathbb{Z}_d$ quotient. This discrete group acts freely on the $(x,y)$ plane and the $(c,d)$ quotient geometry is hence also smooth. For $\alpha_+ = \alpha_- = 2\pi$, in which case KdS becomes empty S$^3$, this quotient defines the smooth lens spaces L$(d,c)$ \cite{Castro:2011xb}. Including spin, the smooth metrics satisfying the constraint \eqref{2.29smooth} are:\footnote{To argue that this defines a smooth horizon, we perform the change of coordinate $\tau_\text{hor}=\pi/2-\tau$ in \eqref{2.14classical metric}, and rewrite the near horizon metric in the form \eqref{2.1metric} of a static patch near a worldline with energy/spin $(E,J)$, but with the horizon playing the role of the worldline. Motivated by \eqref{2.33Stori}, we introduce coordinates $x_\text{hor}=y$, $y_\text{hor}=-x$. The metric \eqref{2.14classical metric} then becomes
\begin{align}
    \d s^2=\d \tau_\text{hor}^2 + \frac{1}{4}(\beta_+^2 + \beta_-^2 + &2 \beta_+ \beta_- \cos(2 \tau_\text{hor}) )\, \d x_\text{new}^2 + \,\frac{1}{4}(\alpha_+^2 + \alpha_-^2 + 2 \alpha_+ \alpha_+ \cos(2 \tau_\text{hor})) \d y_\text{hor}^2\label{2.57snew}\\&\qquad \qquad\quad + \frac{1}{2}(\beta_+ \alpha_+ + \beta_- \alpha_- + (\beta_+ \alpha_- + \beta_- \alpha_+) \cos(2 \tau_\text{hor}) )\, \d x_\text{hor}\, \d y_\text{hor}\,.\nonumber
\end{align}
This is the usual spacetime \eqref{2.14classical metric} with parameters $\alpha_{\pm\,\text{hor}}=\pm \beta_\pm$ and $\beta_{\pm\,\text{hor}}=\pm \alpha_\pm$. The substitution $\varphi=2\pi x_\text{hor}$ and $t=Ty_\text{hor}$ brings the spacetime \eqref{2.57snew} into the form describing a KdS static patch \eqref{2.1metric}
\begin{equation}
    \d s^2=-N^2\d t^2+\frac{\d r^2}{N^2}+r^2(\d \varphi-J_\text{hor} \d t/ 2 r^2)^2\,,\quad N^2=E_\text{hor}-r^2+\frac{J_\text{hor}^2}{4 r^2}\,,
\end{equation}
with modified energy/spin equations $E_\text{hor}= (\beta_+^2+\beta_-^2)/8\pi^2$ and $J_\text{hor} =-{\i}({\beta_+^2}-{\beta_-^2})/8\pi^2$,
see equation \eqref{2.8EJ}. A smooth euclidean horizon  has a trivial defect with mass and spin  $(E_\text{hor},J_\text{hor})=(1,0)$. This implies our previously claimed smoothness condition $\beta_\pm=\pm2\pi$, but now for general energy/spin $(E,J)$. The argument for general $(c,d)$ runs similarly. One performs a coordinate transformation of the type \eqref{2.36z}/\eqref{2.37w} with $(w,z)$ replacing the roles of $(x_\text{hor},y_\text{hor})$, bringing the metric in the static patch form \eqref{2.1metric}, and impose the absence of a particle.}
\begin{equation}
    \d s^2 = \d \tau^2 +\frac{1}{4}(\alpha_++\alpha_-)^2 \sin(\tau)^2 \d x^2 +  4 \pi^2 \cos(\tau)^2 (\d y +(\alpha_+-\alpha_-)\d x/2)^2\,,\label{2.30xx}
\end{equation}
subject to a quotient $(x, y) \sim (x+1,y) \sim (x+c/d,y+1/d)$. We call these spacetimes Kerr-lens spaces.

An insightful geometric construction of these solutions is obtained as follows. Consider the smooth Euclidean KdS solution, with $(c,d)=(0,1)$. Slice the geometry at an intermediate radius $\tau=\tau_\text{c}$. The regions $\tau<\tau_\text{c}$ and $\tau>\tau_\text{c}$ are solid doughnuts with identical torus boundary metrics. The space $\tau<\tau_\text{c}$ contains an $(E,J)$ particle running along the $y$ direction. The doughnut $\tau>\tau_\text{c}$ is characterized by the smoothness constraint \eqref{2.27smooth}: a circle running around the $y$ direction contracts smoothly. Patching the doughnuts together creates the usual KdS Euclidean spacetime, shown in equation \eqref{2.11torusuniverseV1}:
\begin{equation}
    \begin{tikzpicture}[baseline={([yshift=-.5ex]current bounding box.center)}, scale=0.9]
    \pgftext{\includegraphics[scale=1]{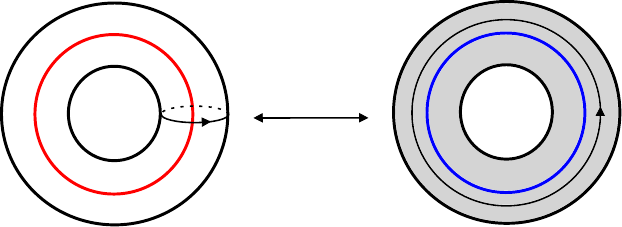}} at (0,0);
    \draw (0.7,1) node {\color{blue}pode};
    \draw (-7,1) node {\color{red}smooth horizon};
    \draw (-3.3,2.3) node {$r_\text{c}<r<r_+$};
    \draw (3.3,2.3) node {$r_-<r<r_\text{c}$};
    \draw (0,-0.6) node {glue};
    \draw (-1.8,-0.6) node {$y$};
    \draw (4.35,-0) node {$y$};
    \draw (-3.3,-2.4) node {\color{red}$(c,d)=(0,1)$};
    \end{tikzpicture}\label{2.33Stori}
\end{equation}
The complement of the gray doughnut in \eqref{2.11torusuniverseV1} in S$^3$ is indeed the white doughnut $\tau>\tau_\text{c}$. Such torus gluing were nicely explained for instance in \cite{witten1989quantum}.

The other $(c,d)$ smooth metrics \eqref{2.30xx} correspond with replacing the white doughnut with a different ``Dehn filling'' of the torus boundary conditions at $\tau>\tau_\text{c}$. These smooth solutions are very similar to the SL(2,$\mathbb{Z}$) family of Dehn fillings studied in \cite{Maloney:2007ud}. Here we are considering $\Lambda>0$ and study Dehn fillings of the Euclidean continuation of an observer's worldline. Picking a Dehn filling corresponds to letting a specific one-cycles of the torus be contractible, for instance we can choose the $y$-cycle to wind $(c,d)$ times around the $(A,B)$ cycles around the boundary of the doughnut:  \cite{Belin:2026pko}
\begin{equation}
    \begin{tikzpicture}[baseline={([yshift=-.5ex]current bounding box.center)}, scale=0.9]
    \pgftext{\includegraphics[scale=1]{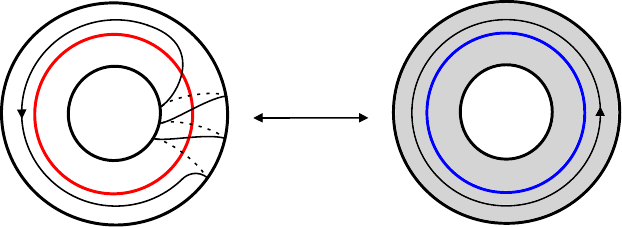}} at (0,0);
    \draw (0.7,1) node {\color{blue}pode};
    \draw (-7,1) node {\color{red}smooth horizon};
    \draw (-3.3,2.3) node {$r_\text{c}<r<r_+$};
    \draw (3.3,2.3) node {$r_-<r<r_\text{c}$};
    \draw (0,-0.6) node {glue};
    \draw (-4.35,-0) node {$y$};
    \draw (4.35,-0) node {$y$};
    \draw (-3.3,-2.4) node {\color{red}$(c,d)=(1,3)$};
    \end{tikzpicture}\label{2.34dehn}
\end{equation}
Imposing the smoothness condition \eqref{2.29smooth} implies that the (red) horizon has no defect. We now explain the map between this picture, and equation \eqref{2.30tausmooth}/\eqref{periodicity kerr-L(p,q)}. In equation \eqref{periodicity kerr-L(p,q)}, the metric at the horizon takes the form:
\begin{equation}
    \frac{1}{\alpha_+}\d s\rvert_\text{horizon}=\d x+\frac{c}{d}\, \d y\,\quad \to \quad \tau_\text{torus}\rvert_\text{horizon}=\frac{c}{d}\,.\label{2.34dshor}
\end{equation}
This torus modular parameter indeed is read off from equation \eqref{2.13torus}. With this metric, which $S^1$ cycle in the $(x,y)$ plane is smoothly contractible in the doughnut \eqref{2.34dehn}? We may introduce the coordinate along the height (A cycle) of the white doughnut:
\begin{equation}
    z=\frac{d}{d^2+c^2}x+\frac{c}{d^2+c^2}y\,,\quad z\sim z+1\,.\label{2.36z}
\end{equation}
The contractible cycle runs on the waist (B cycle) or the doughnut and is orthogonal to the $z$ coordinate. This means $\d z=0$, or $\d x/c=-\d y/d$. The associated (unit period) coordinate along this curve reads:
\begin{equation}
    w=-\frac{c}{d^2+c^2}x+\frac{d}{d^2+c^2}y\,,\quad w\sim w+1\,.\label{2.37w}
\end{equation}
Equations \eqref{2.36z}/\eqref{2.37w} show that indeed, the $y$ $S^1$ curve winds $d$ times around the $B$ cycle and $c$ times around the $A$ cycle, as anticipated in the picture \eqref{2.34dehn}: $y=dw+cz$.

Before proceeding we make several comments.
\begin{enumerate}
    \item Oftentimes, the Dehn fillings \eqref{2.34dehn} are phrases as acting with a mapping class group element on one of the two tori at $\tau_\text{c}$, prior to gluing. The torus mapping class group is SL(2,$\mathbb{Z}$). Say
    \begin{equation}
    \boxed{\gamma={\small\begin{pmatrix}
        \,a &\! b\, \\[-.5mm]
        \, c &\! d\,
    \end{pmatrix}}  \,}\quad ad-bc=1\,,\quad \mathsf{S}  = {\small \begin{pmatrix}
       \, 0 \!& \! -1 \\[-.5mm]
       \, 1\! & 0
    \end{pmatrix}} \,.
    \end{equation}
    The relevant SL(2,$\mathbb{Z}$) element that matches our previous equations acts on $\tau_\text{torus}\rvert_\text{horizon}$ as follows:
    \begin{equation}
        \gamma\cdot \mathsf{S}\cdot \tau_\text{torus}\rvert_\text{horizon}= \frac{-b\,\tau_\text{torus}\rvert_\text{horizon}+a}{-d\,\tau_\text{torus}\rvert_\text{horizon}+c}=\infty\quad \to \quad \tau_\text{torus}\rvert_\text{horizon}=\frac{c}{d}\,.\label{2.40taumap}
    \end{equation}
    The $\mathsf{S}$ matrix exchanges $A$ and $B$ cycles. It is the relevant SL(2,$\mathbb{Z}$) element to reproduce ordinary KdS, see equation \eqref{2.33Stori}. The smoothness condition on $\tau_\text{torus}\rvert_\text{horizon}$ \eqref{2.30tausmooth} follows from imposing that the modular transformed $\tau_\text{torus}\rvert_\text{horizon}$ equals $+\infty$, meaning it becomes a line running along the (modular transformed) $y$ direction. In the KdS case, the $\mathsf{S}$ transform means the horizon torus becomes a circle running along the $x$ direction. This maps $+\infty$ to $0$, reproducing $\tau_\text{torus}\rvert_\text{horizon}=0$. 
    \item Consider the identifications $(x, y) \sim (x+1,y) \sim (x+c/d,y+1/d)$ of the Kerr-lens space \eqref{2.30xx}. We see that $c\to c+n d$ generated the same identification. Our final quantum amplitude $\rho_{(c,d)}(E,J)$ does not respect this symmetry. Rather, summing over $c\to c+n d$ implements spin quantization \eqref{spinquantization}, as in AdS$_3$ \cite{Maxfield:2020ale}.  More generally we can ask which $\gamma$ must be summed over. The answer is:
     \begin{equation}
         \gamma\in \mathbb{Z} \backslash \TT{PSL}(2,\mathbb{Z})\,.
     \end{equation}
     The $\mathbb{Z}$ quotient acts by left multiplication of
     \begin{equation}
         \mathsf{T}  = {\small \begin{pmatrix}
       \, 1 \!& \! 1 \\[-.5mm]
       \, 0\! & 1
    \end{pmatrix}} \,,
     \end{equation}
     sending $(a,b)\to (a+c,b+d)$. Neither $a$ nor $b$ appear in our geometric construction, so that this should be a redundancy. Our exact amplitude \eqref{5.42final} indeed respects this redundancy. Note that right multiplication of $\gamma\cdot \mathsf{S}$ by $\mathsf{T}$ sends $(a,c)\to (a+b,c+d)$ thus implementing spin quantization as mentioned above. The metric \eqref{2.30xx} for $d<0$ is equivalent to $d>0$ upon changing coordinates $(x,y)\to (-x,-y)$. Summarizing, schematically:
     \begin{equation}
         \boxed{\ \sum_\gamma\quad \to\sum_{d =1}^\infty \sum_{\substack{c = -\infty \\ (c,d) = 1}}^{+\infty}\; }\label{sumreplace}
     \end{equation}
     \item Changing coordinates $x \to -x$ in the metric \eqref{2.30xx} shows the equivalence
     \begin{equation}
         \alpha_+\leftrightarrow \alpha_-\,,\quad c\to -c\,.\label{sym from diff}
     \end{equation}
     The condition $\det(\gamma)=1$ also requires $b\to -b$. In terms of energy and spin: flipping $c$ flips spin. So our final answers should exhibit this symmetry.
     
    \item dS$_3$ gravity has a formulation as SL(2,$\mathbb{Z}$) CS theory \cite{Witten:1988hc}, which we discuss in section \ref{sect4.1CS}. In this language, $(\alpha_\pm,\beta_\pm)$ label monodromies of gauge fields when going around respectively the $x$ and $y$ circles. The contractible curve around the horizon winds $d$ times along the $y$ direction and $-c$ times around the $x$ direction. The monodromy picked up along this $w$ cycle is thus:
    \bea
        \beta_{\pm}^\text{contractible}\is d\beta_\pm - c \alpha_\pm\,.\label{2.41contract}
    \eea
    Smoothness is a property of the contractible cycle, and imposes according to \eqref{2.27smooth}:
    \begin{equation}
        \beta_{+\,\text{smooth}}^\text{contractible}=2\pi\,,\quad \beta_{-\,\text{smooth}}^\text{contractible}=-2\pi\,.\label{2.44contsmooth}
    \end{equation}
    Combined with equation \eqref{2.41contract} this reproduces our smoothness conditions \eqref{2.29smooth}.
\end{enumerate}

\section{Torus quantization}\label{sect:3torusquant}

In this section, we present aspects of the canonical quantization of KdS, slicing the path integral open on torus universes \eqref{2.13torus}. Some quantum features of the phase space are discussed in \textbf{section \ref{sect:3.1kdstoruscosmology}}. In \textbf{section \ref{sect:idea}} we introduce the idea of the calculation: to compute the gravitational spectral density as the overlap between two states in the (physical) Hilbert space of torus universes, one state representing the pode and another representing the horizon:
\begin{equation}
    \rho_{(c,d)}(E,J)=\braket{\text{smooth}_{(c,d)}\rvert E,J}\,.\label{3.1xx}
\end{equation}
In \textbf{section \ref{sect2.6boundaryconditions}} we discuss the boundary conditions associated with physical states, which are covariant (unlike for instance Dirichlet boundaries). In \textbf{section \ref{sect:2.5TD}} we compute \eqref{3.1xx} via naive canonical quantization, recovering the semiclassical gravity answer \eqref{2.54rhoJ} and discuss the associated thermodynamics. 

When quantizing a classical phase space, $\hbar$ corrections are ambiguous, and these affect the quantization. This is not an unimportant detail, and finding a correct quantization requires robust and well-founded principles. 
In \textbf{sections \ref{sect4guidance} and \ref{sect:5cls}} we will develop a natural proposal for an exact quantization scheme of 3d dS quantum gravity based on its connection with complex Virasoro CFT and $\mathbb{C}$LS. Henceforth we will oftentimes use the shorthand notation:
\bea
   \ \  \boxed{\, \hbar = 16 \pi G\, \strut }
\eea
This multiplies to usual Planck constant in all places (which we put to unity).

\subsection{KdS as torus quantum cosmology}\label{sect:3.1kdstoruscosmology}

We discussed in section \ref{sect2xx} the space of classical solutions, which consist of torus-shape universes \eqref{2.13torus}:
\begin{equation} \label{torus minisuperspace}
    \d s^2=\d\tau^2+e^\phi\,\frac{ \abs{\d x + (m_1 + \i m_2) \d y}^2}{m_2}  \,, \quad x \sim x+1 \,, y \sim y +1\,.
\end{equation}
The torus moduli $(m_1,m_2)$ and the conformal factor $e^\phi$ evolve as a function of $\tau$. We uniquely specify a classical solution by specifying $(m_1,m_2,e^\phi)$ and their time derivatives at some given Cauchy slice. 

In quantum mechanics, one fixes half of the parameters. For instance, one can study the Euclidean gravitational pathintegral with Dirichlet boundary conditions at some given torus shaped slice where we fix the induced boundary metric to be of the form:
\begin{equation}
    \d s^2\rvert_\text{bdy}=\d s^2_\text{torus}(m_1,m_2,e^\phi). 
\end{equation}
The associated gravitational amplitude are then wavefunctions of $(m_1,m_2,e^\phi)$:
\begin{equation}
    \raisebox{1mm}{\begin{tikzpicture}[baseline={([yshift=-.5ex]current bounding box.center)}, scale=0.9]
    \pgftext{\includegraphics[scale=1]{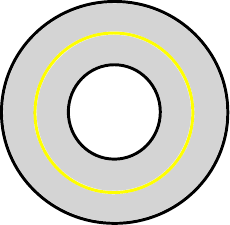}} at (0,0);
    \draw (1.05,0) node {\color{yellow}$\psi$};
    \draw (0,-2.4) node {$\d s^2_\text{torus}(m_1,m_2,e^\phi)$};
    \end{tikzpicture}}\ \raisebox{5mm}{$\; =\; \psi_\text{grav}(m_1,m_2,e^\phi)\,.$}\label{2.70wave}
\end{equation}
To compute $\rho_{(c,d)}(E,J)$ we will need different types of wavefunctions. Therefore, let us dive a bit deeper into the canonical quantization of the torus universes, following \cite{Moncrief:1989dx, Hosoya:1989sy, Carlip:1994ap,Carlip:2004ba}.

After a somewhat lengthy gauge-fixing procedure, one concludes that in the Hamiltonian treatment of dS$_3$ gravity the restriction to the torus minisuperspace \eqref{torus minisuperspace} is exact \cite{Moncrief:1989dx, Hosoya:1989sy, Carlip:1994ap,Carlip:2004ba}. We consider the ADM decomposition:
\begin{equation}
    \d s^2 = N^2 \d \tau^2 + h_{ij}\d x^i\d x^j\,,
\end{equation}
with the metric on fixed Cauchy slices describing torus universes:
\begin{equation}
    h_{ij}\d x^i\d x^j =\d s^2_\text{torus}= e^{\phi}\frac{1}{m_2} \abs{\d x + (m_1 + \i m_2) \d y}^2  \,, \qquad x \sim x+1 \,, y \sim y +1 \,.
\end{equation}
In minisuperspace, one restricts $(m_1,m_2,e^\phi)$ to be only dependent on time. The Einstein-Hilbert action then evaluates to:
\begin{equation} \label{einstein hilbert action}
    I = - \frac{1}{\hbar} \int \d^3 x \sqrt{g} (R-2)=-\frac{1}{\hbar}\int_0^{\pi/2} \d \tau\bigg\{\frac{e^\phi}{2 N}\bigg(\dot{\phi}^2-\frac{\dot{m_1}^2+\dot{m_2}^2}{m_2^2}\bigg)-2 N e^\phi\bigg\}\,.
\end{equation}
Defining Euclidean conjugate momenta to $(m_1,m_2,e^\phi)$ through
\begin{equation}
    K=\frac{\dot{\phi}}{N}\,,\quad p_1=-\frac{e^\phi}{N}\frac{\dot{m_1}}{m_2^2}\,,\quad p_2=-\frac{e^\phi}{N}\frac{\dot{m_2}}{m_2^2}\,,
\end{equation}
one recovers the Einstein-Hilbert action in first order variables:
\begin{equation}
    I=-\frac{1}{\hbar}\int_0^{\pi/2} \d \tau \bigg(p_1\dot{m_1}+p_2\dot{m_2}+K\dot{e^\phi}-N\bigg\{ \frac{1}{2}e^{\phi}(K^2 + 4) - \frac{1}{2}e^{-\phi} m_2^2\,(p_1^2+p_2^2) \bigg\}\bigg)\,.\label{2.75action}
\end{equation}
We deduce from this action the standard algebra between coordinates and (Euclidean) momenta, used for canonical quantization:
\begin{equation}
    [m_1,p_1]=-\hbar\,,\quad [m_2,p_2]=-\hbar\,,\quad [e^\phi,K]=-\hbar\,.
\end{equation}
As per usual, the lapse function ($N$) appears in the action as a Lagrange multiplier enforcing the WDW Hamiltonian constraint
\bea \label{hamiltonian constraint}
    H_\text{WDW} =  \frac{1}{2}e^{\phi}(K^2 + 4) - \frac{1}{2}e^{-\phi} m_2^2\,(p_1^2+p_2^2)=0\,.
\eea
One usually implements this constraint by noticing that the WDW Hamiltonian acts as a second order differential operators on wavefunctions $\psi_\text{grav}(m_1,m_2,e^\phi)$. Physical wavefunctions solve the differential equation\footnote{Physical operators commute with the constraint $
    [a,H_\text{WDW}]=0\,.$
One simple example is the $m_1$ momentum $p_1$. Another simple example is the upper-half plane Laplacian: $\Box=m_2^2\,(p_1^2+p_2^2)$. }
\bea
H_\text{WDW}\,\psi_\text{grav}(m_1,m_2,e^\phi)= 0\, .
\eea

We now show that the physical phase space of the theory consists of the variables $(\alpha_\pm,\beta_\pm)$ introduced in section \ref{sect2xx}. Physical phase space variables parametrize the solution manifold to \eqref{hamiltonian constraint}. The volume (or scale) factor $e^\phi$ and the torus modulus take the following form \eqref{2.21tau} (see also equation \eqref{2.xxm1m2}):
\begin{equation}
     e^{\phi} = \frac{1}{4} (\alpha_- \beta_+ - \alpha_+ \beta_-)\sin(2 N\tau)\,,\quad \tau_\text{torus}=m_1+\i m_2=\frac{\beta_+e^{-\i N\tau}-\beta_-e^{\i N\tau}}{\alpha_+e^{-\i N\tau}-\alpha_-e^{\i N\tau}}\,.\label{2.82metric}
\end{equation}
Furthermore, the conjugate Euclidean momenta take the following form:\footnote{Recall that $
    K = \frac{1}{\sqrt{g}} \frac{\d}{\d \tau} \sqrt{g}\,.$
Note furthermore that these variables are not necessarily real. Indeed, the Lorentzian reality condition is \eqref{2.9reality}. Reality as given by equation \eqref{2.9reality} is consistent with the operator algebra.
}
\begin{equation} \label{canonical transformation momenta}
    K=\frac{2}{\tan(2N\tau)}\,,\quad p_1   =  -\frac{1}{2}(\alpha_+^2 - \alpha_-^2)    \,, \qquad 
    p_2 = \frac{1}{2}\frac{\alpha_-^2 + \alpha_+^2}{\tan(2 N\tau)} - \frac{\alpha_+ \alpha_-}{\sin(2 N\tau)} \, . 
\end{equation}
These equations represent a 5 dimensional space of solutions to $H_\text{WDW}=0$ labeled by $(\alpha_\pm,\beta_\pm,\tau)$.\footnote{If one thinks of \eqref{2.82metric}/\eqref{canonical transformation momenta} as classical solutions rather than configurations at a Cauchy slice, then one must replace $\tau\to \tau+\tau_0$ and think of $\tau_0$ as phase space variable (being the space of solutions at some initial Cauchy slice).}

As discussed below equation \eqref{hamiltonian constraint}, one of these 5 variables $(\alpha_\pm,\beta_\pm,\tau)$ has to be redundant, in the sense that it does not appear in the symplectic form. Roughly speaking, this redundant variable is the variable generated by (conjugate to) $H_\text{WDW}$. Unsurprisingly, we see that this redundant variable is $\tau$. We check this by inserting \eqref{2.82metric}/\eqref{canonical transformation momenta} into the $p\d q$ symplectic term of the gravitational action \eqref{2.75action}:
\begin{equation}
    p_1\d m_1+p_2\d m_2+K\d e^\phi=\beta_+\d\alpha_+-\beta_-\d\alpha_-\,.\label{2.85algebra}
\end{equation}
This indeed does not feature $\tau_0$. Equivalently, one may use \eqref{2.82metric}/\eqref{canonical transformation momenta} to solve $(\alpha_\pm,\beta_\pm,\tau)$ as function of the original phase space variables. The claim is that the functions $(\alpha_\pm,\beta_\pm)$ commute with $H_\text{WDW}$ in equation \eqref{hamiltonian constraint} (with appropriate operator ordering). We previously checked this for the combination $p_1$. It would be interesting to check this explicitly for all the $(\alpha_\pm,\beta_\pm)$. On the other hand, the function $\tau$ should have a non-trivial commutator with $H_\text{WDW}$, and therefore is not a physical operator that one could diagonalize and use to label physical states. It would be interesting to develop this more, in light of understanding which boundary conditions we are putting in terms of local variables, see section \ref{sect2.6boundaryconditions}.

Equation \eqref{2.85algebra} confirms what we aimed to prove: physical phase space is spanned by $(\alpha_\pm,\beta_\pm)$ and the symplectic form matches the one claimed in equation \eqref{2.7omega}:
\bea
    \boxed{\ [\alpha_+,\beta_+]=-\hbar\,,\quad  [\alpha_-,\beta_-]=\hbar\ \strut}\quad \hbar =16\pi G\,.\label{2.86algebra}
\eea
The Einstein-Hilbert action \eqref{einstein hilbert action} reduces to a topological quantum mechanics (with zero Hamiltonian)
\begin{equation} \label{2.68instein hilbert action holonomies}
    I=-\frac{1}{\hbar}\int \beta_+\d \alpha_+ + \frac{1}{\hbar}\int \beta_-\d \alpha_-\,.
\end{equation}

\subsection{Idea of the computation}\label{sect:idea}

Physical phase space consists of the variables $(\alpha_\pm,\beta_\pm)$. In quantum mechanics, one then considers for instance wavefunctions $\psi_\text{gravity}(\alpha_+,\alpha_-)$ - associated with physical states $\ket{\alpha_+,\alpha_-}$. Dirichlet boundary conditions \eqref{2.70wave} correspond with considering states $\ket{m_1,m_2,e^\phi}$ in the pre-Hilbert space. One computes a gravitational $T^2\times I$ amplitude with (for instance) Dirichlet $(m_1,m_2,e^\phi)$ at one end and the Neumann conditions  $(p_1,p_2,K)$ at the other by inserting a propagator that projects on physical states \cite{halliwell1988derivation,Marolf:1996gb,DiazDorronsoro:2017hti,Banihashemi:2024aal,Blommaert:2025bgd}.\footnote{The lapse contour is a question of active investigation, see for instance \cite{DiazDorronsoro:2017hti,Feldbrugge:2017kzv,Banihashemi:2024aal,Dittrich:2024awu,Blommaert:2025bgd}. The real axis is natural in quantum  mechanics and gives reasonable answers, however it is not obviously equivalent with rigorous calculations for instance in 2d quantum gravity and in matrix models. It is also not obviously inequivalent. We thank Dionysios Anninos for comments on this. Timelike Liouville might provide one rigid framework to make progress on this \cite{Anninos:2024iwf,Anninos:2025fer}.}
Practically, this amplitude would be computed by diagonalizing $H_\text{WDW}$, and finding the correct inner product on the associated physical states (solutions of constraint). Instead, we will end up considering gravitational propagation between physical states, by specifying (for instance) $(\alpha_+,\alpha_-)$ at one end, and $(\beta_+,\beta_-)$ at the other end. This gravitational amplitude does not require a projector, with the understanding that inner products are evaluated in the physical Hilbert space:
\begin{equation}
    \mathcal{Z}_\text{grav}( \alpha_+,\alpha_-\to \beta_+,\beta_-)=\braket{\beta_+,\beta_-\rvert \alpha_+,\alpha_-}=\,\,\,\raisebox{2.1mm}{\begin{tikzpicture}[baseline={([yshift=-.5ex]current bounding box.center)}, scale=0.7]
 \pgftext{\includegraphics[scale=1]{KdS6.pdf}} at (0,0);
    \draw (3,-1.1) node {\color{blue}$(\alpha_+,\alpha_-)$};
    \draw (0.7,1.7) node {\color{blue}pode};
    \draw (-4.2,0) node {\color{red}horizon $(\beta_+,\beta_-)$};
  \end{tikzpicture}}\label{grav-amp}
\end{equation}
For more background see for instance \cite{Held:2024rmg,Witten:2022xxp,Chandrasekaran:2022cip,Blommaert:2025avl}. In the picture we anticipate that $\braket{\beta_+,\beta_-\rvert \alpha_+,\alpha_-}$ corresponds with the gluing procedure explained in equation \eqref{2.33Stori} - where we patch two tori together as follows
\begin{equation}
    \begin{tikzpicture}[baseline={([yshift=-.5ex]current bounding box.center)}, scale=0.9]
    \pgftext{\includegraphics[scale=1]{KdS4.pdf}} at (0,0);
    \draw (0.5,1) node {\color{blue}$(\alpha_+,\alpha_-)$};
    \draw (-6,1) node {\color{red}$(\beta_+,\beta_-)$};
    \draw (-3.5,2.3) node {$\tau_\text{c}<\tau<\pi/2$};
    \draw (3.5,2.3) node {$0<\tau<\tau_\text{c}$};
    \draw (0,-0.6) node {glue};
    \draw (-1.8,-0.6) node {$y$};
    \draw (4.35,-0) node {$y$};
    \end{tikzpicture}
\end{equation}
The doughnut $0<\tau<\tau_\text{c}$ prepares in a gauge-invariant way an initial state $\ket{\alpha_+,\alpha_-}$ by having a defect running along the $y$ direction, with variables $(\alpha_+,\alpha_-)$. Similarly, $\tau_\text{c}<\tau<\pi/2$ prepares the final state $\bra{\beta_+,\beta_-}$ by having a defect along the $x$ direction, specified by $(\beta_+,\beta_-)$. In section \ref{sect2.6boundaryconditions} we explain that these defects are implemented covariantly by dynamical particle path integrals.
 
 We can compute the overlap \eqref{grav-amp} using the fact that on wavefunctions $\psi(\alpha_+,\alpha_-)$, $(\beta_+,\beta_-)$ act as
\bea
    \beta_+=\hbar \frac{\d}{\d \alpha_+}\,,\quad \beta_-=-\hbar \frac{\d}{\d \alpha_-}\,.\label{2.88betaDE}
\eea
One therefore obtains:
\begin{equation}
    \,\,\,\raisebox{2.2mm}{\begin{tikzpicture}[baseline={([yshift=-.5ex]current bounding box.center)}, scale=0.7]
 \pgftext{\includegraphics[scale=1]{KdS6.pdf}} at (0,0);
    \draw (3,-1.1) node {\color{blue}$(\alpha_+,\alpha_-)$};
    \draw (0.7,1.7) node {\color{blue}pode};
    \draw (-4.2,0) node {\color{red}horizon $(\beta_+,\beta_-)$};
  \end{tikzpicture}}=e^{\frac{\alpha_+\beta_+-\alpha_-\beta_-}{\hbar}}\,.\label{2.89link}
\end{equation}
This indeed matches the correct on-shell action for the spacetime \eqref{2.14classical metric} cartooned in the left \cite{Hikida:2021ese}. As we recall in section \ref{sect4.1CS}, this expression \eqref{2.89link} also immediately follows as the classical action associated with two linked line operators corresponding with the dynamical path integral of particles in dS$_3$. 

Crucially, the $(c,d)$ smooth Euclidean horizons are implemented by the smoothness condition \eqref{2.29smooth}. Quantum mechanically, one associated the horizons thus with the following smooth states:
\bea
    \boxed{\ \ket{d \beta_+ -c \alpha_+ =+ 2\pi}\otimes \ket{d \beta_- -c \alpha_- =- 2\pi}=\ket{\text{smooth}_{(c,d)}}\;\strut}\label{2.92state}
\eea
We thus arrive at the following proposal for computing the observer's spectral density, for a fixed $(c,d)$:
\begin{equation}
\rho_{(c,d)}(E,J)=
     \begin{tikzpicture}[baseline={([yshift=-.5ex]current bounding box.center)}, scale=0.7]
 \pgftext{\includegraphics[scale=1]{KdS6.pdf}} at (0,0);
    \draw (2.65,-1.1) node {\color{blue}$(E,J)$};
    \draw (-3.8,0) node {\color{red}smooth $(c,d)$};
    \draw (0.7,1.7) node {\color{blue}pode};
  \end{tikzpicture}=\braket{\text{smooth}_{(c,d)}\rvert E,J}\,,\label{2.93rhocd}
\end{equation}
The minimal sum over SL(2,$\mathbb{Z}$) geometries then leads to the following proposal for an observer's spectral density:
\bea
    \boxed{\ \rho(E,J)\, =\sum_{\gamma \, \in\, \mathbb{Z}\backslash \TT{PSL}(2,\mathbb{Z})} \braket{\text{smooth}_{\gamma}\rvert E,J}\large{\strut} \,}\label{2.94main}
\eea
We will compute the matrix elements $\braket{\text{smooth}_{\gamma}\rvert E,J}$ in subsequent levels of accuracy. A naive attempt at quantization in section \ref{sect:2.5TD} reproduces the gravitational on-shell action \eqref{2.54rhoJ}.

We remark that associating the no-boundary gravitational path integral with matrix elements of a particular state in the physical Hilbert space (rather than the pre-Hilbert space of boundary conditions) is standard in 2d dilaton gravity, for instance in dS JT gravity \cite{Held:2024rmg} and in sine dilaton gravity \cite{Blommaert:2025avl,Blommaert:2025eps,Blommaert:2024ydx,Okuyama:2024eyf}. The association \eqref{2.93rhocd} should in this sense come as no surprise. The less standard part of the dictionary \eqref{2.94main} is the association of $(E,J)$ directly with a gauge-invariant state, rather than with a set of Dirichlet boundary conditions (pre-Hilbert states) as is standard for instance in AdS/CFT \cite{Witten:1998qj,Gubser:1998bc,Maldacena:1997re}. We propose that static patch quantum gravity works in a fundamentally new way: with gauge-invariant boundary conditions/states. We now discuss this point in more detail.

\subsection{Dynamical boundary conditions}\label{sect2.6boundaryconditions}
A natural way to describe physics in dS$_3$ from an observer's perspective, is to fix boundary conditions on a timelike tube surrounding the pode. This gives non-trivial thermodynamics \cite{hayward1990euclidean,Coleman:2021nor,Banihashemi:2022jys,Svesko:2022txo}. One could fix Dirichlet boundary conditions, which selects the boundary state $\ket{m_1,m_2,e^\phi}$. Dirichlet boundary conditions on a timelike tube in gravity are not great, a better option is to fix conformal boundary conditions $\ket{m_1,m_2,K}$ \cite{Anninos:2024wpy,An:2021fcq,anderson2008boundary}. Another option is to adopt Neumann boundary conditions, $\ket{p_1,p_2,K}$, or perhaps more naturally (given experience with AdS/CFT), one can consider the state $
    \ket{p_1,p_2,K\to +\infty}.$
According to equation \eqref{canonical transformation momenta}, this state has fixed energy and spin $(E,J)$ determined by the momenta:
\begin{equation}
    \frac{p_1}{4\pi^2}=\i J, \,\qquad \frac{p_2}{4\pi^2}=-\sqrt{E^2+J^2}\,.
\end{equation}
$K\to +\infty$ imposes that the torus degenerates and so naturally implements the insertion of a line defect. 

Physical states like $\ket{\alpha_+,\alpha_-}$ decompose into linear combinations of states $\ket{m_1,m_2,e^\phi}$. However, this decomposition need not be simple. The best case scenario would be that physical states $\ket{\alpha_+,\alpha_-}$ correspond with an equivalence class of pre-Hilbert states $\ket{\alpha_+,\alpha_-,0}\sim \ket{\alpha_+,\alpha_-,T}$ with $T$ canonically conjugate to $H_\text{WDW}$. One calls this equivalence class co-invariant states \cite{Held:2024rmg,Witten:2022xxp}. One could then determine what are the gravitational boundary conditions associated with $\ket{\alpha_+,\alpha_-,T}$ and claim that this are the boundary conditions imposed when computing the amplitude \eqref{grav-amp}. The state $\ket{p_1,p_2,K\to\infty}$ seems close to achieving this goal, but the WDW Hamiltonian acts non-trivial on it. Perhaps the group averaged version of this state implements a state $\ket{\alpha_+.\alpha_-}$. Pinpointing the precise boundary conditions associated with $\ket{\alpha_+.\alpha_-}$ is an interesting open question for future work. Does $\psi_{(\alpha_+,\alpha_-)}(p_1,p_2,K\to +\infty)$ delta localize?

Rather than fixing boundary conditions directly, physical states $\ket{\alpha_+,\alpha_-}$ have a gauge invariant path integral interpretation, which we believe may be closer to how dS$_3$ quantum gravity may fundamentally work. We claim that one can construct worldline operators $\mathcal{W}(E,J)$ such that the torus quantum state $\ket{E,J}$ is obtained by acting with the corresponding line operator on the trivial state:
\bea
    \ket{E,J}=\mathcal{W}(E,J) |1,0\rangle\,. \label{3.xxx}
\eea
For the special case $J=0$, this line operator is the worldline path integral of a particle of mass $m(E)$:
\bea
    \mathcal{W}(E,J=0)  = \oint\mathcal{D}x\,\exp\bigg(-m(E)\oint \d \tau \sqrt{g_{\mu\nu}\frac{\d x^\mu}{\d\tau}\frac{\d x^\mu}{\d\tau}}\; \bigg)\,,\quad m(E)=\frac{1-E}{8G}\,.
    \label{2.xxpath}
\eea
More generally, $\ket{E,J}$ corresponds with a particle of mass and spin $(m,s)$ with $s=-J/8G$ where we path integrate over the particles trajectory in dS$_3$, with a homology constraint on the particle trajectory.

A way to understand this is via the first order formalism for dS$_3$ quantum gravity, which can be  cast in the form of SL(2,$\mathbb{C}$) CS \cite{Witten:1988hc, Witten:1989ip}.\footnote{To perform our calculations, we will not assume that SL(2,$\mathbb{C}$) CS is equivalent to dS$_3$ gravity. As discussed in \textbf{section \ref{sect4.1CS}}, the close relationship between the phase space of dS$_3$ gravity  and SL(2,$\mathbb{C}$) CS still gives useful guidance. } A standard fact \cite{elitzur1989remarks,Belaey:2023jtr,Castro:2018srf,Ammon:2013hba,Iliesiu:2019xuh} is that particle worldlines can be represented in terms of Wilson line operators $\mathcal{W}_{\Delta_+,\Delta_-}(A_+,A_-)\leftrightarrow\mathcal{W}(E,J)$,
with $m(\Delta)$ in \eqref{2.52dic}. The natural way to specify data $(E,J)$ on a worldline is to specify the representation $(\Delta,s)$ of the particle under the dS isometry group. Another fundamental fact \cite{witten1989quantum} is that line operators prepare states in the physical Hilbert space on any torus homologous to the line operator via $
  \ket{\alpha_+,\alpha_-}  =  \mathcal{W}_{(\alpha_+,\alpha_-)}|1,0\rangle$.
This implies the claimed fact that inserting a single particle path integral \eqref{2.xxpath} prepares a physical state:\footnote{As compared to for instance equation \eqref{2.70wave} we removed a depiction of the black boundary of the doughnut, to indicate that we are not imposing boundary conditions at some arbitrary bulk surface, when preparing a physical state like $\ket{E,J}$.}
\bea
    \ket{E,J}\ \longleftrightarrow\raisebox{1mm}{\begin{tikzpicture}[baseline={([yshift=-.5ex]current bounding box.center)}, scale=0.75]
    \pgftext{\includegraphics[scale=1]{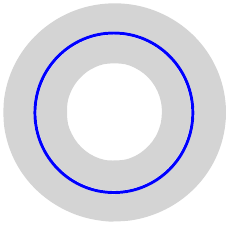}} at (0,0);
    \draw (-2.3,1.5) node {\color{blue}$(E,J)$};
    \end{tikzpicture}}\label{2.110torusstate}
\eea

So, $(E,J)$ are parameters in the fully dynamical action. This seems appropriate, since the worldline of the observer is not a place where gravity is turned off. The particle action \eqref{2.xxpath} is the action of the dynamical observer \cite{Chandrasekaran:2022cip}, whose entropy we aim to compute. 

\subsection{Observer's spectral density V1}\label{sect:2.5TD}
As a warm-up for an exact calculation in section \ref{sect:5cls}, we first aim to compute the gravitational amplitude
\bea
\rho_{(c,d)}(E,J) = \braket{\text{smooth}_{(c,d)}\rvert \alpha_+,\alpha_-}\,, 
\eea
using a naive quantization. The states $\ket{\text{smooth}_{(c,d)}}$ are defined in \eqref{2.92state}. Using \eqref{2.88betaDE}, the quantum implementation of the smoothness condition \eqref{2.29smooth} takes the form of two decoupled differential equations
\bea
   \Big[ \hbar \frac{\d}{\d \alpha_+} - \frac{c}{ d}\alpha_+ -\frac{2\pi}{d}\Big]\, \rho_{(c,d)}(\alpha_+,\alpha_-) 
    =  \Big[ \hbar \frac{\d}{\d \alpha_-} + \frac{c}{ d}\alpha_- -\frac{2\pi}{d}\Big]\, \rho_{(c,d)}(\alpha_+,\alpha_-) 
   =0\,.
\eea
Solving these gives an answer for the observer's spectrum associated with a given Kerr-lens spacetime: 
\begin{equation}
    \rho_{(c,d)}(\alpha_+,\alpha_-)\, =\, e^{\frac{2\pi}{d\hbar}(\alpha_++\alpha_-)+\frac{c}{2d\hbar}(\alpha_+^2-\alpha_-^2)}=\rho_{(c,d)}(\alpha_+)\rho_{(-c,d)}(\alpha_-)\,.\label{2.96wave}
\end{equation}
Notice that according to equation \eqref{2.9reality} we have $\alpha_+^*=\alpha_-$. One can introduce real variables $(\alpha,\delta)$ via
\begin{equation}
    \alpha_+=\alpha+\i \delta
\end{equation}
From equation \eqref{2.30xx} we see that $\alpha$ parameterizes the conical defect, while $\delta$ represents a time shift. 

To obtain the spectral density, we transform from $(\alpha_+,\alpha_-)$ to $(E,J)$ variables using equation \eqref{2.8EJ}. In the first exponential, one recognizes in particular the area of the cosmological horizon \eqref{2.16Ar}:
\begin{equation}
    A(E,J)=2\pi r_+=\frac{\alpha_++\alpha_-}{2}\,.
\end{equation}
The second exponential is proportional to $J$. Therefore an observer's spectral density for fixed $(c,d)$ is:
\begin{equation}
    \begin{tikzpicture}[baseline={([yshift=-.5ex]current bounding box.center)}, scale=0.7]
 \pgftext{\includegraphics[scale=1]{KdS6.pdf}} at (0,0);
    \draw (2.65,-1.1) node {\color{blue}$(E,J)$};
    \draw (-3.8,0) node {\color{red}smooth $(c,d)$};
    \draw (0.7,1.7) node {\color{blue}pode};
  \end{tikzpicture}\hspace{-4mm} =\ \boxed{\; \rho_{(c,d)}(E,J)=\braket{\text{smooth}_{(c,d)}\rvert E,J}=e^{\frac{1}{d}\frac{A(E,J)}{4 G}+2\pi \i \frac{c}{d}\frac{J}{8G}}\large\strut \, }\label{2.79I}
\end{equation}
The above answer for the Kerr-lens amplitude was claimed in equation \eqref{2.54rhoJ} in the introduction. We will improve on this result in the following two sections, where
we will obtain the exact quantum answer.
The $1/d$ in front of the area reflects that $(c,d)$ lens spaces are $\mathbb{Z}_d$ quotients of KdS, see equation \eqref{periodicity kerr-L(p,q)}.

The on-shell action \eqref{2.79I} allows one to calculate the thermodynamic relations. This is achieved by Legendre transforming the canonical ensemble $(\beta,\theta)$. The action becomes:
\bea
    -I(\beta,\theta) =\frac{A(E,J)}{4 G}+2\pi \i \frac{c}{d}\frac{J}{8G}-\beta \frac{E}{8G}-\i\theta \frac{J}{8G}\,.
\eea
Here $(E,J)$ have become dynamical variables, to be extremized. Variation with respect to $(E,J)$ gives
\bea
    \beta=-\frac{4\pi^2}{d}\frac{\alpha}{\alpha^2+\delta^2} \, =\, -\frac{2\pi}{d}\frac{r_+}{r_+^2-r_-^2} \,,\quad \theta= 2\pi \frac{c}{d}-\i \frac{4\pi^2}{d}\frac{\delta}{\alpha^2+\delta^2}\,=2\pi \frac{c}{d}-\frac{2\pi}{d}\frac{r_-}{r_+^2-r_-^2}\,.
\eea
These equations pass several checks in known regimes. 

Firstly, for the case of empty dS $(E,J)=(1,0)$ this reduces to the known thermodynamic relations for the lens space L$(d,c)$, found in equation (2.10) of \cite{Castro:2011xb} (our conventions for $\theta$ differs by the factor $\i$):
\bea
    \beta=-\frac{2\pi}{d}\,,\quad \theta=2\pi \frac{c}{d}\,.
\eea

A second comparison is to equation (2.16) in \cite{Maxfield:2020ale}, which describes a similar thermodynamics of AdS$_3$ SL(2,$\mathbb{Z}$) black hole horizons. Finally, for the case $(c,d)=(0,1)$ one recovers the usual thermodynamics of standard KdS spacetimes:
\bea
    \beta=-\spc 4\pi^2\frac{\alpha}{\alpha^2+\delta^2}\,,\quad  \theta=-
    \spc\Omega\beta\,,\quad \Omega=\frac{\delta}{\beta}\label{2.89}
\eea
This matches exactly with equation (7.6) and equation (7.16) in \cite{Bousso:2001mw} (their energy is (proportional to) minus our energy \eqref{2.8EJ}, resulting in opposite temperature signs). These comparisons support the fact that we derived the correct thermodynamics of Kerr-lens spaces KL$(c,d)$, and support our approach to find gravitational amplitudes via canonical quantization of torus universes.

\section{{Holographic triality}}\label{sect4guidance}
In this section, we collect guidance for an exact quantization of the $(c,d)$ KdS spacetimes, in the form of three dual perspectives on 3d dS quantum gravity. These are (i) complex Virasoro TFT in \textbf{section \ref{sect4.1CS}}, (ii) the $\mathbb{C}$LS in \textbf{section \ref{sect4.2CLS}} and (iii) DSSYK in \textbf{section \ref{sect4.3DSSYK}}. The first two dual perspectives are closely related: $\mathbb{C}$LS is the gravitational edge dynamics of $\mathbb{V}$TFT, in analogy to how gauged WZW theory arises as the edge theory of 3d CS theory. More importantly, as we will argue, they can be assembled into an exact quantum description of pure dS$_3$ with concrete rules for computing all amplitudes of interest. 

To illustrate the usefulness of these rules, we already quote here our proposed  exact quantum answer for  the (c,d) Kerr-lens amplitude with an observer with energy and spin $(E,J)$:
\begin{equation}
    \rho_{(c,d)}(\alpha_+,\alpha_-)=\quad \raisebox{2mm}{\begin{tikzpicture}[baseline={([yshift=-.5ex]current bounding box.center)}, scale=0.7]
 \pgftext{\includegraphics[scale=1]{KdS6.pdf}} at (0,0);
    \draw (2.95,-1.1) node {\color{blue}$(\alpha_+,\alpha_-)$};
    \draw (-3.2,0.45) node {\color{red} smooth};  
    \draw (-3.2,-0.45) node {\color{red} $(c,d)$};
    \draw (0.7,1.7) node {\color{blue}pode};
\end{tikzpicture}}=\mathsf{M}^{\gamma_+}_{\mathbb{1}\alpha_+}\mathsf{M}^{\gamma_-}_{\mathbb{1}\alpha_-}\,.\label{kl-amp-exact}
\end{equation}
The explicit wavefunctions are:
\begin{equation}
    \mathsf{M}^{\gamma}_{\mathbb{1} \alpha} =
\frac{1}{\sqrt{d}} \, e^{\frac{c}{2 d\hbar}\alpha^2}\biggl[ e^{  \frac{2\pi}{d \hbar}\alpha} \sin\Bigl(\frac{2\pi b-\alpha}{2 d}\Bigr)  + e^{-\frac{2\pi}{d \hbar}\alpha} \sin\Bigl(\frac{2\pi b+\alpha}{2 d}\Bigr)\,\biggr]\,,\quad \gamma_\pm = \biggl(\!\begin{array}{cc} a\! & \! \pm b \\[-.75mm] \! \pm c\!& d \end{array}\!\biggr)\,.\label{m-exact}
\end{equation}
This expression is one of our main results. It reduces to \eqref{2.79I} in the semiclassical limit.  We summarize the $\mathbb{V}$TFT derivation for \eqref{kl-amp-exact}/\eqref{m-exact} in section \ref{sect4.1CS}, deferring the more detailed calculation based on the $\mathbb{C}$LS perspective to section \ref{sect:5cls}.

\subsection{Representation theory}\label{sec4.1:representations}

We first recall how the parametrization of $(E,J)$ in terms of holonomy variables $(\alpha_+,\alpha_-)$ relates with SL(2,$\mathbb{C})$ representation theory.

SL(2,$\mathbb{C}$) representations are labeled by mass and spin. 
A common definition of mass $M$ and spin $S$ in KdS spacetimes \cite{Bousso:2001mw,Balasubramanian:2001nb} is related with our equations \eqref{2.8EJ} through a constant shift so that $M=0$ for empty dS and (following equation \eqref{2.7omega}) rescaling $E$ and $J$ by a factor of $1/8G$ so that $M$ and $S$ become normalized canonical conjugates to the time-shift and opening angle variables $T$ and~$\theta$: 
\begin{equation}
[M,T]=[S,\theta]=\i\,,\quad M = \frac {1-E}{8G}\, \quad S =\frac{J}{8G}\,.
\end{equation}
Hence, assuming we define $\theta$ as a periodic variable with period $2\pi$, $S$ takes on integer values. We have:
\bea
    M= 
    \frac{1}{16G}\bigg(2 - \frac{\alpha_+^2}{4\pi^2}- \frac{\alpha_-^2}{4\pi^2} \bigg)\,,\ & & \  S=\frac{\i}{16G} \bigg(\frac{\alpha_+^2}{4\pi^2}-\frac{\alpha_-^2}{4\pi^2} \bigg)\,.\label{2.35MJ}
\eea
We would like to compare these equations with formulas that appear in the context of Liouville CFT in section \ref{sect:5cls}, with SL(2,$\mathbb{C}$) representation theory, and with dS/CFT \cite{Anninos:2011af,Collier:2025lux,Collier:2024kmo,Strominger:2001pn}. 
We consider two weights in this theory, with parameters $\alpha_+$ respectively $\alpha_-$:
\begin{equation}
    h_+=\frac{1}{2}
    -\frac{\i}{16 G}\biggl(1- \frac{\alpha_+^2}{4\pi^2}\biggr)\,,\quad h_-=\frac{1}{2}
    -\frac{\i}{16 G}\biggl(1- \frac{\alpha_-^2}{4\pi^2}\biggr)\,.\label{2.39CLSweights} 
\end{equation}
Combining this with equation \eqref{2.35MJ}, we find the relations 
\bea
    h_+=\frac{1+S-\i M}{2}\,,\ & & \  h_-=\frac{1-S-\i M}{2}\,.
\eea
which are reminiscent of SL(2,$\mathbb{C}$) representation theory \cite{Anninos:2011af,Collier:2025lux}.
Principle series representations of SL(2,$\mathbb{C}$) are labeled by two real parameters $(\Delta,s)$ related to the conformal weights $(h_+,h_-)$ via
\bea
    \Delta=h_++h_-=1-\i M\,,\quad s=h_+-h_-=S\,.
\eea
The SL(2,$\mathbb{C}$) quadratic Casimir is related with $\Delta$ via the familiar dS/CFT dictionary \cite{Strominger:2001pn,Collier:2025lux}:
\begin{equation}
    m^2=\Delta(2-\Delta)=1+M^2\,.\label{2.52dic}
\end{equation}
As $M$ is order $1/G$, this is the correct classical equation. In conclusion, one associates SL(2,$\mathbb{C}$) weights $(h_+,h_-)$ with the KdS variables $(\alpha_+,\alpha_-)$. We will argue that KdS gravity is dual to two copies of  $\mathbb{C}$LS theory with complex Virasoro symmetry. In this dictionary, $h_+$ is a conformal weight in one $\mathbb{C}$LS while $h_-=h_+^*$ is a weight in the second  $\mathbb{C}$LS.\footnote{In the exact quantum dictionary, $h_+$ and $h_-$ receive an additional perturbative contribution equal to $-2G$.} Note that $M$ and $-M$ are the same SL$(2,\mathbb{C})$ representation, but this symmetry is not visible to us in the gravitational description.\footnote{We thank Lorenz Eberhardt for pointing this out.} 
\subsection{Perspective 1. $\mathbb{V}$irasoro TFT}\label{sect4.1CS}

The discussion of the torus universe in  section \ref{sect:3torusquant} naturally connects to the first order formalism of 3d dS in which $(\alpha_\pm,\beta_\pm)$ represent holonomies of SL(2,$\mathbb{C}$), the isometry group of dS$_3$. Equivalently, $\alpha_\pm$ and $\beta_\pm$ are labeling eigenvalues of closed Verlinde line operators of a complex Virasoro TFT ($\mathbb{V}$TFT) \cite{Blommaert:2025eps}. Complex $\mathbb{V}$TFT, while still in an early stage of development, is a cousin of SL(2,$\mathbb{C}$) CS theory that aims to modify standard Virasoro TFT, so that it becomes suitable for application to dS$_3$ quantum gravity. Some work on $\mathbb{V}$TFT and its relation to dS$_3$ gravity was reported in \cite{Cotler:2019nbi,Verlinde:2024zrh,Collier:2025lux, Blommaert:2025eps}. The results provide hints that $\mathbb{V}$TFT man be quantized via methods developed in \cite{Gaiotto_Teschner:2024osr} and allows for exact computations of amplitudes like \eqref{2.89link} and \eqref{2.93rhocd} via the usual rules of 3d TQFT \cite{witten1989quantum}. It remains to be made precise whether this indeed works in full generality, which is beyond our scope.

The Hilbert space of $\mathbb{V}$TFT on a Riemann surface $\Sigma$ times some interval $I$ is obtained by quantizing the space of (a chiral half of) \slt{} holonomies on non-trivial cycles of $\Sigma$. In the torus case, this is $(\alpha_+,\beta_+)$. If $\Sigma$ admits a hyperbolic constant curvature metric, this space is Teichm\"uller space ${\mathcal{T}}_\Sigma$, and the symplectic form is $\i$ times the Weil-Peterson symplectic form:
\bea
\omega_{\mathbb{V}\rm TFT} = \frac{\i}{\hbar} \, \omega_{\sf WP}\,, \quad \hbar = 16\pi G\,.
\eea 
This $\i$ is the new ingredient for dS. Without this $\i$, quantization of Teichm\"uller space is well developed \cite{Verlinde:1989ua, Kashaev:1998fc, Chekhov:1999tn, Kashaev:2000ku,Teschner:2003em,Terashima:2011qi, EllegaardAndersen:2011vps,Mikhaylov:2017ngi,Collier:2023fwi}. The Hilbert space is the space of Virasoro blocks with real $c=3/2G$.
A reasonable claim (which we will not prove) is that the vector space of $\mathbb{V}$TFT is the space of Virasoro blocks with complex central charge $c_\pm = 13 \pm \i 3/2G$ (in dS units and for small~$G$). The phase space of dS$_3$ quantum  gravity on  $\Sigma$ is the product of two copies on $\mathcal{T}_\Sigma$ modulo the mapping class group $\text{MCG}(\Sigma)$ of large diffeos:
\bea
\frac{\mathcal{T}_\Sigma \times {\mathcal T}_\Sigma}{\text{MCG}(\Sigma)}\,.\label{4.10}
\eea
The two Teichm\"uller factors each opposite sign symplectic forms
\bea
\omega \is \frac{\i}{\hbar} \,\omega^{(1)}_{\sf WP}\, -\, \frac{\i}{\hbar} \,
\omega^{(2)}_{\sf WP}\,.
\eea
So dS$_3$ gravity can be thought of as the tensor product of two $\mathbb{V}$TFTs, with the constraint that physical states must be invariant under large diffeos. In path integral language, this means to prepare dS$_3$ states on $\Sigma$, we must sum over Dehn fillings. For $\mathbb{V}$TFT every single Dehn filling would create a physical state.

One can study eigenstates of holonomy operators in $\mathbb{V}$TFT, which when written as Virasoro blocks are labeled by Virasoro representations in the corresponding channels.
Holonomy eigenstates are created by line operators running along the dual cycle \cite{elitzur1989remarks,witten1989quantum}.
As discussed in section \ref{sect:2.1solutions}, line operators indeed represent the world lines of particles which \textbf{create} curvature sources, and thus non-trivial holonomies along the dual cycles. Line operators also {\textbf{measure}} this holonomy created by line operators along dual cycles. The basis change between eigenstates of different sets of $\mathbb{V}$TFT line operators are governed by unitary modular matrices.\footnote{These matrices preserve the inner product defined used in the $\mathbb{C}$LS \cite{Verlinde:1989ua,Verlinde:2024zrh,Collier:2023fwi,Collier:2025lux}.}

\subsubsection*{Torus universes}

In the following, we'll propose that Dehn surgery rules for topological QFT \cite{witten1989quantum} may be generalized to $\mathbb{V}$TFT on $T^2\times I$.\footnote{The torus does not have a hyperbolic metric. Yet, independent $\mathbb{C}$LS calculations in section \ref{sect:5cls} support this assumption.} As explained in section \ref{sect:3torusquant}, the torus analog of the two Teichm\"uller spaces $\mathcal{T}_{\Sigma}$ is the phase space consisting of two pairs of holonomies $(\alpha_\pm,\beta_\pm)$, with symplectic form (see equation \eqref{2.86algebra})
\bea
\omega = \frac{\i}{\hbar}\d\alpha_+\wedge \d\beta_+ - \frac{\i}{\hbar}\d\alpha_- \wedge \d\beta_-\,.
\eea

The geometric idea of the Dehn surgery procedure has been explained in section \ref{sect:idea}. For computing the KdS amplitudes of interest, the implementation of this procedure in $\mathbb{V}$TFT relies on three ingredients (i) the claim that complex Virasoro torus blocks form a linear vector space, on which (ii) eigenstates of Verlinde lines form an orthonormal basis, and (iii) the modular matrices that implement the action of $\gamma \in$ SL(2,$\mathbb{Z}$) of the torus modular group on the space of complex Virasoro torus blocks are known \cite{Collier:2024kwt}.

For the $\mathsf{S}$ and $\mathsf{T}$ generators of  the SL(2,$\mathbb{Z})$ modular group 
\begin{equation}
    \mathsf{S} = \Bigl(\!\begin{array}{cc}
        0\! &\! -1 \\[-.75mm]
        1 \! & 0
    \end{array} \! \Bigr) \,, \quad  \mathsf{T} = \Bigl(\! \begin{array}{cc}
        1 & 1 \\[-.75mm]
        0 & 1
    \end{array}\!\Bigr)\,, 
\end{equation}
the complex Virasoro modular matrices are\footnote{In this and the following section, we drop irrelevant overall factors. More detailed formulas are found in Appendix \ref{app: conventions}.}
\begin{equation}
    \mathsf{S}_{\alpha \beta} = \, 
    \cosh\Big(\frac{\alpha \beta}{ \hbar}\Big)  \,, \quad   \mathsf{S}_{\mathbb{1} \beta} \, = \, 
    \sin\Big(\frac{\beta}{2}\Big) \sinh\Big(\frac{2 \pi \beta}{\hbar}\Big) \,,
\quad 
    \mathsf{T}_{\alpha\beta} =\, 
    e^{-   \frac{\beta^2}{2\hbar}} \, \delta_{\alpha\beta} \, .
\end{equation}
Here $\hbar = 16\pi G  = 4\pi \sfb^2$ with $\sfb$ the $\mathbb{C}$LS coupling related to the central charge via $c_\pm = 13 \pm 6\i(\sfb^2-\sfb^{-2})$. The modular matrix for general $\gamma$ are read off from the transformation of complex torus characters
\footnote{Here the contour of integration over $\beta$ needs to be chosen such that the integral converges. }
\begin{equation}\label{modular transformation character generic}
    \boxed{\chi_{\alpha_+} \left(\frac{a \tau + b}{c \tau + d} \right) = \int_\mathcal{C} \d \beta_+ \, \mathsf{K}^\gamma_{\alpha_+\beta_+}  \,\chi_{\beta_+}(\tau)\,}\quad \chi_{\alpha_+}(\tau) =\frac{e^{-\frac{\alpha_+^2}{2 \hbar}\tau}}{\eta(\tau)}\,,\quad \gamma \, = \spc \bigg(\!\! \begin{array}{cc}
        a &\! b \\[-.75mm]
        c &\! d
    \end{array}\!\!\bigg)\,.
\end{equation}
One finds (suppressing prefactors which will not affect our final Kerr-lens spectral density):
\begin{equation}\label{4.16K}
    \mathsf{K}^\gamma_{\alpha \beta} = 
    \frac 1 {\sqrt{c}}\, e^{-\frac{a \alpha^2 + d \beta^2}{2 c \hbar}} \cosh\Big(\frac{\alpha \beta}{c \hbar}\Big)\,,\quad \mathsf{K}^\gamma_{\mathbb{1}\beta} = \mathsf{K}^\gamma_{2\pi + \frac {\i \hbar} 2\, \beta} - \mathsf{K}^\gamma_{2\pi - \frac {\i\hbar} 2\, \beta}\,.
\end{equation}
A basic application of TFT rules states that the amplitude associated with the link \eqref{2.89link} is an $\mathsf{S}$-matrix element \cite{witten1989quantum}:
\begin{equation}
    \raisebox{2.2mm}{\begin{tikzpicture}[baseline={([yshift=-.5ex]current bounding box.center)}, scale=0.7]
 \pgftext{\includegraphics[scale=1]{KdS6.pdf}} at (0,0);
    \draw (3,-1.1) node {\color{blue}$(\alpha_+,\alpha_-)$};
    \draw (0.7,1.7) node {\color{blue}pode};
    \draw (-3.4,0.45) node {\color{red}horizon};  
    \draw (-3.4,-0.45) node {\color{red} $(\beta_+,\beta_-)$};
  \end{tikzpicture}}\hspace{-7mm} =\, \mathsf{S}_{\beta_+\alpha_+}\mathsf{S}_{\beta_-\alpha_-}= \cosh\Big(\frac{\beta_+\alpha_+}{\hbar}\Big)\cosh\Big(\frac{\beta_-\alpha_-}{\hbar}\Big)\,.\label{S}
\end{equation}
The right-hand side is our proposed exact quantum answer for this amplitude. Its leading order behavior for small $G$ agrees with the semiclassical gravity answer given in equation \eqref{2.89link}. At subleading order, it also includes a sum over other saddles obtained by $\alpha_+\to -\alpha_+$ and $\alpha_-\to-\alpha_-$. The gravitational meaning of the other saddles is not well understood, but can presumably be explained by the presence of an inner horizon \cite{Kruthoff:2024gxc,Caminiti:2026efx}.
 
The line operators $\mathcal{W}_{(\alpha_+,\alpha_-)}$ and $\mathcal{W}_{(\beta_+,\beta_-)}$ in \eqref{S} run along the $y$ circle near the pode and the $x$ circle near the horizon. More generally, consider a line $\mathcal{W}^{(c,d)}_{(\beta_+,\beta_-)}$ along the $w$ circle defined in equation \eqref{2.37w}.  Requiring that this operator acts as the identity on $\ket{\text{smooth}_{(c,d)}}$ is the quantum version of the smoothness condition \eqref{2.92state}. This condition is solved by projecting the state on the identity conformal block in the channel specified by the $w$-cycle. The overlap between the $(\beta_+,\beta_-)$ line along the $w$ cycle and the $(\alpha_+,\alpha_-)$ line is thus the modular matrix associated with the transformation $\gamma\cdot S$ in equation \eqref{2.40taumap}. The identity line $\beta_+\to\mathbb{1}$ gives the wavefunction:
\bea\label{kl-amp-chiral}
\rho_{(c,d)}(\alpha_+)  = \quad   \raisebox{2mm}{\begin{tikzpicture}[baseline={([yshift=-.5ex]current bounding box.center)}, scale=0.7]
 \pgftext{\includegraphics[scale=1]{KdS6.pdf}} at (0,0);
    \draw (2.15,-1.1) node {\color{blue}$\alpha_+$};
    \draw (-3.2,0.35) node {\color{red} smooth};  
    \draw (-3.2,-0.35) node {\color{red} $(c,d)$};
    \draw (0.7,1.7) node {\color{blue}pode};
\end{tikzpicture}}  =\mathsf{M}^{\gamma(c,d)}_{\mathbb{1}\alpha_+} \equiv \int_\mathcal{C}\d \beta_+\; \mathsf{K}^{\gamma(c,d)}_{\mathbb{1} \beta_+}\,\mathsf{S}_{\beta_+\,\alpha_+}\,.
\eea
The exact expression for this modular matrix is given in \eqref{m-exact}. This produced our proposed equation \eqref{kl-amp-exact} for the amplitude of KL$(c,d)$ with the $(E,J)$ line operator. The factorization of $\rho_{(c,d)}(\alpha_+,\alpha_-)$ into chiral amplitudes has a a representation theoretic origin: the generators of the dS$_3$ isometry group split into two commuting sets of generators $J_i \pm \i K_i$. The gravitational line operators can thus formally be decomposed into products of two chiral operators.\footnote{In the complexified chiral theory, the holonomy variable $\alpha_+$ can take on any complex value.}

In section \ref{sect:5cls}, we reproduce equation \eqref{kl-amp-chiral} from a different dual perspective: $\mathbb{C}$LS on a disk crosscap with crosscap boundary conditions labeled by $\gamma(c,d)$. We discuss this perspective in section \ref{sect4.2CLS}. Before doing do, we discuss one important subtlety.

\subsubsection*{Periodic holonomies versus semiclassical gravity}

We have treated $\alpha_+=\alpha+\i\delta$ as a complex number labeling the opening angle $\alpha$ created by an observer. In gravity, an opening angle can indeed be more than $2\pi$. However, another point of view is to instead interpret $\alpha$ as labeling an SL(2,$\mathbb{C}$) holonomy matrix associated with a closed loop around the observer:
\bea
\biggl(\! \begin{array}{cc} e^{\i \alpha_+/2}\! &\!\! 0 \\ 0 \! &\! e^{-\i\alpha_+/2}\! \end{array}\! \biggr) \; \in \; \mathrm{ SL(2,}\mathbb{C})
\eea
In CS, the holonomy is the only information about the geometry, so one would treat $\alpha$ as an angle with period $4 \pi$. To prepare a state with any given holonomy, one sums over all local defects compatible with this holonomy. This leads to a sum over shifts of $\alpha\to \alpha+4\pi n$. As we will show in section \ref{sect:5cls} by direct computation in $\mathbb{C}$LS, this is indeed what naturally happens:
\begin{equation}\label{kl-vtft-amp}
    \boxed{\; \rho^{(c,d)}_{\mathbb{V}\rm TFT}(\alpha_+)=\sum_{n=-\infty}^{+\infty} \mathsf{M}^{\gamma}_{\mathbb{1},\alpha_++4\pi n}\,}
\end{equation}
From a gravitational perspective, the right-hand side sums over an infinite number of Kerr-lens saddles, all associated with the same horizon geometry specified by the lens space L$(c,d)$ but with distinct conical opening angles differing by integer multiples of $4\pi$. This sum restricts the range of inequivalent values of $\alpha$ to lie between $-2\pi$ and $2\pi$, in $\mathbb{V}$TFT and $\mathbb{C}$LS.\footnote{Geometrically, the Kerr-lens space $(c,d)$ with angle $\alpha$ and the Kerr-lens space $(-c,d)$ with angle $-\alpha$ are mirror images. Hence the shift symmetry group is generated by $(\alpha,c,d) \to (\alpha + 4\pi,c, d)$ and $(\alpha,c,d) \to (-\alpha, -c,d)$.} 

In gravity, summing over $n$ is an unnecessary complication. Yet, asides from the $\mathbb{C}$LS calculation of section \ref{sect:5cls}, there are two pieces of evidence that this is the right definition of $\mathbb{V}$TFT . 
A first motivation comes from recent studies of SL(2,$\mathbb C$) CS theory and its connection to Schur indices in 4d gauge theory \cite{Gaiotto_Teschner:2024osr}. The latter duality is a generalization of the celebrated AGT correspondence that, meaningfully, can be used to provide a concrete quantization scheme for SL(2,$\mathbb{C}$) CS theory, called Schur quantization \cite{Gaiotto_Teschner:2024osr}. When applied to one of the geometries of our interest --- a line defect in L$(-2,1)$ --- Schur quantization produces a quantity of the form of equation \eqref{kl-vtft-amp}, with a summation over $n$ \cite{Gaiotto:2024osr,Gaiotto:2024kze}:\footnote{In the context of \cite{Gaiotto_Teschner:2024osr}, this amplitude describes the Schur half-index of pure Seiberg-Witten theory. }
\bea
\label{rhotwoone}
\rho^{(-2,1)}_{\mathbb{V}\rm TFT\tiny\strut}(\alpha)=\quad   \raisebox{2.2mm}{\begin{tikzpicture}[baseline={([yshift=-.5ex]current bounding box.center)}, scale=0.7]
 \pgftext{\includegraphics[scale=1]{KdS6.pdf}} at (0,0);
    \draw (2.3,-1.1) node {\color{blue}$\alpha_+$};
    \draw (-3.25,.35) node {\color{red} smooth};
    \draw (-3.25,-.35) node {\color{red} $(-2,1)$};
    \draw (0.7,1.7) node {\color{blue}pode};
\end{tikzpicture}}\hspace{-2mm} =\sum_{n=-\infty}^\infty\, \mathsf{M}^{\gamma(-2,1)}_{\mathbb{1},\alpha+4\pi n} =\big(e^{\pm \i\alpha};\sfq \big)_\infty\,,\quad \sfq = e^{-\hbar/2}\,.
\eea

More relevant for our story is that this amplitude exactly matches the well-known expression of the DSSYK spectrum as function of the spectral parameter $\theta=\alpha_+/2$ \cite{Berkooz:2018jqr}. As shown in \cite{Blommaert:2025eps} and reviewed in section \ref{sect4.2CLS} and section \ref{sect4.3:open}, this match is not coincidental. It was the initial inspiration for our study and search for a potential holographic dual description of 3d KdS, and can be derived via a 3d/2d/1d holographic triality that centers around $\mathbb{C}$LS on a crosscap.

So, 3 rigid quantization schemes all predict a summation over $n$: Schur quantization, $\mathbb{C}$LS and SYK.

For later reference, we quote the result for the $\mathbb{V}$TFT amplitude on a general (c,d) Kerr-lens spacetime. Writing the sum \eqref{kl-vtft-amp} with $\mathsf{M}^\gamma_{\mathbb{1}\alpha}$ given in \eqref{m-exact} in terms of theta-function and using the familiar decomposition of theta functions in terms of Pochhammer symbols gives (with $\sfq = e^{d \hbar/ 4 c}$):
\bea \label{cls spec amplitude}
     \rho_{\mathbb{V}\rm TFT}^{(c,d)}(\alpha)=\frac{1}{\sqrt{c}} \Bigl[ e^{- \frac{\i \pi a}{c}} \big( \nspc -\nspc \sfq^{\frac 1 2\mp\frac 1 d}  \spc e^{\pm \frac \i 2  (\alpha + \frac{2 \pi} c) };\sfq \big)_\infty \!  - {e^{\frac{\i \pi a}{c}}}
     \big(\nspc -\nspc\sfq^{\frac 1 2\pm\frac 1 d}  e^{\pm \frac\i  2 (\alpha + \frac {2\pi} c) };\sfq\big)_\infty \Bigr]+ ({\small \alpha\! \to -\alpha} )\,.
\eea

\smallskip

\subsection{Perspective 2. $\mathbb{C}$LS $\otimes$ $\mathbb{C}$LS}\label{sect4.2CLS}

In this section we introduce the $\mathbb{C}$LS perspective. The $\mathbb{C}$LS worldsheet theory consists of two Liouville scalar fields (plus gauge-fixing ghosts) with the following action \cite{Collier:2024kmo,Collier:2025lux}:
\begin{align}
    I_{\mathbb{C}\rm LS} &=-\frac{\i}{\hbar}\int\d^2x \,\Bigr( \de^\mu \phi_+ \de_\mu \phi_++  e^{2\phi_+}\Bigr)-\frac{2\i}{\hbar}\,\oint \d x \, \mu_{\text{B}+} \spc e^{\phi_+} 
    \nonumber\\[2mm]&\qquad \qquad\qquad\qquad +\frac{\i}{\hbar}\int\d^2x \,\Bigr( \de^\mu \varphi_- \de_\mu \varphi_- +  e^{2\varphi_-} \Bigr)+\frac{2\i}{\hbar}\oint \d x \,\mu_{\text{B}-} \spc e^{\varphi_-}\,
    \label{2.17liouac}
    \end{align}
Note that the action is imaginary, indicating that the two Liouville CFTs have complex central charge
    \bea
    \label{cpm}
c_\pm=13 \pm 6\i \Bigl(\frac{1}{\mathsf{b}^2}-\mathsf{b}^2\Bigr)\,, 
\quad 
\boxed{\hbar = 4\pi \sfb^2 = 16\pi G\,\strut}
\eea
The boundary actions implement FZZT boundary conditions. For the application of $\mathbb{C}$LS to our problem, we will choose the 
 boundary cosmological constants $\mu_{\text{B}\pm}$ to be specified by
\begin{equation}
\label{mubee}
\mu_{\text{B}+} = \cos(\alpha_+/2)\,, \quad \quad \mu_{\text{B}_-} = 1\,.
\end{equation}

The spectrum of Virasoro weights of the two complex Liouville CFTs is conveniently parametrized in terms of momenta $P_+$ via
\bea
\label{clsspec}
     \label{2.46c}h_+=\frac{c_+-1}{24}+P_+^2\,,\quad
     P_+ = \frac{\sqrt{i}}{4\pi \sfb}\,  \alpha_+\,.
\eea
Physical states satisfy\footnote{Null states are labeled by $\alpha_{+\,(n,m)} = 2\pi n + 2\i \pi \sfb^2 m.$}
\begin{equation} 
h_+^* = h_-\,,\quad h_+ + h_- = 1\,.
\end{equation}
When the FZZT boundary is paired with an ZZ boundary on two sides of a strip, $\alpha_+$ labels the Liouville momentum in the open string channel via \eqref{clsspec}. 

In our application to 3d KdS gravity, we will instead consider the tensor product $\mathbb{C}$LS $\otimes$ $\mathbb{C}$LS of two decoupled complex Liouville worldsheet theories, each placed on a disk with the above FZZT boundary conditions and with a (generalized) crosscap boundary condition at the center.  

We will parametrize the non-trivial ($\mu_\text{B} \neq 1$) boundary cosmological constant of each $\mathbb{C}$LS theories by $\alpha_+$ and $\alpha_-$, respectively. So in total we will have two pairs of two complex Liouville CFTs $(\phi_+,\varphi_-)$ and $(\phi_-,\varphi_+)$, each placed on a disk with boundary conditions at the outer disk boundary specified by $|\alpha_+\rangle \otimes |\alpha_-\rangle$ 
with\footnote{The state $\ket{\text{FZZT}_{\bullet}(\alpha)}$ is defined carefully in section \ref{sect:5cls} and in appendix \ref{app: states}. Classically, it is an FZZT brane \cite{Fateev:2000ik,Zamolodchikov:2001ah,Teschner:2000md} with boundary cosmological constant as indicated in the action \eqref{2.17liouac}. } 
\bea
\label{fzztpair}
    \ket{\alpha_\pm}  = \ket{\text{FZZT}_{\bullet}(\alpha_\pm)}\otimes \ket{\TT{FZZT}(0)} \,.
\eea
Each pair of Liouville CFTs is assembled (with a ghost sector) into a covariant $\mathbb{C}$LS worldsheet theory.
Since the two $\mathbb{C}$LS are independent and decoupled, there is no Virasoro constraint relating the weights $\alpha_+$ and $\alpha_-$. So we can consider $\alpha_+ = \alpha_-^*\in \mathbb{C}$, without restrictions. Combining $(\alpha_+,\alpha_-)$, we propose that $\mathbb{C}$LS $\otimes$ $\mathbb{C}$LS is the dual description of the 3d KdS static patch. As a check, the tensor product $\mathbb{C}$LS theory has the same parameters $\ket{\alpha_+,\alpha_-}$ labeling physical states as does KdS (in torus quantization).
 
The choice of boundary conditions \eqref{fzztpair} is motivated by our earlier study of the duality between $\mathbb{C}$LS and the $G\Sigma$-formulation of DSSYK \cite{Blommaert:2025eps}.  
In \cite{Blommaert:2025eps} it was shown that there is a precise duality, at the level of the partition function, between the collective field theory of DSSYK and the $\mathbb{C}$LS  worldsheet theory on a disk with a crosscap. 
The duality was derived by considering $\mathbb{C}$LS in a lightcone gauge. In this gauge, one of the Liouville fields $\varphi_-$ takes the meaning of a reference clock while the other Liouville field $\phi_+$ evolves with respect to this clock \cite{Blommaert:2025eps}. This field $\phi_+$ becomes the collective field $g$ of DSSYK. A connection between $\mathbb{C}$LS with boundary states \eqref{fzztpair} and the spectrum of KL$(-2,1)$ was anticipated in \cite{Blommaert:2025eps}, which motivates our study of similar boundary conditions in relation to the general Kerr-lens spacetimes KL$(c,d)$.

We claim that the spectrum of KL$(c,d)$ is computed by inserting a specific boundary state in the center of the disk, implementing geometrically a generalized crosscap. The boundary state is the tensor product   $\ket{\mathbb{1}_{(c,d)}}\otimes   \ket{\mathbb{1}_{(-c,d)}}$ with (conform \cite{Blommaert:2025eps}):
\begin{equation}\label{cls states}
    \boxed{\; \ket{\mathbb{1}_{(c,d)}} = \ket{\TT{C}_{(c,d)}} \otimes e^{-2\pi i \frac{d}{c} L_0}\ket{\TT{ZZ}}\large \strut \,}
\end{equation}
Here $\ket{\TT{C}_{(c,d)}}$ denotes an SL(2,$\mathbb{Z})$ generalization of the standard crosscap boundary state $\ket{\TT{C}_{(-2,1)}}$ in complex Liouville CFT. Details on these states and their geometric construction is contained in section~\ref{sect4.1Cstates}. The relation between the SL(2,$\mathbb{Z}$)-twisted crosscap states and the $(c,d)$ Kerr-lens geometries is understood as follows. Instead of splitting the Kerr-lens geometry along a torus slicing, we introduce a disk-shaped 2d boundary by cutting the geometry along the disk surrounded by the circular observer worldline, as shown below:
\bea
    \begin{tikzpicture}[baseline={([yshift=-.5ex]current bounding box.center)}, xscale=-0.68,yscale=.68]
 \pgftext{\includegraphics[scale=1]{KdS9.pdf}} at (0,0);
    \draw (-4.5,-2) node {\color{blue}$(\alpha_+,\alpha_-)$};
    \draw (5.8,0) node {\color{red}horizon $(c,d)$};
    \draw (-2,0.6) node {\color{red}$\ket{\text{C}_{(c,d)}}$};
    \draw (-4.8,1.7) node {\color{blue}worldline};
    \draw (-1.4,-1.9) node {\color{black}brane};
    \draw (-1.4,-1.2) node {\small \color{black}$\mathbb{C}$LS $\otimes$ $\mathbb{C}$LS};
    \draw (-2.5,-0.3) node {\color{black}$r$};
    \draw (6.6,2.3) node {\color{black}KL$(c,d)$};
  \end{tikzpicture}\qquad \qquad \\[-2mm] \notag
\eea
Here the gray disk indicates the radial $r$ plane at some fixed angular coordinate. The $(c,d)$ cosmological horizon pierces through the center of the disk. The gravitational path integral over the 3d space with fixed boundary conditions on the disk produces a density matrix with the bra- and ket states identified with $\mathbb{C}$LS blocks on the disk with boundary conditions specified by $|\alpha_+\rangle |\alpha_-\rangle$ and with a $(c,d)$ crosscap placed at center of the disk.

In section \ref{sect:5cls} we compute the $\mathbb{C}$LS amplitude on an annulus with FZZT \eqref{fzztpair} and smooth$_{(c,d)}$ \eqref{cls states} boundary states. We find that it matches the Kerr-lens amplitude \eqref{kl-vtft-amp} predicted by $\mathbb{V}$TFT:  
\begin{equation}\label{cls-vtft-match}
    \boxed{\; \rho_{\mathbb{C}\rm LS}^{(c,d)}(\alpha_+) = \int_0^{\i \infty} \d \tau \, Z_\TT{ghost}(\tau) \bra{\alpha_+} e^{\i \pi \tau H} \ket{\mathbb{1}_{(c,d)}} = \sum_{n=-\infty}^{+\infty}\, \mathsf{M}^{(c,d)}_{\mathbb{1}\alpha_++4\pi n}\,}
\end{equation}
This correspondence is our main quantitative evidence in favor of the KdS/$\mathbb{C}$LS duality and forms the basis of our proposed exact answer of the KdS amplitude. 
The $\mathbb{C}$LS amplitude \eqref{cls-vtft-match} is periodic in $\alpha_+$. The $\mathbb{C}$LS and $\mathbb{V}$TFT amplitudes are sums over an infinite family of gravitational Kerr-lens amplitudes:
\begin{equation}
\rho_{\mathbb{C}\rm{LS}\tiny\strut}^{(c,d)}(\alpha_+) = \sum_{n=-\infty}^{\infty} \rho_{(c,d)}(\alpha_+ + 4\pi n)\label{4.33claim}
\end{equation}
In the semiclassical limit indeed different terms in the sum encode clearly quite different gravitational saddles. We make two brief comments:
\begin{enumerate}
    \item The match between $\mathbb{C}$LS \eqref{cls-vtft-match} and $\mathbb{V}$TFT \eqref{kl-vtft-amp} and the application to 3d KdS is a new result but not an unexpected one. It has been known for some time that the boundary theory of 3d dS gravity takes the form of a Liouville CFT with a complex central charge \cite{Klemm:2002ir} of the form \eqref{cpm}. A somewhat more recent insight is that in physical amplitudes of 3d gravity, the dual CFT should be coupled to 2d gravity and turned into a generally covariant $\mathbb{C}$LS worldsheet theory \cite{Verlinde:2024zrh,Collier:2025lux}. 
    \item Amplitudes of  $\mathbb{C}$LS are known to compute ``big-bang'' wavefunctions at $\mathscr{I}_+$ of 3d cosmology \cite{Collier:2025lux}, with reflecting boundary condition that SL(2,$\mathbb{C}$) CS gauge fields obey $A_+=A_-$. This essentially projects the gravity theory to a chiral subsector $\alpha_+=\alpha_-$ of dS$_3$ gravity (on a doubled spacetime, because of the reflection). As we do not have such a reflecting boundary in the KdS static patch, it is reasonable to land on a dual theory that combines the two chiral sectors into a tensor product  $\mathbb{C}$LS $\otimes$ $\mathbb{C}$LS  of two complex Liouville string theories.
\end{enumerate}

\smallskip

\subsection{Perspective 3. SYK}\label{sect4.3DSSYK}

One ambition of our work is to uncover direct hints of a microscopic realization of worldline holography, in which KdS gravity arises as the dual of a 1d quantum many body system. A promising candidate of such a 1d quantum system is the double scaled SYK  model \cite{Berkooz:2018jqr}, which already has known connections to the $\mathbb{C}$LS worldsheet theory \cite{Verlinde:2024zrh,Blommaert:2024ydx,Blommaert:2025eps} and 3d dS gravity \cite{Verlinde:2024znh,Gaiotto:2024kze}. In this work we propose the precise embedding of DSSYK in 3d quantum cosmology, at the level of the spectral density. We suggest logical next steps towards developing this microscopic worldline hologram \cite{Anninos:2011af} in the discussion section \ref{sect5.3gsigma}.

Recall the SYK Hamiltonian \cite{kitaev2015simple,Maldacena:2016hyu}
\bea
    H_\text{SYK}\is \i^{p/2}\sum_{i_1<\dots <i_p}J_{i_1\dots i_p}\psi_{i_1}\dots \psi_{i_p}\,,\quad \hbar=\frac{4 p^2}{N}\,.
\eea
In DSSYK \cite{Cotler:2016fpe,Berkooz:2018jqr}, one scales $N\to \infty$, $p\to\infty$ with the ratio $p^2/N$ finite. In the DSSYK/dS$_3$ dictionary proposed in \cite{Narovlansky:2023lfz}, this ratio is identified with the Newton constant $\hbar=16 \pi G$ (in de Sitter units) in 3d dS gravity.
We will provide direct quantitative evidence for this relation. 

 Parameterizing the DSSYK energy as $E=\cos(\alpha_+/2)$, one derives the following exact DSSYK spectral density \cite{Berkooz:2018jqr,Berkooz:2024lgq}:
\bea
    \boxed{\, \rho_\text{SYK}(\alpha_+) = \big(e^{\pm \i\alpha_+};\sfq \big)_\infty\,, \quad \sfq = e^{-\hbar/2}\large \strut \,}\label{3.xxzsyk}
\eea
The DSSYK model admits an exact description in terms of the $G\Sigma$ collective field theory \cite{Maldacena:2016hyu}. In the double scaling limit, the $G\Sigma$ field theory takes on the form of a complex Liouville theory with central charge given in \eqref{cpm} with $4\pi \sfb^2 = \hbar$. This theory lives on the kinematic space of the DSSYK thermal circle, which for standard DSSYK takes the form of a M\"obius strip (a disk with a crosscap) \cite{Lin:2023trc}.

The duality between DSSYK and $\mathbb{C}$LS is supported \cite{Blommaert:2025eps} via the match between the DSSYK partition function and the  $\mathbb{C}$LS amplitude on a disk with a standard crosscap:
\bea\label{3.14cc}
    \rho_{\mathbb{C}\rm LS\tiny\strut}^{(-2,1)}(\alpha_+)\, =\quad \begin{tikzpicture}[baseline={([yshift=-.95ex]current bounding box.center)}, xscale=-0.67,yscale=.67]
 \pgftext{\includegraphics[scale=1]{KdS10.pdf}} at (0,0);
    \draw (-2.6,-2) node {\color{blue}$\alpha_+$};
    \draw (-0.6,0.6) node {\color{red}\small $\ket{\text{C}_{(-2,1)}}$};
    \draw (-3.4,1.7) node {\color{blue}\small worldline};
    \draw (-3.4,2.4) node {\color{blue}$G\spc \Sigma$};
    \draw (0,-1.8) node {\color{black}\small crosscap};
    \draw (0,-1.1) node {\color{black}\small $\mathbb{C}$LS};
    \draw (-1.1,-0.3) node {\color{black}$\tau$};
  \end{tikzpicture} 
\eea
The proof of this duality \cite{Blommaert:2025eps}, via lightcone gauge fixing of $\mathbb{C}$LS, was discussed below equation \eqref{fzztpair}.
The associated $\mathbb{C}$LS crosscap amplitude was  derived in \cite{Blommaert:2025eps} (see also section \ref{sect:5cls}) with the result:
\begin{align}
    \rho_{\mathbb{C}\rm LS}^{(-2,1)}(\alpha_+) &=  \int_0^{\i \infty} \d \tau \, Z_\TT{ghost}(\tau) \bra{\alpha_+} e^{\i \pi \tau H} \ket{\mathbb{1}_{(-2,1)}}\nonumber\\
    &=\sum_{n=-\infty}^{+\infty} e^{-\frac{1}{\hbar}(\alpha_+ + 4 \pi n)^2} \sin(\alpha_+)\sinh\bigl(2\pi(\alpha_++4\pi n)/\hbar\bigr)=\big(e^{\pm \i\alpha_+};\sfq\big)_\infty\,.\label{rho21sum}
\end{align}
This indeed exactly matches with the exact DSSYK spectral density \eqref{3.xxzsyk}:
\begin{equation}
    \boxed{\, \rho_{\mathbb{C}\rm LS}^{(-2,1)}(\alpha_+)  = \rho_{\rm SYK}(\alpha_+)\,\large \strut }\label{3.16string}
\end{equation}

The embedding of the $\mathbb{C}$LS disk \eqref{3.14cc} in the 3d KL$(-2,1)$ geometry \eqref{2.110torusstate} is pictorially the following
\begin{equation}
    \begin{tikzpicture}[baseline={([yshift=-.5ex]current bounding box.center)}, xscale=-0.67,yscale=.67]
 \pgftext{\includegraphics[scale=1]{KdS9.pdf}} at (0,0);
    \draw (-4.8,-2) node {\color{blue}$(\alpha_+,\alpha_-)$};
    \draw (6,0) node {\color{red}horizon $(-2,1)$};
    \draw (-2,0.6) node {\color{red}$\ket{\text{C}_{(-2,1)}}$};
    \draw (-5,1.7) node {\color{blue}worldline};
    \draw (-5,2.4) node {\color{blue}$G\Sigma\otimes G\Sigma$};
    \draw (-1.4,-1.9) node {\color{black}brane};
    \draw (-1.4,-1.2) node {\color{black}CLS $\otimes$ CLS};
    \draw (-2.5,-0.3) node {\color{black}$\tau$};
    \draw (6.6,2.3) node {\color{black}KL$(-2,1)$};
  \end{tikzpicture}
\end{equation}

\noindent
Whilst we have not derived a detailed 3d embedding of the above amplitude, the above results suggest that we can think of the DSSYK model as placed on the circular worldline in KL$(-2,1)$. The boundary of the $\mathbb{C}$LS disk becomes the pode while the location of the $\mathbb{C}$LS crosscap state $\ket{\text{C}_{(-2,1)}}$ is geometrically related with the (-2,1) cosmological horizon where the 2d torus degenerates. 

Formula \eqref{zdouble} (given below) as well as the equivalence between  $\mathbb{C}$LS and the $G\Sigma$ theory \cite{Blommaert:2025eps} recalled above, indicate that that KL$(-2,1)$ can be described by a $G\Sigma\otimes G\Sigma$-like theory where the second $G\Sigma$ theory has an opposite action (reflecting $\hbar\to-\hbar$). 
In similar spirits, it is tempting to look for a possible microscopic dual to the ${\mathbb{C}}$LS theory on the SL(2,$\mathbb{Z}$) crosscap, generalizing DSSYK. We comment further on both possible dualities in the concluding section \ref{sect5.3gsigma}.

\subsection{Overview}

In this section, we have presented three dual perspectives on Kerr-de Sitter gravity. Along the way, we have introduced several notions of a spectral density:
\bea
\rho_{(c,d)}(\alpha_+) \is \mbox{chiral spectral density \eqref{kl-amp-chiral} of 3d gravity on $(c,d)$ Kerr-lens}\notag\\[2mm]
\rho^{(c,d)}_{\mathbb{V}\rm TFT \tiny\strut}(\alpha_+) \is \mbox{line amplitude \eqref{kl-vtft-amp} of $\mathbb{V}$TFT on $(c,d)$  Kerr-lens}\notag \\[2mm]
\rho^{(c,d)}_{\mathbb{C}\rm LS \tiny\strut}(\alpha_+)\is \mbox{amplitude \eqref{cls-vtft-match} of $\mathbb{C}$LS on a disk with $(c,d)$ crosscap insertion}\notag \\[2mm]
\rho_{\rm SYK \tiny\strut}(\alpha_+)\is \mbox{spectral density  \eqref{3.xxzsyk} of DSSYK} \notag
\eea
We have argued, based on our earlier exposition in sections \ref{sect:2.1solutions} and \ref{sect:3torusquant} and the (anticipated) results of \cite{Blommaert:2025eps} and section \ref{sect:5cls} that these quantities exactly agree for $(c,d)=(-2,1)$, with the explicit formula in \eqref{rho21sum}:
\bea
\boxed{\rho_\text{SYK\tiny\strut}(\alpha_+)\,\, = \, \rho_{\mathbb{C}\rm LS\tiny\strut}^{(-2,1)}(\alpha_+)
\, = \, \rho_{\mathbb{V}\rm TFT\tiny\strut}^{(-2,1)}(\alpha_+) \, = \, \sum_{n=-\infty}^{\infty} \rho_{(-2,1)}(\alpha_+ + 4\pi n)\,}\label{3.23proposal}
\eea

Based on this, we propose an exact duality between DSSYK and a chiral half of pure 3d cosmology on the $(-2,1)$ Kerr-lens spacetime. The non-chiral $\mathbb{V}$TFT amplitude on KL$(-2,1)$ becomes the product:
    \bea
    \label{zdouble}
\rho_{\mathbb{V}\text{TFT}}^{(-2,1)}(\alpha_+,\alpha_-) \is
        \rho_{\mathbb{C}\rm LS\tiny\strut}^{(-2,1)}(\alpha_+)\, \rho_{\mathbb{C}\rm LS\tiny\strut}^{(2,1)}(\alpha_-) \, =\, \big(e^{\pm \i\alpha_+};\sfq \big)_\infty\big(e^{\pm \i\alpha_-};\sfq^{-1} \big)_\infty
    \eea
We propose that this amplitude is an exact quantum gravity contribution associated to the KL$(-2,1)$ geometry. By replacing the Pochhammers by the infinite sum \eqref{rho21sum}, we find it matches the expression \eqref{2.96wave} for semiclassical KL$(-2,1)$ plus a sum over shifted saddles $(\alpha_+ + 4\pi k_+, \alpha_- + 4\pi  k_-)$
\bea
\rho_{\mathbb{V}\text{TFT}}^{(-2,1)}(\alpha_+,\alpha_-) &  \underset{G\to 0}{\to} &\, e^{-\frac{1}{\hbar}(\alpha_+^2-\alpha_-^2) + \frac{2\pi}{\hbar}(\alpha_++\alpha_-)} + \dots 
\eea
Shifted angles have identical holonomy and therefore are indistinguishable for probes that do not directly measure the curvature singularity at the observer's trajectory. This suggests that it may be reasonable to sum over such saddles in gravitational calculations, even though we do not commit to this right now.

\begin{figure}[t]
    \centering
\quad    \includegraphics[width=.92 \linewidth]{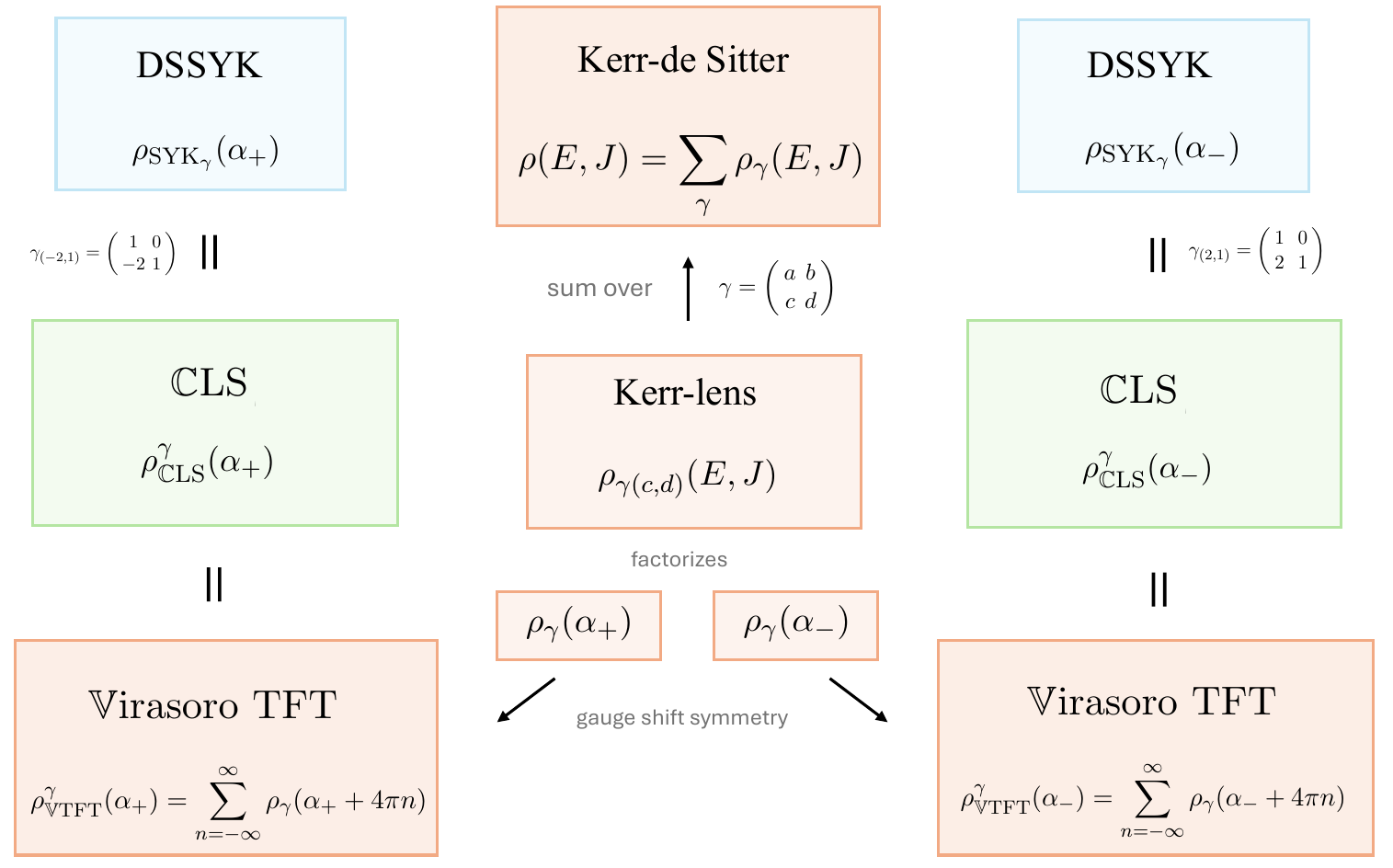}
    \caption{KdS triality. Our proposed Kerr-de Sitter spectral density is a Poincar\'e sum of Kerr-lens amplitudes labeled by $\gamma \in$ SL(2,$\mathbb{Z})$. The spectral densities of complex Virasoro TFT on Kerr-lens, the complex Liouville string with a generalized crosscap and DSSYK are all equal (the latter only for a specific choice of $\gamma$) and given by a shift symmetric summation of chiral Kerr-lens amplitudes.} 
    \label{fig:kds-duality}
\end{figure}

Figure \ref{fig:kds-duality} displays an overview of this Kerr-de Sitter duality web.

\section{$\mathbb{C}$LS $\otimes$ $\mathbb{C}$LS quantization}\label{sect:5cls}

In this section we consider two copies of the complex Liouville string or $\mathbb{C}$LS \cite{Collier:2024kmo} as a proposed starting point for constructing the exact spectrum of pure 3d quantum cosmology. In \textbf{section \ref{sect4.1Cstates}}, we describe generalized crosscap states $\ket{\text{C}_{(c,d)}}$ from the perspective of 2d CFT. Their key feature is that they define a smooth geometry. The wavefunction of the standard crosscap $\ket{\text{C}_{(-2,1)}}$ is a Virasoro identity modular matrix element. The wavefunction of $\ket{\text{C}_{(c,d)}}$ is a Virasoro identity matrix elements associated with the SL(2.$\mathbb{Z}$) element $\gamma(c,d)$. We were not able to identity prior reference for such generalized crosscaps. In \textbf{section \ref{subsec: slz crosscap amplitudes}} we derive equations \eqref{m-exact}/\eqref{cls-vtft-match}. We find, as announced, that this amplitude decomposes into a sum of gravitational saddles associated with Kerr-lens spaces. In \textbf{section \ref{sect4.3:open}} we discuss a dual channel in which the amplitudes have an interpretation as a twisted trace over an open string Hilbert space, generalizing the Mobius band dual of the usual crosscap \cite{blumenhagen2009boundary}. Finally, in \textbf{section \ref{sect5.5entropy}}, we study the $\mathbb{C}$LS $\otimes$ $\mathbb{C}$LS amplitudes. We do the sum over $\gamma$ to obtain our final equation \eqref{1.26final} for the observer's spectral density, and discuss some of its properties. Details are gathered in \textbf{appendix \ref{app:stringdetails}}.

\subsection{SL(2,$\mathbb{Z}$) crosscaps}\label{sect4.1Cstates}
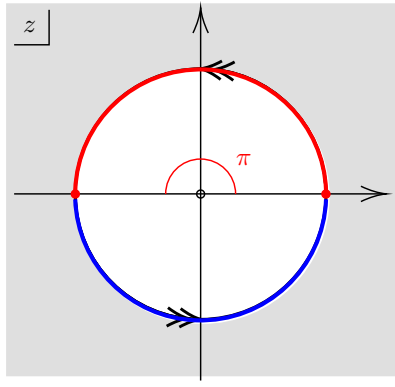
\begin{figure}[t]
    \centering
\raisebox{2mm}{\resizebox{!}{5cm}{\tikzset{every picture/.style={line width=0.75pt}} 

\tikzset{
    partial ellipse/.style args={#1:#2:#3}{
        insert path={+ (#1:#3) arc (#1:#2:#3)}
    }
}   
\begin{tikzpicture}[x=0.75pt,y=0.75pt,yscale=-1,xscale=1]

\draw[lightgray, fill=lightgray, opacity=.5] (175,380) -- (175,635) -- (445,635) -- (445,380) -- cycle;
\draw[white,fill=white][line width=.75mm]  (309,512) [partial ellipse=0:360:2.27cm and 2.27cm];

\draw [line width=0.75]    (308,638.4) -- (308,384.4) ;
\draw [shift={(308,382.4)}, rotate = 90] [color={rgb, 255:red, 0; green, 0; blue, 0 }  ][line width=0.75]    (17.49,-5.26) .. controls (11.12,-2.23) and (5.29,-0.48) .. (0,0) .. controls (5.29,0.48) and (11.12,2.23) .. (17.49,5.26)   ;
\draw [line width=0.75]    (180,510.3) -- (433.5,510.3) ;
\draw [shift={(435.5,510.3)}, rotate = 180] [color={rgb, 255:red, 0; green, 0; blue, 0 }  ][line width=0.75]    (17.49,-5.26) .. controls (11.12,-2.23) and (5.29,-0.48) .. (0,0) .. controls (5.29,0.48) and (11.12,2.23) .. (17.49,5.26)   ;
\draw  [line width=1.5]  (394,510.4) .. controls (394,462.9) and (355.5,424.4) .. (308,424.4) .. controls (260.5,424.4) and (222,462.9) .. (222,510.4) .. controls (222,557.9) and (260.5,596.4) .. (308,596.4) .. controls (355.5,596.4) and (394,557.9) .. (394,510.4) -- cycle ;
\draw   (306.63,512.77) .. controls (307.94,513.52) and (309.61,513.07) .. (310.37,511.77) .. controls (311.12,510.46) and (310.67,508.79) .. (309.37,508.03) .. controls (308.06,507.28) and (306.39,507.73) .. (305.63,509.03) .. controls (304.88,510.34) and (305.33,512.01) .. (306.63,512.77) -- cycle ;
\draw  [draw opacity=0][line width=0.75]  (284,510.4) .. controls (284,510.4) and (284,510.4) .. (284,510.4) .. controls (284,510.4) and (284,510.4) .. (284,510.4) .. controls (284,497.15) and (294.75,486.4) .. (308,486.4) .. controls (321.25,486.4) and (332,497.15) .. (332,510.4) -- (308,510.4) -- cycle ; \draw [color={rgb, 255:red, 255; green, 0; blue, 0 }  ,draw opacity=1 ][line width=0.75]    (284,510.4) .. controls (284,510.4) and (284,510.4) .. (284,510.4) .. controls (284,497.15) and (294.75,486.4) .. (308,486.4) .. controls (321.25,486.4) and (332,497.15) .. (332,510.4) ;  
\draw   (204,383.97) -- (204,407.97) -- (180,407.97) ;
\draw  [draw opacity=0][line width=1.5]  (222,510.4) .. controls (222,557.9) and (260.5,596.4) .. (308,596.4) -- (308,510.4) -- cycle ; \draw [line width=1.5]    (222,510.4) .. controls (222,557.9) and (260.5,596.4) .. (308,596.4) ; \draw [shift={(308,596.4)}, rotate = 184.18] [color={rgb, 255:red, 0; green, 0; blue, 0 }  ][line width=1.5]    (22.93,-6.37) .. controls (17.76,-2.99) and (13.02,-0.87) .. (8.72,0) .. controls (13.02,0.87) and (17.76,2.99) .. (22.93,6.37)(14.21,-6.37) .. controls (9.04,-2.99) and (4.3,-0.87) .. (0,0) .. controls (4.3,0.87) and (9.04,2.99) .. (14.21,6.37)   ; 
\draw  [draw opacity=0][line width=1.5]  (394,510.4) .. controls (394,462.9) and (355.5,424.4) .. (308,424.4) -- (308,510.4) -- cycle ; \draw [line width=1.5]    (394,510.4) .. controls (394,462.9) and (355.5,424.4) .. (308,424.4) ; \draw [shift={(308,424.4)}, rotate = 4.18] [color={rgb, 255:red, 0; green, 0; blue, 0 }  ][line width=1.5]    (22.93,-6.37) .. controls (17.76,-2.99) and (13.02,-0.87) .. (8.72,0) .. controls (13.02,0.87) and (17.76,2.99) .. (22.93,6.37)(14.21,-6.37) .. controls (9.04,-2.99) and (4.3,-0.87) .. (0,0) .. controls (4.3,0.87) and (9.04,2.99) .. (14.21,6.37)   ; 
\draw  [color={rgb, 255:red, 255; green, 0; blue, 0 }  ,draw opacity=1 ][fill={rgb, 255:red, 255; green, 0; blue, 0 }  ,fill opacity=1 ] (224.73,510.4) .. controls (224.73,508.89) and (223.51,507.67) .. (222,507.67) .. controls (220.49,507.67) and (219.27,508.89) .. (219.27,510.4) .. controls (219.27,511.91) and (220.49,513.13) .. (222,513.13) .. controls (223.51,513.13) and (224.73,511.91) .. (224.73,510.4) -- cycle ;
\draw  [color={rgb, 255:red, 255; green, 0; blue, 0 }  ,draw opacity=1 ][fill={rgb, 255:red, 255; green, 0; blue, 0 }  ,fill opacity=1 ] (396.73,510.4) .. controls (396.73,508.89) and (395.51,507.67) .. (394,507.67) .. controls (392.49,507.67) and (391.27,508.89) .. (391.27,510.4) .. controls (391.27,511.91) and (392.49,513.13) .. (394,513.13) .. controls (395.51,513.13) and (396.73,511.91) .. (396.73,510.4) -- cycle ;

\draw[blue][line width=.75mm]  (308,511) [partial ellipse=2:178:2.27cm and 2.27cm];
\draw[red][line width=.75mm]  (308,511) [partial ellipse=182:358:2.27cm and 2.27cm];
\draw (330.67,481.13) node [anchor=north west][inner sep=0.75pt]  [font=\normalsize,color={rgb, 255:red, 255; green, 0; blue, 0 }  ,opacity=1 ,xscale=1.2,yscale=1.2]  {$\pi $};
\draw (184.51,389.8) node [anchor=north west][inner sep=0.75pt]  [xscale=1.2,yscale=1.2]  {$z$};

\end{tikzpicture}}}
    \caption{The quotient $z \bar{z}=-1$ implements a crosscap. Indeed at $\abs{z}=1$ we identify the antipodal points. The quotiented geometry is the region $\abs{z}\geq 1$ with antipodal boundary conditions at $\abs{z}=1$.}
    \label{fig:crosscap regular}
\end{figure}

We first recall the definition of the crosscap state $\ket{\text{C}_{(-2,1)}}$ in 2d CFT \cite{blumenhagen2009boundary}, and more specifically \cite{Hikida:2002bt, Blommaert:2025eps} in a complex Liouville CFT with central charge $c_+=13+6\i\, (\sfb^2-\sfb^{-2})$. In the following we often write
\begin{equation}
    \hbar =4\pi \sfb^2\,,
\end{equation}
and leave implicit $+$ sub/superscripts. Topologically a crosscap is made by cutting out a hole in a surface and identifying antipodal points on the boundary. In 2d CFT in radial quantization one implements a crosscap by imposing the following $\mathbb{Z}_2$ quotient:
\begin{equation} \label{crosscap quotient}
    z \to -1/{\bar{z}}\,.
\end{equation}
At $\abs{z}=1$, the quotient \eqref{crosscap quotient} indeed amounts to an antipodal $\mathbb{Z}_2$ identification, as shown in figure \ref{fig:crosscap regular}.

In radial quantization, from the perspective of the closed string channel, the crosscap boundary condition at $\abs{z}=1$ corresponds with a boundary state $\ket{\text{C}_{(-2,1)}}$. What are the properties of this state? The antipodal identification relates left-and right moving modes in the initial state. At the level of the stress tensor (and its Fourier modes) it implements the following constraints on so-called crosscap Ishibashi states $\ishiket{\beta_{(-2,1)}}$:
\begin{equation} \label{crosscap local}
   T(z) = \bar{T}(-1/\bar{z})   \quad \to  \quad  (L_n - e^{\i \pi n} \bar{L}_{-n})\ishiket{\beta_{(-2,1)}}   = 0
\end{equation}
For a pedagogical introduction  to crosscap Ishibashi states, we refer the reader to \cite{blumenhagen2009boundary}. Using the Virasoro algebra we see that crosscap Ishibashi state can be expressed in terms of the ordinary Ishibashi state $|\beta\rangle\!\rangle$ (the boundary state associated with a circular reflecting boundary without antipodal identification, projected onto a given Virasoro primary sector) as follows:
\begin{equation}
    \ishiket{\beta_{(-2,1)}}=e^{\i\pi L_0}\ishiket{\beta}\,,\quad (L_n-\bar{L}_{-n})\ishiket{\beta}=0\,.
\end{equation}
For future purposes, we quote the overlap between an Ishibashi state $\ishiket{\alpha}$ and a crosscap Ishibashi state
\begin{equation}
    \ishibra{\alpha}e^{\i \pi \tau H}\ishiket{\beta_{(-2,1)}}=\delta(\alpha-\beta)\,\chi_\beta \Bigr( \tau +\frac{1}{2}\Bigr) +\,(\alpha \to -\alpha)\,,\quad \chi_\beta(\tau)=\frac{e^{-\frac{1}{2 \hbar}\alpha^2\tau}}{\eta(\tau)}\,.
\end{equation}
The shift $\tau+1/2$ encodes the $\pi$ rotation on the $\abs{z}=1$ circle between identified points.

The local condition \eqref{crosscap local} does not uniquely specify this state $\ket{\TT{C}_{(-2,1)}}$. An additional requirement follows from open-closed duality \cite{blumenhagen2009boundary}: an annulus with a crosscap on one end is a M\"obius strip from an open string perspective. In Liouville CFT, this imposes that the closed amplitude between the crosscap $\ket{\TT{C}_{(-2,1)}}$ and an identity $\ket{\TT{ZZ}}$ brane matches with the open channel trace of the identity representation over a M\"obius strip \cite{blumenhagen2009boundary}. This trace computes a Virasoro vacuum character:\footnote{The contour is such that this integral converges.}
\bea \label{crosscap open closed}
    \bra{\TT{ZZ}} e^{\i \pi \tau H} \ket{\TT{C}_{(-2,1)}} \is \chi_{\mathbb{1}}\Bigl(\!- \frac{1}{4 \tau} - \frac{1}{2}\Bigr)=\int_\mathcal{C}\d \beta\,\,\mathsf{K}^{(-2,1)}_{\mathbb{1}\beta}\,  \chi_\beta \Bigl(\tau+\frac{1}{2}\Bigr)\,.
\eea
Here, $\mathsf{K}^{(-2,1)}_{\mathbb{1}\beta}$ denotes the modular matrix associated with the SL(2,$\mathbb{Z}$) matrix element
\bea
    \gamma(-2,1)=\biggl(\!{\begin{array}{cc}
        1\! & 0 \\[-.5mm]
        -2\! & 1
    \end{array}}\!\biggr)\,.
\eea
This maps $\tau+1/2$ onto $-1/4\tau-1/2$. Here, $-1/4\tau$ is the propagation time on the M\"obius band \cite{blumenhagen2009boundary} in the open string channel. The shift by $-1/2$ implements the twist to create the M\"obius band. So, the crosscap state $\ket{\TT{C}_{(-2,1)}}$ is uniquely fixed by the fact that it prepares the vacuum in the open channel. The fact that this is the vacuum representation implements that, after lifting the geometry to 3d, the horizon is smooth \eqref{2.29smooth} in the dual KL$(-2,1)$ geometry. The explicit form of the modular matrix $\mathsf{K}^{(-2,1)}_{\mathbb{1}\beta}$  reads\cite{Hikida:2002bt}:
\bea
\mathsf{K}^{(-2,1)}_{\mathbb{1}\beta}\is e^{\frac{\beta^2}{4\hbar}} \cos({\beta}/{4}) \cosh({\pi \beta}/{\hbar})\,.
\eea
We will think of $\mathsf{K}^{(-2,1)}_{\mathbb{1}\beta}$ as the wavefunction of the crosscap state $\ket{\TT{C}_{(-2,1)}}$. 

\begin{figure}[t]
    \centering
\raisebox{2mm}{\resizebox{!}{5cm}{\input{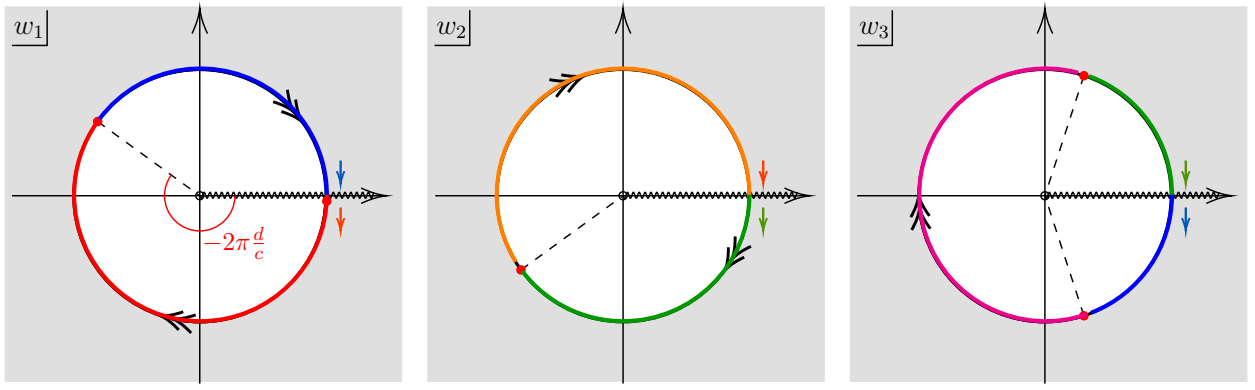}}}
    \caption{To make a $(c,d)$ crosscap we start from a $d$-fold cover $z^{1/d}$ with sheets $w_1, w_2, \dots, w_d$. We take out the unit circle and identify points at $\abs{z}=1$ under a $-2\pi d/c$ rotation. We display the case $(c,d) = (5,3)$.}
    \label{fig:crosscap generalized}
\end{figure}

The above story can be readily generalized to define what we will call an \slz{} crosscap associated with a general modular matrix\footnote{We consider $c\neq 0$. The case $c=0$ only occurs when $d=1$ which describes a particle in ordinary $S^3$ (or the disk in 2d).}
\begin{equation}
    \gamma(c,d) = \biggl(\!{\begin{array}{cc}
        a\! & b\\[-.5mm]
        c\! & d
    \end{array}}\!\biggr)\, \,\in \, \TT{SL}(2,\mathbb{Z})   \, .
\end{equation}
We define a \slt{} crosscap by considering a $d$-fold cover $z^{1/d}$ of the complex plane with a $\mathbb{Z}_{c}$ quotient:
\bea
    z^{1/d} \; \to \; \frac{e^{-2\pi \i/c}} {\bar{z}^{1/d}} \,  . \label{4.15quotient}
\eea
This quotient is shown in figure \ref{fig:crosscap generalized}. 
We can use the identification to restrict to the exterior region $|z|>1$ of the unit circle. On the unit circle $\abs{z}=1$, this identifies points under a rotation $-2\pi d/c$.

This generalized crosscap boundary at $\abs{z}=1$ defines again a boundary state $\ket{\text{C}_{(c,d)}}$. Let us deduce the properties of this state, following the same steps as before. The quotient in \eqref{4.15quotient} implies a relation between left-and right moving stress tensors (and their Fourier modes):
\bea
    \boxed{\; T(z)=\bar{T}(e^{-2\pi \i \frac{d}{c}}/\bar{z})\quad \to \quad \bigl(L_n - e^{-2 \pi \i \frac{d}{c}n} \bar{L}_{-n}\bigr)\ishiket{\beta_{(c,d)}}=0 \Large \strut \,}
\eea
One might call the solutions $\ishiket{\beta_{(c,d)}}$ to this set of conditions the $(c,d)$-crosscap Ishibashi states:
\bea
    \ishiket{\beta_{(c,d)}}  = e^{- 2 \pi \i \frac{d}{c} L_0} \ishiket{\beta}\,.\label{5.26ccI}
\eea
Their overlap with an ordinary Ishibashi state $\ishiket{\alpha}$ reads:
\begin{equation}
    \ishibra{\alpha}e^{\i \pi \tau H}\ishiket{\beta_{(c,d)}}=\delta(\alpha-\beta)\,\chi_\beta \Bigl( \tau -\frac{d}{c}\spc \Bigr) +\,(\alpha \to -\alpha)\,.\label{4.18}
\end{equation}
Notice the appearance of the shift by $-d/c$. 

Once again, to completely determine the \slz{} crosscap state $\ket{\TT{C}_{(c,d)}}$, we impose that its annulus amplitude with the $\ket{\TT{ZZ}}$ state leads to an identity state propagating in the open string channel. In this case, the correct equation reads:
\bea
    \boxed{\; \bra{\TT{ZZ}} e^{\i \pi \tau H} \ket{\TT{C}_{\gamma}} \, =\, \chi_{\mathbb{1}}\Bigl(- \frac{1}{c^2 \tau} + \frac{a}{c}\, \Bigr)=\int_\mathcal{C}\d \beta\,\,\mathsf{K}^{(c,d)}_{\; \mathbb{1}\beta}\, \,  \chi_\beta \Bigl(\tau-\frac{d}{c}\spc\Bigr)\,. \Large \strut}\label{5.28chi}
\eea
The appearance of the argument $-1/c^2\tau+a/c$ in this character is most directly understood as follows. Following equation \eqref{4.18}, the annulus amplitude decomposes into characters of $\tau-d/c$. An open string character should depend on some combination ${\rm const}_1/\tau+{\rm const}_2$. This determines the modular matrix to be $K^{\gamma(c,d)}$. This leads to the character on the left with $-1/c^2\tau$ as a propagation time in the open string channel. The interpretation of the shift by $a/c$ is discussed in section \ref{sect4.3:open}. The modular matrix is found from equation \eqref{5.28chi} to be:
\bea\label{modular p matrix}
    \boxed{\ \mathsf{K}^{(c,d)}_{\; \mathbb{1}\beta}\, =\, e^{\i \pi \phi(c,d)}\frac{1}{\sqrt{c}}e^{-\frac{d}{2c\hbar} \beta^2 + \frac{2\pi}{c \hbar}\beta} \sin(\frac{2\pi a - \beta}{2c})+ \,(\beta\to-\beta){}_{\strut}^{\strut}\ }
\eea
The phase factor involves the Dedekind sum $s(d,c)$, given in Appendix A:\footnote{Our eventual application of the above formulas is the calculation of the non-chiral Kerr-lens amplitude $\rho_{(c,d)}(E,J)$. This amplitude involves the product of two $\mathbb{C}$LS amplitudes, and the above phase will then cancel. For notational simplicity, we will henceforth drop this phase, with the understanding that if one were to track it throughout the calculations in the remainder of this section, one would land on the same final gravitational amplitudes $\rho_{(c,d)}(E,J)$.}
\begin{equation}
   \phi(c,d)=s(d,c) -\frac{a+d}{12 c} \label{normalization constant}
\end{equation}
Again, we will interpret $\mathsf{K}^{(c,d)}_{\; \mathbb{1}\beta}$ as an exact wavefunction associated with crosscap boundary state $\ket{\text{C}_{(c,d)}}$.

\subsection{SL(2,$\mathbb{Z}$) crosscap amplitude}\label{subsec: slz crosscap amplitudes}

\begin{figure}[h]
    \centering
\raisebox{2mm}{\resizebox{0.3\linewidth}{!}{\input{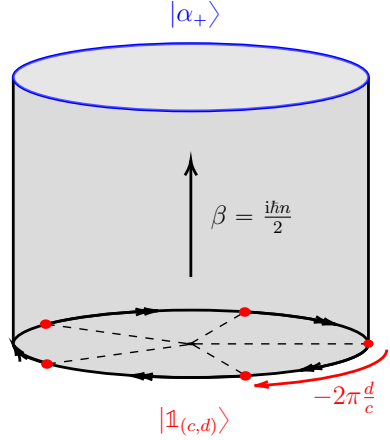}}}
    \caption{The \slz{} crosscap amplitude \eqref{5.31cc} is an annulus amplitude between a generalized \slz{} crosscap state and an FZZT boundary, corresponding to the horizon and observer respectively. The crosscap state depends on a modular matrix $\gamma(c,d)$. Physical states in the closed channel are on-shell primaries with discretized imaginary momenta $\beta \in \frac{\i \hbar}{2} \mathbb{Z}$.}
    \label{fig: string amplitude closed}
\end{figure}

We study $\mathbb{C}$LS amplitudes 
\begin{equation}
    \rho^{\mathbb{C}\rm LS}_{(c,d)}(\alpha_+) = \int_0^{\i \infty} \d \tau \, Z_\TT{ghost}(\tau) \bra{\alpha_+} e^{\i \pi \tau H_\text{string}} \ket{\mathbb{1}_{(c,d)}}\,,\label{5.31cc}
\end{equation}
where the important boundary states are:
\bea \label{5.23boundary states}
\ket{\mathbb{1}_{(c,d)}} = \ket{\text{C}_{(c,d)}}\otimes e^{-2\pi \i \frac{d}{c}L_0^-}\ket{\text{ZZ}}\,,\quad \ket{\alpha_+}  = \ket{\text{FZZT}_{\bullet}(\alpha_+)}\otimes \ket{\TT{FZZT}(0)}\,.
\eea
The FZZT boundary state $\ket{\TT{FZZT}_\bullet(\alpha_+)}$ fixes the holonomy $\alpha_+$ associated to the Kerr black hole, while the crosscap state $\ket{\TT{C}_{(c,d)}}$ encodes the geometry of the horizon, i.e. which lens space we are considering. The precise definition of the boundary states \eqref{5.23boundary states} can be consulted in appendix \ref{app: states}. Our construction parallels the one of \cite{Blommaert:2025eps}, with the regular crosscap replaced by its generalized \slz{} counterpart. The \slz{} crosscap amplitude is pictured in figure \ref{fig: string amplitude closed}.

As we noted in equation \eqref{sym from diff}, the classical $\TT{KL}(c,d)$ metric is symmetric under 
\begin{equation}
    \alpha_+ \leftrightarrow \alpha_- \,, \qquad  b \to -b \,, \qquad c \to -c \, .\label{sym}
\end{equation}
This is promoted to an exact symmetry for KL$(c,d)$ if the anti-holomorphic part of the gravity amplitude $\rho_{(c,d)}(\alpha_+,\alpha_-)$ in equation \eqref{kl-amp-exact} is computed using the following $\mathbb{C}$LS amplitude:
\begin{equation}
    \rho^{\mathbb{C}\rm LS}_{(c,d)}(\alpha_-) = \int_0^{\i \infty} \d \tau \, Z_\TT{ghost}(\tau) \bra{\alpha_-} e^{\i \pi \tau H_\text{string}} \ket{\mathbb{1}_{(-c,d)}}\,,
\end{equation}
where the relevant modular matrix is modified according to the symmetry \eqref{sym}:
\begin{equation}  \label{g tilde definition}
     \boxed{\gamma_+ = \biggl(\!{\begin{array}{cc}
        a\! & b\\[-.5mm]
        c\! & d
    \end{array}}\!\biggr) \quad \to \quad   \gamma_- = \biggl(\!{\begin{array}{cc}
        a\! & \!-b\\[-.5mm]
       \! - c\! & d
    \end{array}}\!\biggr)\,}
\end{equation}

We now compute the chiral wavefunction \eqref{5.31cc}. The contribution from the $\phi_+$ Liouville theory is:
\bea\label{cls + amplitude}
    \bra{\TT{FZZT}_\bullet(\alpha_+)} e^{\i \pi \tau H} \ket{\TT{C}_{(c,d)}}  =   \int_{\mathcal{C}} \d \beta \, \frac{\beta}{\sinh(2\pi\beta/\hbar)}\, \mathsf{K}^{{(c,d)}}_{\mathbb{1} \beta} \, \mathsf{S}_{\beta\,\alpha_+ + 2\pi} \, \chi^+_{\beta}\bigg(\tau - \frac{d}{c} \bigg)\,.
\eea
The wavefunction $\mathsf{K}^{{(c,d)}}_{\mathbb{1} \beta}$ is that of the crosscap state \eqref{5.28chi}. The modular $\mathsf{S}$ matrix $\mathsf{S}_{\beta\,\alpha_+ + 2\pi}$ combined with an extra $\beta$-dependent measure factors is the wavefunction of the spectral state $\ket{\TT{FZZT}_\bullet(\alpha_+)}$. The contribution of the $\varphi_-$ Liouville theory is much simpler:
\bea
    \bra{\TT{FZZT}(0)} e^{\i \pi \tau H} e^{-2\pi \i \frac{d}{c} L_0}\ket{\TT{ZZ}} =\int_\mathcal{C} \d \bar{\beta}  \, \chi^-_{\bar{\beta}}\bigg(\tau - \frac{d}{c} \bigg)\,.\label{barbeta}
\eea
The state $e^{-2\pi \i \frac{d}{c} L_0}\ket{\TT{ZZ}}$ satisfies according to equation \eqref{5.26ccI} generalized crosscap Ishibashi boundary conditions. A side comment for the present purposes, is that these $\varphi_-$ states have the interpretation as fixing a physical time coordinate for the $\mathbb{C}$LS, which becomes a physical coordinate in the gauge-fixed $G\Sigma$ formulation \cite{Blommaert:2025eps}. See section \ref{sect4.3DSSYK} and section \ref{sect5.3gsigma}. Here $\ket{\text{FZZT}(0)}$ has time $T=0$ and $e^{-2\pi \i \frac{d}{c} L_0}\ket{\TT{ZZ}}$ implements time $T=d/2c$. Finally, the ghost cancel the non-zero-mode contributions in \eqref{5.31cc}:
\bea
    Z_\TT{ghost}(\tau) = \eta\Bigl(\tau - \frac{d}{c}\Bigr)^2 \,.
\eea
Integrating over $\tau$ in \eqref{5.31cc} results in a pole at $\beta=\bar{\beta}$. The purpose of the $\bar{\beta}$ integral in equation \eqref{barbeta} is to pick up that pole, effectively projecting on on-shell states \cite{Collier:2024mlg_zz, Blommaert:2025eps}. A single integration remains:
\bea
    \rho^{\mathbb{C}\rm LS}_{(c,d)}(\alpha_+) = \int_\mathcal{C}\frac{\d\beta}{\sinh(2\pi\beta/\hbar)} \,\mathsf{K}^{\gamma}_{\mathbb{1} \beta} \, \mathsf{S}_{\beta\,\alpha_+ + 2\pi}= \sum_{n \in \mathbb{Z}}  \mathsf{K}^{\gamma}_{\mathbb{1}, \frac{\i \hbar n}{2}}\, \mathsf{S}_{\frac{\i \hbar n}{2}\,\alpha_+} \, . \label{5.39}
\eea
The details of the contour $\mathcal{C}$ were discussed in \cite{Blommaert:2025eps}. The result is that we pick up the poles at $\beta=\i\hbar n/2$. This is the physical spectrum of $\mathbb{C}$LS. In sine dilaton gravity language, this is quantization of geodesic lengths \cite{Blommaert:2025avl,Blommaert:2025eps}. Using 
\bea
    \mathsf{S}_{\frac{\i \hbar n}{2}\,\alpha_+}= \cos(n \alpha_+/2)\,,
\eea
we recognize equation \eqref{5.39} as a discrete Fourier transform. A continuum Fourier transform yields our formula for the modular matrix associated with $\gamma\cdot S$:
\bea
\mathsf{M}^\gamma_{\mathbb{1}\alpha_+} \equiv \int_\mathcal{C}\d \beta\,\mathsf{K}^{\gamma}_{\mathbb{1} \beta}\,\mathsf{S}_{\beta\,\alpha_+}\,.\label{5.27MM}
\eea
The discrete Fourier transform sums this quantity over periodic images (indeed giving a Dirac comb in the Fourier domain):
\bea \label{cls sum over images}
    \rho^{\mathbb{C}\text{LS}}_{(c,d)}(\alpha_+) \is  \sum_{n \in \mathbb{Z}}  \, \mathsf{M}^\gamma_{\mathbb{1}\alpha_++4\pi n} = \sum_{n\in \mathbb{Z}} \rho_{(c,d)}(\alpha_++4\pi n)\, .
\eea
Here $\rho_{(c,d)}(\alpha_+)$ is the result of carrying out the integral \eqref{5.27MM}, plugging in equations \eqref{modular p matrix} and \eqref{modularS}:
\bea \label{cls z spec chiral}
    \boxed{\; \rho_{(c,d)}(\alpha_+)= \mathsf{M}^{\gamma}_{\mathbb{1} \alpha_+}=
    \frac{1}{\sqrt{d}} \, e^{\frac{c}{2 d\hbar}\alpha_+^2}\biggl[ e^{  \frac{2\pi}{d \hbar}\alpha_+} \sin\Bigl(\frac{2\pi b-\alpha_+}{2 d}\Bigr)  + e^{-\frac{2\pi}{d \hbar}\alpha_+} \sin\Bigl(\frac{2\pi b+\alpha_+}{2 d}\Bigr)\,\biggr]
    \, \LARGE\strut }
\eea
As anticipated in section \ref{sect4.3DSSYK}, we interpret this ``basic'' $\mathbb{C}$LS amplitude as computing the chiral gravitational wavefunction of the $(c,d)$ Kerr-lens space time with a line defect with general complex holonomy angle $\alpha_+$. This ``basic'' amplitude is obtained by projecting on a single FZZT momentum sector\footnote{The projection onto a single FZZT momentum \eqref{cls z spec chiral} has the added benefit of being well defined for any value of $c$ and $d$, contrary to the sum \eqref{cls sum over images} which formally converges only when $c/d<0$.}, an operation which is well defined in the open string channel, see section \ref{sect4.3:open}.

Comparing with the semiclassical gravitational wavefunction \eqref{2.96wave} we note that in the $\hbar\to 0$ limit we recover the classical answer, including a second saddle (obtained by $\alpha_+\to-\alpha_+$). The other factors are one-loop corrections. One way of interpreting this amplitude is that the smoothness condition \eqref{2.27smooth} is quantum corrected in a way which is not surprising given the relation with the Virasoro identity block, replacing equation \eqref{2.44contsmooth} with:
\begin{equation} \label{smoothness quantum}
   \pm\beta^\text{contractible}_{\pm\,\text{smooth}} =  2\pi \quad \to \quad \pm\beta^\text{contractible}_{\pm\,\text{smooth}} = \mathbb{1} = [2\pi + \i \hbar/2] - [2\pi - \i \hbar/2]  \, . 
\end{equation}

Let us briefly comment on the trivial but important KdS case $(c,d)= (0,1)$. Despite the generalized crosscap being ill-defined as $c=0$, the partition function \eqref{cls z spec chiral} still makes sense and indeed reproduces the disk spectral density of \cls{} \cite{Collier:2024mlg_zz} (see also \cite{Blommaert:2025avl, Blommaert:2025eps}):
\bea \label{disk spectral density}
    \mathsf{S}_{\mathbb{1} \alpha_+}  = \sin\Big(\frac{\alpha_+}{2}\Big)\sinh\Big(\frac{2\pi \alpha_+}{\hbar}\Big)=  \rho_{(0,1)}(\alpha_+)\,.
\eea
The leading $\hbar \to 0$ reproduces the Gibbons-Hawking entropy formula for KdS \cite{Collier:2025lux, Tietto:2025oxn}. One connection which would be interesting to understand better is that \cite{Collier:2025lux} provide evidence that the full $S^3$ amplitude is computed by integrating $\rho_\text{disk}(\alpha_+)\rho_\text{disk}(\alpha_-)$ independently over $(\alpha_+,\alpha_-)$ and $\in [0,2\pi]$. Formula (5.7) in \cite{Collier:2025lux}, reporting the final answer of this integration, is completely consistent with our equations, with the understanding as already discussed below equation \eqref{1.26final}, that empty dS is implemented by $\alpha_\pm=\mathbb{1}$, consistent with equation \eqref{smoothness quantum}:
\begin{equation}
    \rho_{(0,1)}(\alpha_+\to\mathbb{1},\alpha_-\to \mathbb{1}) = 16 \sinh\Big(\frac{\hbar}{4}\Big)^2 \sinh\Big(\frac{4\pi^2}{\hbar}\Big)^2 =\mathsf{S}_{\mathbb{1}\mathbb{1}}^2 =e^{S_\text{GH}}\,.\label{empty}
\end{equation}


\subsection{Open string interpretation}\label{sect4.3:open}
\par
\begin{figure}[t]
    \centering
\raisebox{2mm}{\resizebox{!}{5cm}{\input{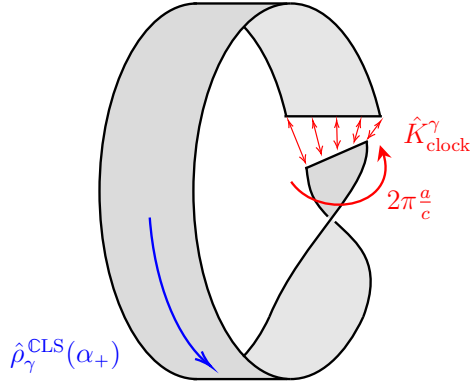}}}
    \caption{In the open string channel, the generalized crosscap is a generalized M\"obius strip trace where we insert an operator that morally rotates the strip by $2\pi a/c$. The trace is twisted by the operator $\hat{\mathsf{K}}^\gamma_\TT{clock}$.}
    \label{fig: string amplitude open}
\end{figure}

 It is instructive to recast the closed string channel computations of section \ref{subsec: slz crosscap amplitudes} in the open string channel. The crosscap amplitude can be computed as a twisted trace over a generalized \slz{} M\"obius strip. One starts from a regular strip of length
\begin{equation}
    \tau_\text{open} = -\frac{1}{c^2 \tau} \,,
\end{equation}
and includes in the trace an operator that roughly speaking rotates by $2\pi a/c$, before integrating over the modulus $\tau_\text{open}$:
\begin{equation} \label{cls open channel integral}
\begin{split}
    \rho^{\mathbb{C}\text{LS}}_\gamma (\alpha_+) = 
     \int_0^{\i \infty} \frac{\d \tau_\text{open}}{\tau_\text{open}} \, \eta\bigg(\tau_\text{open} + \frac{a}{c}\bigg)^2 \,  & \underset{\ca{H}_+}{\Tr_\bullet} \LL[\hat{\rho}^{\,\mathbb{C}\text{LS}}_\gamma(\alpha_+) \,  e^{\frac{2\pi \i a}{c} (L_0 - h)} e^{2\pi \i \tau_\text{open} \big(L_0 - \frac{c_+}{24}\big) }  \RR] \\
     & \qquad \underset{\ca{H}_-}{\Tr} \LL[\hat{\mathsf{K}}^{\gamma}_\TT{clock} e^{\frac{2\pi \i a}{c} (L_0 - h)} e^{2\pi \i \tau_\text{open} \big(L_0 - \frac{c_-}{24}\big) }  \RR]\,.
\end{split}  
\end{equation}
The worldsheet is shown schematically in figure \ref{fig: string amplitude open}.
The operator $e^{\frac{2\pi \i a}{c} (L_0 - h)}$ twists the non-zero modes of the Liouville fields. The traces includes the operator:
\bea \label{clock twist}
    \boxed{\hat{\mathsf{K}}^{\gamma}_\TT{clock} =  \frac{1}{\sqrt{d}} e^{\frac{2 \pi \i}{c d} \big(h - \frac{c_- -1}{24}\big)}\,}
\eea
The $_\bullet$ denotes a derivative of the trace with respect to the momentum propagating in the open channel, see appendix \ref{appA.3open}. The Hilbert spaces $\ca{H}_\pm$ are spanned by all open channel primaries and descendants. The states propagating in the open string channel are encoded in a superposition
\bea \label{open channel states}
    \hat{\rho}^{\,\mathbb{C}\text{LS}}_\gamma(\alpha_+) \is  \sum_{n \in \mathbb{Z}} \hat{\rho}_{\gamma}(\alpha_+ + 4 \pi n) \,.
\eea    
Here $\hat{\rho}_\gamma(\alpha_+)$ are ``basic'' density matrix projectors
\bea \label{open channel basic states}
    \hat{\rho}_\gamma(\alpha_+)\is  e^{- \i \pi \frac{a}{c}} \big\vert c\alpha_+ + 2\pi + \mbox{\large $\frac{\i \hbar}{2}$} \bigl\rangle  \big\langle c \alpha_+ + 2\pi + \mbox{\large $\frac{\i \hbar}{2}$} \big\vert - e^{\i \pi \frac{a}{c}} \big\vert c \alpha_+  + 2\pi - \mbox{\large $\frac{\i \hbar}{2}$} \big\rangle \big\langle c \alpha_+  + 2\pi - \mbox{\large $\frac{\i \hbar}{2}$} \big\vert \qquad  \notag\\[-.35cm]\\[-.25mm]\notag
    & & \qquad  +\, (\alpha_+ \to -\alpha_+)\, 
\eea
This is a discrete set of primaries and their descendants, related by shifting $\alpha_+ \to \alpha_+ + 4\pi n$. This shift symmetry is the open-closed dual to the statement that the Ishibashi momenta $\beta = \i \hbar n/2$ are quantized. This is also manifest in the symmetry of the FZZT boundary cosmological constant: $\mu_{\TT{B}+} = \cos(\alpha_+/2)$. Fixing $\mu_{\text{B}+}$ fixes only an equivalence class of Liouville momenta.

Integrating over $\tau_\text{open}$, one can reduce the \slz{} M\"obius amplitude \eqref{cls open channel integral} to a topological twisted trace over only on-shell Virasoro primaries:
\begin{equation}
   \rho^{\mathbb{C}\text{LS}}_\gamma (\alpha_+) =  \underset{\ca{H}_\text{phys}}{\Tr} \LL[ \hat{\rho}^{\,\mathbb{C}\text{LS}}_\gamma(\alpha_+)\, \hat{\mathsf{K}}^{\gamma}_\TT{clock}  \RR]\,,\quad  \ca{H}_\TT{phys} = \bigoplus_{\alpha} \ket{\alpha}_+ \otimes \ket{\alpha}_-\,.\label{5.46trace}
\end{equation}
This reproduces equation \eqref{cls sum over images}. In the open channel we can naturally project on one term in the sum of \eqref{open channel states}, allowing us to identify the chiral gravitational partition function \eqref{cls z spec chiral} as the twisted trace associated with a Liouville momentum $\alpha_+$, rather than an FZZT bdy cosmological constant $\cos(\alpha_+/2)$:
\begin{equation} \label{5.46trace projected}
    \rho_\gamma (\alpha_+) =  \underset{\ca{H}_\text{phys}}{\Tr} \LL[ \hat{\rho}_\gamma(\alpha_+)\, \hat{\mathsf{K}}^{\gamma}_\TT{clock}  \RR]\,,\quad  \ca{H}_\TT{phys} = \bigoplus_{\alpha} \ket{\alpha}_+ \otimes \ket{\alpha}_-
\end{equation}
Indeed, with each classical FZZT brane (determined by $\mu_\text{B}$) corresponds several quantum mechanically inequivalent FZZT branes, see for instance \cite{Maldacena:2004sn,Saad:2019lba,Blommaert:2019wfy}.

\subsection{Observer's spectral density V2}\label{sect5.5entropy}
We now combine the elements and reach our final answer \eqref{1.26final} for the observer's spectral density, and discuss some of its properties.

\subsubsection*{Step 1. One geometry}
The \slz{} crosscap amplitudes we computed in section \ref{subsec: slz crosscap amplitudes} can be interpreted as chiral gravitational partition functions. Including the anti-holomorphic answer \eqref{g tilde definition}, following equation \eqref{2.96wave}, one obtains the KdS spectral density:
\bea
    \rho_{(c,d)}(E,J) \is \mathsf{M}^{\gamma_+}_{\mathbb{1} \alpha_+} \,  \mathsf{M}^{\gamma_-}_{\mathbb{1} \alpha_-} = \rho
    _{(c,d)}(\alpha_+)\, \rho
    _{(-c,d)}(\alpha_-) 
\eea
Here the modular matrices $\gamma_\pm$ in equation \eqref{g tilde definition} are uniquely fixed by $(c,d)$:
\begin{equation}
    \boxed{b=c^*: c\, c^* = 1\, \text{mod}\, d\,,\quad a=\frac{1+c\, c^*}{d}\,}
\end{equation}
and the relation $(E,J)\leftrightarrow(\alpha_+,\alpha_-)$ is found in equation \eqref{2.8EJ}. From equation \eqref{cls z spec chiral} one then finds:
\begin{equation}
    \boxed{\; \rho_{(c,d)}(E,J) =\sum_{\sigma_+,\sigma_-=\pm 1}    \frac{1}{d}\sin(\frac{\sigma_+\alpha_+ - 2 \pi c^*}{2d}) \sin(\frac{\sigma_-\alpha_- + 2 \pi c^*}{2d})\,e^{\frac{1}{d}\frac{A(E,J)}{4 G}+2\pi \i \frac{c}{d}\frac{J}{8G}}\,}\label{5.42final}
\end{equation}
This spectral density is defined with respect to the integration measure (ignoring prefactors):
\begin{equation}
    \i\, \d \alpha_+ \d \alpha_-=
     \frac{\d J \d E}{\sqrt{E^2 + J^2}}\label{meassure}
\end{equation}
Notice for future reference that
\begin{equation}
    \rho_{(c,d)}{(E,J)}^*=\rho_{(-c,d)}{(E,-J)}\,.
\end{equation}
In the open string channel, one would naturally interpret the KdS spectral density as the twisted trace over $\mathbb{C}\text{LS} \otimes \mathbb{C}\text{LS}$ primaries on the \slz{} M\"obius strip (generalizing equation \eqref{5.46trace projected}):
\begin{equation}
    \rho_{(c,d)}{(E,J)} \, =\, \underset{\ca{H}_\text{phys}}{\Tr} \LL[ \hat{\rho}_{\gamma_+}(\alpha_+)\, \hat{\mathsf{K}}^{\gamma_+}_\TT{clock} \RR] \,
    \underset{\ca{H}_\text{phys}}{\Tr}\LL[ \hat{\rho}_{\gamma_-}(\alpha_-)\, \hat{\mathsf{K}}^{\gamma_-}_\TT{clock} \RR]\,.
\end{equation}

\subsubsection*{Step 2. Farey tail sum}
As discussed in section \ref{sect:summary} and in section \ref{sect:idea}, we now sum over the set of Kerr-lens geometries:
\begin{equation} \label{farey tail sl(2,z)}
    \rho(E,J) = \sum_{d =1}^\infty \sum_{\substack{c = -\infty \\ (c,d) = 1}}^{+\infty} \rho_{(c,d)}(E,J) \, .
\end{equation}
To perform this dS Farey tail sum \cite{Maloney:2007ud, Keller:2014xba, Benjamin:2020mfz}, we first sum over $c\to c+nd$. As we observe in equation \eqref{5.42final}, the results in spin quantization, analogous to what happens in AdS$_3$ \cite{Maxfield:2020ale}:
\begin{equation}\label{angular momentum quantization}
    \sum_{c\to c+n d}e^{2\pi \i \frac{c}{d}\frac{J}{8G}}=\sum_{j=-\infty}^{+\infty}\delta(J/8G-j)\,e^{2\pi \i \frac{c}{d}j}\,. 
\end{equation} 
This leaves us with a more restricted summation to perform
\bea
    \rho_j(E) = \sum_{d = 1}^{\infty} \sum_{\substack{c = 0 \\ (c,d)=1}}^{d-1} \rho_{j\,(c,d)}(E) \, .
\eea
The summation over $c$, using equation \eqref{5.42final}, can be carried out using Kloosterman sums:
\bea
   K\LL(j,s; d \RR) = \sum_{\substack{c = 0 \\ (c,d) = 1}}^{d-1} e^{\frac{2\pi \i}{d} (j c + s c^*)}
\eea
The final answer takes a satisfying form: a reasonable continuation of the AdS$_3$ answer (see for instance equation (2.6) in \cite{Benjamin:2020mfz}) to $\Lambda>0$:
\begin{equation}
    \boxed{\; \rho_j(E) = -\!\!\! \sum_{\sigma_+,\sigma_- = \pm 1}\sigma_+\sigma_- \sum_{d=1}^\infty 
     \frac{1}{d}K\Bigl(j,\mbox{\Large $\frac{\sigma_+-\sigma_-}{2}$},d\Bigr)
     \cosh\biggl(\frac{\alpha_+}{d \hbar}\Bigl(2\pi\nspc +\nspc \mbox{\Large{$\frac {\i \hbar\sigma_+}{2}$}}
     \Bigr)\biggr)
     \cosh\biggl(\frac{\alpha_-}{d \hbar}\Bigl(2\pi\nspc +\nspc \mbox{\Large{$\frac{\i \hbar\sigma_-}{2}$}}\Bigr)\biggr)\,\LARGE \strut}\label{mwk sum divergent}
\end{equation}
It would be interesting to inspect the harmonic properties of this expression \cite{DiUbaldo:2023qli,Benjamin:2022pnx,Collier:2022emf,Haehl:2023tkr}, but we leave this for future work.

\begin{figure}[h]
    \centering
\begin{tikzpicture}[baseline={([yshift=-.5ex]current bounding box.center)}, xscale=0.75,yscale=.75]
 \pgftext{\includegraphics[scale=1]{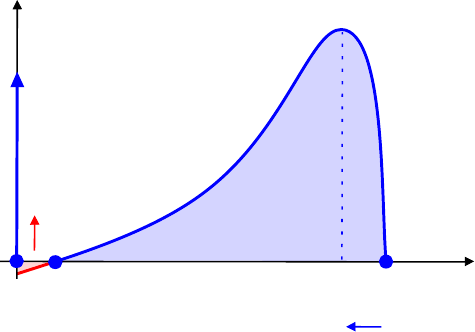}} at (0,0);
    \draw (2.2,-3.3) node {\color{blue}$O(G)$};
    \draw (1.9,2.9) node {\color{blue}maximum};
    \draw (3.9,-2.2) node {$E$};
    \draw (2.5,-2.2) node {\color{blue}$1$};
    \draw (-2.8,-2.2) node {\color{blue}$E_\text{min}$};
    \draw (-4.7,1) node {\color{blue}delta};
    \draw (-5.2,2) node {$\rho_{j=0}(E)$};
    \draw (-2.45,-0.35) node {\color{red}negative};
  \end{tikzpicture}
    \caption{Sketch of the scalar spectrum $\rho_{j=0}(E)$ where $A(E) = 2 \pi \sqrt{E}$. Empty dS space is defined as the Virasoro identity representation, which leads to a quantum correction in the energy of the maximum entropy state. This leads to a question: should we include a quantum correction in our definition of real energy? 
    The spectrum goes negative for $A(E)\to 0$. In this regime, subleading saddles compete, and off-shell geometries and other potential  saddles might have to be taken into account. Notice the delta function for $E\to 0$.}
    \label{fig: spectral density discussion}
\end{figure}

The spectral density $\rho_j(E)$ is real and depends only on $\abs{j}$. This gives hope that it may be physically reasonable to interpret is as a genuine observer's spectral density in 3d cosmology. The sum in equation \eqref{mwk sum divergent} is divergent due to large $d$, but may be regularized as in AdS$_3$ \cite{Benjamin:2020mfz}. For $j=0$, one finds that this results in a delta function
\begin{equation}
    \rho_j(E)\; \supset \; 3 E\, \delta (E)\,.
\end{equation}
Recall from equation \eqref{meassure} that this is to be integrated with a measure proportional to $\d E/E$. Unlike in AdS$_3$ this is positive. This is where the good news stops. Like in AdS$_3$, the spectrum has regions of negativity close to the spectral edge. In the current cosmological case, negativity arises also for $j=0$. 
A sketch of $\rho_{j=0}(E)$ is drawn in figure \ref{fig: spectral density discussion}.

The convergent part of the scalar spectral density reads:
\begin{equation} \label{mwk spectral convergent}
\begin{split}
    \rho^\TT{conv\tiny\strut}_{j=0}(E)  = \sum_{d=1}^\infty \frac{\mu(d)}{d} \Big[\cosh\Bigl(\frac{A(E)}{4 G d}\Bigr)+\cos\Bigl(\frac{A(E)}{d}\Bigr)-2 \Big]-\frac{\phi(d)}{d} \Big[\cosh\Bigl(\frac{A(E)}{4 G d}\Bigr) \cos\Bigl(\frac{A(E)}{d}\Bigr)-1\Big] 
\end{split}
\end{equation}
where $\phi(d)$ is Euler's totient function, $\mu(d)$ is the M\"obius function. We would like to determine whether $\rho^\TT{conv}_{j=0}(E)$ is positive everywhere or has regimes where it turns negative. In the semiclassical limit:
\bea
    \rho^\TT{conv\tiny\strut}_{j=0}(E)  \quad \to \quad  \sum_{d=1} \frac{\mu(d) - \phi(d) \cos(A(E)/d)}{2 d}  \, e^{\frac{1}{d}\frac{A(E)}{4 G}} \,.
\eea
The term $\mu(d) - \phi(d) \cos(A(E)/d)$ is positive for $d=1$ (the S$^3$ contribution), but is negative for $d \ge 2$ (all other lens spaces). The $d=1$ saddle dominates except for $A(E)\to 0$, when quantum  effects become important. One estimates that the density turns negative for $A(E)\sim G\log G$. One interpretation could be that the KdS spacetime has a bounded regime of stability, like near BPS black holes \cite{Heydeman:2020hhw}. It may be that off-shell spaces cure this negativity. We make comments aimed towards computing their potential contributions in section \ref{sect:6xx}.

As discussed already in the introduction below equation \eqref{1.26final}, the maximum is shifted away from naive empty dS, because of the feature \eqref{empty} that quantum empty dS is defined via $\alpha_\pm=\mathbb{1}$. This also results in negativity on the real $\abs{\alpha_\pm}$ axis, when $2\pi-\abs{\alpha_\pm}\sim e^{-\frac{\#}{G}}$. 
Note that the maximal value for the scalar density is perturbatively smaller than the GH density \eqref{empty}:
\begin{equation}
    \rho_{j=0}(E)<e^{S_\text{GH}}=\mathsf{S}_{\mathbb{1}\mathbb{1}}^2\,.
\end{equation}

\section{Concluding remarks}\label{sect:concl}
Our approach and results leave many questions unanswered. We start the concluding remarks by listing several seemingly important ones. We provide suggestions to make progress towards answering them.
\begin{enumerate}
    \item \textbf{Derivation of the duality} We proposed a holographic duality between the dS$_3$ static patch and $\mathbb{C}$LS $\otimes$ $\mathbb{C}$LS living on a disk bounded by the observer's worldline. It would be interesting to give a first-principle derivation of this duality, along the lines of \cite{Verlinde:1989ua,Collier:2023fwi,Verlinde:2024zrh,Collier:2025lux}. This looks feasible. Here we bypassed this step and instead assumed that it holds and used it as a road to quantization.
    The physical Hilbert space (describing the radial $\tau$ evolution) matches between both theories and is spanned by $\ket{\alpha_+,\alpha_-}$. In minisuperspace gravity, discussed in section \ref{sect:3.1kdstoruscosmology}, the pre-Hilbert space has three parameters, one of which is a redundant time coordinate. In $\mathbb{C}$LS $\otimes$ $\mathbb{C}$LS the pre-Hilbert space naturally has 4 parameters, involving one redundant time coordinate for each independent $\mathbb{C}$LS \cite{Blommaert:2025eps}. A preliminary step to deriving this duality would be to pinpoint the precise embedding of the gravitational time coordinate $\tau$ in $\mathbb{C}$LS $\otimes$ $\mathbb{C}$LS. 
    \item \textbf{A microscopic worldline hologram} In section \ref{sect5.3gsigma} we recall \cite{Blommaert:2025eps} that gauge-fixing $\mathbb{C}$LS $\otimes$ $\mathbb{C}$LS leads to two copies of SYK's collective field theory living on an observer's worldline. As discussed in section \eqref{sect4.3DSSYK}, DSSYK corresponds with a chiral sector of the density of states for one particular Kerr-lens space $(c,d)=(-2,1)$. Does the anti-chiral sector for $(c,d)=(-2,1)$ have a microscopic interpretation \cite{Narovlansky:2025tpb,Narovlansky:2023lfz}? Do general $(c,d)$ Kerr-lens spaces have a microscopic dual? In section \ref{sect5.3gsigma} we showed that the anti-chiral sector for $(c,d)=(-2,1)$ has a $G\Sigma$ interpretation with $\sfq\to 1/\sfq$. A suitably regularized version of the anti-chiral spectral density has a chance of having a microscopic realization. We speculate that gauge-fixing $\mathbb{C}$LS on a $(c,d)$ crosscap hints at a $G\Sigma$ formulation of a generalized SYK model involving parafermions, for which the $\mathbb{C}$LS calculation predicts the exact amplitude.

    \item \textbf{Off-shell geometries} The spectral negativity for $A(E)\to 0$ in our final answer \eqref{mwk sum divergent}, discussed in section \ref{sect5.5entropy}, might have to be taken seriously. However, if off-shell geometries would contribute to a reasonable 3d static patch path integral, they would also become competitive in this regime. Could one compute their contribution? Could they cure the spectral negativity as in AdS$_3$ \cite{Maxfield:2020ale}?

    Whilst the type of negativity seems different from in AdS$_3$, it may still be interesting to compute off-shell spacetimes. We speculate in section \ref{sect:6xx} that those may be holographically computed by inserting a gas of $(c,d)$ crosscaps in $\mathbb{C}$LS $\otimes$ $\mathbb{C}$LS. We should remark that a similar approach using VMS in AdS$_3$ has not yet proven successful (to the best of our understanding) \cite{Yan:2025usw,Collier:2023cyw}. A place to start would be to compute the ramp in DSSYK and attempt to find a Cotler-Jensen \cite{Cotler:2020ugk} type wormhole for KL$(-2,1)$, perhaps drawing inspiration from \cite{Yang:2025lme}.
    
    \item \textbf{Observer boundary conditions} Could one rephrase our dynamical particle insertion as local boundary conditions in the gravitational path integral? In section \ref{sect2.6boundaryconditions} we learned that Neumann boundary conditions $\ket{p_1,p_2,K\to \infty}$ are closely related to fixing $\ket{\alpha_+,\alpha_-}$. It would be interesting to develop such a local understanding of gauge-invariant states. 
    One could start by figuring out the detailed co-invariant states in minisuperspace associated with $\ket{\alpha_+,\alpha_-}$.
    
    \item \textbf{Entropy of de Sitter} An interesting object to investigate is the sum over lens space amplitudes for pure de Sitter spacetime without any worldline. This problem was studied in \cite{Castro:2011xb} and mapped to a computation in SU(2) CS theory. We can improve their calculation by using the new insights from $\mathbb{V}$FT and $\mathbb{C}$LS. This predicts the following generalization of equation \eqref{empty} as a Farey tail sum:
\begin{equation}
    \boxed{\ e^{S_\text{GH,exact}}=\sum_{d =1}^\infty \sum_{\substack{c = -\infty \\ (c,d) = 1}}^{+\infty} \mathsf{M}^{\gamma(c,d)}_{\mathbb{1} \mathbb{1}} \,  \mathsf{M}^{\gamma(-c,d)}_{\mathbb{1} \mathbb{1}}\; }
\end{equation}
Here $\mathsf{M}^{\gamma}_{\mathbb{1}\mathbb{1}} = \mathsf{M}^{\gamma}_{\mathbb{1},{2\pi + \frac{i\hbar}{2} } } - \mathsf{M}^{\gamma}_{\mathbb{1},{2\pi - \frac{i\hbar}{2} } }$
with $\mathsf{M}^{\gamma}_{\mathbb{1}{\alpha}}$ given in equation \eqref{m-exact}. This expressions matches the GH answer to leading order in $1/G$, and contains contains an infinite number of perturbative and non-perturbative corrections.  We leave investigating this quantity and comparisons with the path integral for future investigations. One check is that the summand is invariant under $c\to c+n d$, as expected of empty dS \cite{Castro:2011xb}, unlike the general $(\alpha_+,\alpha_-)$ case.

\end{enumerate}

We end with some speculation on the $G\Sigma \otimes G\Sigma$ interpretation of the general $(c,d)$ crosscap amplitudes in \textbf{section \ref{sect5.3gsigma}} and on the possibly significant off-shell geometries that may give higher order contributions to the SL(2,$\mathbb{Z}$) amplitudes considered in \textbf{section \ref{sect:6xx}}.

\subsection{Towards a $G\Sigma\otimes G\Sigma$ worldline hologram}\label{sect5.3gsigma}
Recall the duality between \cls{} with a crosscap and DSSYK \cite{Blommaert:2025eps}. The SYK fermion two-point function is encoded in a bilocal field $g(\tau_1,\tau_2)$, with a Liouville action. The Majorana statistics of SYK fermions implies that the $g$ field lives on crosscap. In \cls{} lightcone gauge, one of the two Liouville fields $\varphi_-$ is gauge-fixed to a time coordinate. The physical degrees of freedom are carried by a single Liouville $\varphi_+$, identified with the bilocal field $g$ \cite{Blommaert:2025eps}. The crosscap in  \cls{} directly stems from the crosscap in SYK. The partition function of \cls{} on the crosscap computes the spectral density of the DSSYK model. 

We have proposed \eqref{4.33claim} that \cls{} on the crosscap computes the chiral density of states $\rho_{(-2,1)}(\alpha_+)$ of the Kerr-lens space $(c,d) = (-2,1)$. So, DSSYK provides a concrete worldline hologram for a chiral half of dS$_3$ gravity on KL$(-2,1)$. What about the anti-chiral half? Using $\mathbb{C}$LS $\otimes$ $\mathbb{C}$LS we obtain \eqref{zdouble}:
\begin{equation}
    \boxed{\; \rho_{(-2,1)}(\alpha_+,\alpha_-)=\big(e^{\pm \i\alpha_+};\sfq \big)_\infty\big(e^{\pm \i\alpha_-};\sfq^{-1} \big)_\infty\;\Large\strut}\label{6.1answer}
\end{equation}
Gauge-fixing the second copy of $\mathbb{C}$LS too, one obtains a second bilocal collective field on the worldline. 
Looking at the $\mathbb{C}$LS prediction \eqref{6.1answer} suggest that the second bilocal field $g$ has an opposite sign action.
\begin{equation}
    \boxed{\; \mathbb{C}\text{LS}\otimes \mathbb{C}\text{LS} \quad \leftrightarrow \quad   g_{+\hbar} \otimes g_{-\hbar}\;\strut }
\end{equation}
It is not obvious that this second $G\Sigma$ theory is fully well-defined or can be obtained in a reasonable manner as a double scaled limit of a finite $N$ SYK-like model. Indeed, the spectral density $\big(e^{\pm \i\alpha_-};\sfq^{-1} \big)_\infty$ of the second $G\Sigma$ theory
formally expands as a sum over unsuppressed Gaussians $\alpha_-\to \alpha_-+4\pi n$ and thus looks divergent. To balance this caution with some reason for optimism, we note that divergences of this kind are typical in \slt{} CS theory; they can be regularized via the so-called ``Fock space reorganization'' \cite{Dimofte:2011py}\footnote{To regularize \eqref{6.1answer}, one applies \eqref{fock space reorganization} to the $\sfq$-Pochhammers in \eqref{cls spec amplitude conclusion}, rather than to the ones in \eqref{6.1answer}. This ensures that the regularized answer matches, in the $\hbar \to 0$ limit and for a certain range of $\alpha_-$, with the gravitational answer \eqref{cls z spec chiral}.}
\begin{equation} \label{fock space reorganization}
    \big( z ; \, \sfq^{-1} \big)_\infty  \quad \to \quad \big( \sfq z ; \sfq \big)_\infty^{-1} \, .
\end{equation}
It seems reasonable to assume that a similar procedure may be applied here. Can these ideas be used to find a hologram of dS$_3$ gravity on KL(-2,1) in terms of a two SYK collective field theories with opposite kinetic terms? Besides reproducing the spectral density, a logical next target would be to predict the (exact) bulk two-point function in KL$(-2,1)$ using DSSYK $\otimes$ DSSYK. Some progress in this direction was reported in \cite{Narovlansky:2023lfz,Narovlansky:2025tpb,Verlinde:2024zrh}. We intend to investigate both of these questions in future work.

Another interesting direction would be to find a microscopic description of the $(c,d)$ spectral density. In lightcone gauge \cite{Blommaert:2025eps}, the physical degrees of freedom of $\mathbb{C}\text{LS} \otimes \mathbb{C}\text{LS}$ are carried by 2 Liouville fields, indicating a $G\Sigma\otimes G\Sigma$ worldline hologram. What changes between geometries with different $(c,d)$? In worldsheet lightcone coordinates $\tau_1/\beta = x - t$, $\tau_2/\beta = x+t$, we have an \slz{} crosscap at $t = d/2c$, as suggested below \eqref{barbeta}. Generalizing the usual crosscap story \cite{Lin:2023trc,Cotler:2016fpe}, this suggest that the bilocal fields may be required to respect twisted boundary conditions:
\bea \label{gsigma flip order}
    g(\tau_1,\tau_2) = g(\tau_2 - 2d\beta/c , \tau_1) \, .
\eea
It would be interesting to study the $g$ path integral with such boundary conditions, and to try following \cite{Lin:2023trc,Cotler:2016fpe} to compute the associated amplitude. By writing the discrete sum over $n$ via the Jacobi triple product identity, the $\mathbb{C}$LS prediction \eqref{5.39} becomes:
\begin{equation}\label{cls spec amplitude conclusion}
    \rho_{\mathbb{C}\rm LS}^{(c,d)}(\alpha_+) = \frac{e^{-\i \pi \frac{a}{c}}}{\sqrt{-c}} \Big(\nspc -\nspc \sfq^{\frac 1 2\mp\frac 1 d}  \spc  e^{\pm \frac \i 2  (\alpha_+ + \frac{2 \pi} c) } ; \sfq\Big)_\infty - \frac{e^{+\i \pi \frac{a}{c}}}{\sqrt{-c}}  \Big(\nspc -\nspc \sfq^{\frac 1 2\pm\frac 1 d}  \spc  e^{\pm \frac \i 2  (\alpha_+ + \frac{2 \pi} c) } ; \sfq \Big)_\infty+\,(\alpha_+\to-\alpha_+)\,.
\end{equation}
Here\footnote{When $\sfq>1$ one might regularize the diverging $\sfq$-Pochhammers using \eqref{fock space reorganization}. The regularized spectral density is shift symmetric and reproduces as $\hbar \to 0$ the (non-shift symmetric) gravitational answer \eqref{cls z spec chiral} for a restricted range of values of $\alpha$.}
\begin{equation}
    \sfq = e^{d \hbar/ 4 c}\,.
\end{equation}
Each $\sfq$-Pochhammer satisfies an exponential version of the quantum smoothness condition, analogous to equation \eqref{smoothness quantum}. For instance:
\begin{equation} \label{difference equation pochhammer}
    \LL[ e^{\frac{\i}{2}(d \beta_+ - c\alpha_+)} - e^{\frac{\i}{2}(2\pi + \i \hbar/2)} \RR] \Big(\nspc -\nspc \sfq^{\frac 1 2\mp\frac 1 d}  \spc  e^{\pm \frac \i 2  (\alpha + \frac{2 \pi} c) } ; \sfq\Big)_\infty =0 \,.
\end{equation}
This is a generalization of the Skein algebra vacuum condition \cite{Gaiotto:2024osr, Gaiotto:2024kze}, hinting at some possible duality between the $(c,d)$ model and the Schur half-index of some 4d gauge theory. 

Does Liouville theory with twisted boundary conditions \eqref{gsigma flip order} descend from a microscopic model such as SYK? The following is speculation. Consider a model of $N$ fermions $\chi^j$ with $p$-body interactions on a thermal circle $\tau \in [0,d \beta]$, with anti-periodic boundary conditions
\bea \label{gsigma thermal periodicity}
    \chi^j(\tau) \is -\chi^j(\tau + d \beta) \, . 
\eea
Notice that, as shown in figure \ref{fig:crosscap generalized}, the \slz{} crosscap is indeed a $d$-fold cover of $\mathbb{C}$. Following the standard large $N$ treatement, let us introduce the collective two-point correlator
\bea
\label{collectivetwopt}
    \frac{1}{N}\sum_{j=1}^N \expval{\chi^j(\tau_1) \chi^j(\tau_2)} \is \TT{sign}(\tau_{12}) \, e^{g(\tau_1, \tau_2)/p}\,. 
\eea
In the following, we keep the index $j$ implicit to lighten the notation, and assume that $c$ is even for simplicity. We cut the circle in $c/2$ pieces. Consider a $\mathbb{Z}_{c/2}$ quotient of the thermal circle by restricting $\tau \in [0,2d/c]$ and introduce $c/2$ oscillators $\chi_{n}$ labeling fermions along the different pieces of the big circle shifted by integer multiples of $2d/c$. We can then write equation \eqref{collectivetwopt} as
\bea
    \frac{1}{N}\sum_{j=1}^N \expval{\chi_{n_1}(\tau_1) \chi_{n_2}(\tau_2)} \is \TT{sign}(\tau_{12}) \, e^{g\LL(\tau_1+ n_1 \frac{2d}{c}\beta, \tau_2+n_2 \frac{2d}{c}\beta \RR)/p}  \, .
\eea
Equation \eqref{gsigma thermal periodicity} implements a cyclic identification:
\bea
    \chi_{n}(\tau) \is -\chi_{n+c/2}(\tau) \, .
\eea
The boundary conditions of the $\chi_n$ oscillators on the quotiented thermal circle $\tau \in [0,2d/c]$ are:
\bea
    \chi_{n}(\tau + 2d\beta/c ) \is \chi_{n+1}(\tau)  \, .
\eea
For the collective two-point function $g(\tau_1,\tau_2)$ to satisfy the \slz{} crosscap boundary conditions \eqref{gsigma flip order}, the oscillators $\chi_n$ must apparently obey some form of parafermionic statistics. We leave the further development of this proposed twisted version of the SYK model to future work.

Finally, we comment on the possibility of relating the Kerr-lens geometries to matrix integrals. For $(c,d)=(0,1)$, the chiral amplitude \eqref{disk spectral density} is the disk spectrum of the matrix model dual to \cls{} \cite{Collier:2024lys,Collier:2025lux}. When $(c,d)=(-2,1)$, the chiral spectrum matches the DSSYK spectral density, and the disk amplitude of the ETH matrix model \cite{jafferis2022jt}. For general $(c,d)$, consider decomposing \eqref{5.39} as follows:
\begin{equation} \label{cap amplitude}
    \rho_{\mathbb{C}\rm LS}^{(c,d)}(\alpha_+) = \sum_{n=-\infty}^{+\infty} \psi^{(c,d)}_\TT{cap}(n) \, Z_\text{trumpet}(\alpha_+,n)\,.
\end{equation}
Here, $Z_\text{trumpet}(\alpha_+,n)$ is the sine dilaton trumpet \cite{Blommaert:2025avl,Saad:2019lba}, and $\psi_\TT{cap}(n)$ it called the cap amplitude \cite{Okuyama:2025xwx}:
\begin{equation} \label{cap wavefunction}
    \psi^{(c,d)}_\TT{cap}(n) = \frac{1}{\sqrt{-c}}e^{\frac{d \hbar}{8c} n^2} e^{-\frac{\hbar}{4c}n} \sin(\frac{\pi(a-n)}{c})    \,, \qquad  Z_\TT{trumpet}(\alpha;n) = \cos(\frac{n\alpha}{2})  \, .
\end{equation}
It is possible to associate to a cap amplitude $\psi_\TT{cap}(n)$ a random matrix model computing a set of discrete volumes for higher genus surfaces \cite{Okuyama:2025xwx}. Therefore, each Kerr-lens geometry can potentially be realized as a \slz{} ETH $\otimes$ ETH matrix model, whose spectral density computes the gravitational amplitude.

\subsection{Towards off-shell geometries from $\mathbb{C}$LS $\otimes$ $\mathbb{C}$LS}\label{sect:6xx}

We computed an observer's spectral density using a solipsistic version of the no-boundary proposal \cite{hartle1983wave}:
\begin{equation} \label{spectral density no-boundary inner product}
    \rho(E,J) = \bigl\langle \alpha_+ ,\alpha_- \bigr| \TT{no-boundary} \bigr\rangle\,.
\end{equation}
Here $\bigl\vert \alpha_+ ,\alpha_- \bigr\rangle$ represents the observer's state, and $\ket{\text{no-boundary}}$ is obtained by summing over smooth Dehn fillings (of the complement of the observer's worldline). Each of these Dehn fillings corresponds in terms of torus quantum mechanics with a state $\ket{\text{smooth}_{(c,d)}}$ defined semiclassically in equation \eqref{2.92state}, and in terms of $\mathbb{V}$TFT with two identity torus conformal blocks $\ket{\mathbb{1}_{(c,d)}} \otimes \ket{\mathbb{1}_{(-c,d)}}$:
\begin{equation} \label{vtft horizon}
    \ket{\text{no-boundary}} = \sum_{\mathbb{Z}\backslash \TT{PSL}(2,\mathbb{Z}) } \ket{\mathbb{1}_{(c,d)}} \otimes \ket{\mathbb{1}_{(-c,d)}}
\end{equation}
Pictorially:
\begin{equation} \label{farey tail state vtft}
    \ket{\text{no-boundary}} =\quad \raisebox{-.95cm}{\resizebox{3cm}{!}{\tikzset{every picture/.style={line width=0.75pt}} 

\begin{tikzpicture}[x=0.75pt,y=0.75pt,yscale=-1,xscale=1]

\draw  [draw opacity=0][fill={rgb, 255:red, 219; green, 219; blue, 219 }  ,fill opacity=1 ][line width=1.5]  (139,789.35) .. controls (209.69,789.35) and (267,830.66) .. (267,881.62) .. controls (267,932.59) and (209.69,973.9) .. (139,973.9) .. controls (68.31,973.9) and (11,932.59) .. (11,881.62) .. controls (11,830.66) and (68.31,789.35) .. (139,789.35) -- cycle (90.73,881.62) .. controls (90.73,894.88) and (112.34,905.62) .. (139,905.62) .. controls (165.66,905.62) and (187.27,894.88) .. (187.27,881.62) .. controls (187.27,868.37) and (165.66,857.62) .. (139,857.62) .. controls (112.34,857.62) and (90.73,868.37) .. (90.73,881.62) -- cycle ;
\draw  [dash pattern={on 1.69pt off 2.76pt}][line width=1.5]  (10,881.62) .. controls (10,830.66) and (67.31,789.35) .. (138,789.35) .. controls (208.69,789.35) and (266,830.66) .. (266,881.62) .. controls (266,932.59) and (208.69,973.9) .. (138,973.9) .. controls (67.31,973.9) and (10,932.59) .. (10,881.62) -- cycle ;
\draw  [draw opacity=0][line width=1.5]  (187.8,879.37) .. controls (185.51,894.09) and (164.09,905.62) .. (138,905.62) .. controls (111.91,905.62) and (90.49,894.09) .. (88.2,879.37) -- (138,876.76) -- cycle ; \draw  [line width=1.5]  (187.8,879.37) .. controls (185.51,894.09) and (164.09,905.62) .. (138,905.62) .. controls (111.91,905.62) and (90.49,894.09) .. (88.2,879.37) ;  
\draw  [draw opacity=0][line width=1.5]  (91.43,881.62) .. controls (91.43,881.62) and (91.43,881.62) .. (91.43,881.62) .. controls (91.43,881.62) and (91.43,881.62) .. (91.43,881.62) .. controls (91.43,868.37) and (112.28,857.62) .. (138,857.62) .. controls (163.72,857.62) and (184.57,868.37) .. (184.57,881.62) -- (138,881.62) -- cycle ; \draw  [line width=1.5]  (91.43,881.62) .. controls (91.43,881.62) and (91.43,881.62) .. (91.43,881.62) .. controls (91.43,881.62) and (91.43,881.62) .. (91.43,881.62) .. controls (91.43,868.37) and (112.28,857.62) .. (138,857.62) .. controls (163.72,857.62) and (184.57,868.37) .. (184.57,881.62) ;  
\draw  [color={rgb, 255:red, 255; green, 0; blue, 0 }  ,draw opacity=1 ][line width=1.5]  (46.65,881.62) .. controls (46.65,849.15) and (87.55,822.83) .. (138,822.83) .. controls (188.45,822.83) and (229.35,849.15) .. (229.35,881.62) .. controls (229.35,914.09) and (188.45,940.42) .. (138,940.42) .. controls (87.55,940.42) and (46.65,914.09) .. (46.65,881.62) -- cycle ;

\end{tikzpicture}}} \; + \; \raisebox{-.95cm}{\resizebox{3cm}{!}{\tikzset{every picture/.style={line width=0.75pt}} 

\begin{tikzpicture}[x=0.75pt,y=0.75pt,yscale=-1,xscale=1]

\draw  [draw opacity=0][fill={rgb, 255:red, 219; green, 219; blue, 219 }  ,fill opacity=1 ][line width=1.5]  (473,789.35) .. controls (543.69,789.35) and (601,830.66) .. (601,881.62) .. controls (601,932.59) and (543.69,973.9) .. (473,973.9) .. controls (402.31,973.9) and (345,932.59) .. (345,881.62) .. controls (345,830.66) and (402.31,789.35) .. (473,789.35) -- cycle (424.73,881.62) .. controls (424.73,894.88) and (446.34,905.62) .. (473,905.62) .. controls (499.66,905.62) and (521.27,894.88) .. (521.27,881.62) .. controls (521.27,868.37) and (499.66,857.62) .. (473,857.62) .. controls (446.34,857.62) and (424.73,868.37) .. (424.73,881.62) -- cycle ;
\draw [color={rgb, 255:red, 255; green, 0; blue, 0 }  ,draw opacity=1 ][line width=1.5]    (443.61,825.94) .. controls (456.35,823.74) and (449.8,790.17) .. (458,789.35) ;
\draw [color={rgb, 255:red, 255; green, 0; blue, 0 }  ,draw opacity=1 ][line width=1.5]  [dash pattern={on 5.63pt off 4.5pt}]  (458,789.35) .. controls (463.47,789.27) and (461.47,857.6) .. (467.97,857.18) ;
\draw  [draw opacity=0][line width=1.5]  (510.33,827.95) .. controls (542.17,837.14) and (564.35,857.71) .. (564.35,881.62) -- (473,881.62) -- cycle ; \draw  [color={rgb, 255:red, 255; green, 0; blue, 0 }  ,draw opacity=1 ][line width=1.5]  (510.33,827.95) .. controls (542.17,837.14) and (564.35,857.71) .. (564.35,881.62) ;  
\draw  [draw opacity=0][line width=1.5]  (443.61,825.94) .. controls (407.57,833.82) and (381.65,855.77) .. (381.65,881.62) -- (473,881.62) -- cycle ; \draw  [color={rgb, 255:red, 255; green, 0; blue, 0 }  ,draw opacity=1 ][line width=1.5]  (443.61,825.94) .. controls (407.57,833.82) and (381.65,855.77) .. (381.65,881.62) ;  
\draw [color={rgb, 255:red, 255; green, 0; blue, 0 }  ,draw opacity=1 ][line width=1.5]    (467.97,857.18) .. controls (473.47,856.93) and (477.13,789.02) .. (482.63,789.52) ;
\draw  [draw opacity=0][line width=1.5]  (505.89,936.49) .. controls (495.68,939.03) and (484.59,940.42) .. (473,940.42) .. controls (422.55,940.42) and (381.65,914.09) .. (381.65,881.62) -- (473,881.62) -- cycle ; \draw  [color={rgb, 255:red, 255; green, 0; blue, 0 }  ,draw opacity=1 ][line width=1.5]  (505.89,936.49) .. controls (495.68,939.03) and (484.59,940.42) .. (473,940.42) .. controls (422.55,940.42) and (381.65,914.09) .. (381.65,881.62) ;  
\draw  [draw opacity=0][line width=1.5]  (498.91,938.02) .. controls (536.75,930.83) and (564.35,908.3) .. (564.35,881.62) -- (473,881.62) -- cycle ; \draw  [color={rgb, 255:red, 255; green, 0; blue, 0 }  ,draw opacity=1 ][line width=1.5]  (498.91,938.02) .. controls (536.75,930.83) and (564.35,908.3) .. (564.35,881.62) ;  
\draw [color={rgb, 255:red, 255; green, 0; blue, 0 }  ,draw opacity=1 ][line width=1.5]  [dash pattern={on 5.63pt off 4.5pt}]  (482.63,789.52) .. controls (490.31,790.24) and (478.77,856.86) .. (485.1,858.19) ;
\draw  [dash pattern={on 1.69pt off 2.76pt}][line width=1.5]  (345,881.62) .. controls (345,830.66) and (402.31,789.35) .. (473,789.35) .. controls (543.69,789.35) and (601,830.66) .. (601,881.62) .. controls (601,932.59) and (543.69,973.9) .. (473,973.9) .. controls (402.31,973.9) and (345,932.59) .. (345,881.62) -- cycle ;
\draw  [draw opacity=0][line width=1.5]  (522.8,879.37) .. controls (520.51,894.09) and (499.09,905.62) .. (473,905.62) .. controls (446.91,905.62) and (425.49,894.09) .. (423.2,879.37) -- (473,876.76) -- cycle ; \draw  [line width=1.5]  (522.8,879.37) .. controls (520.51,894.09) and (499.09,905.62) .. (473,905.62) .. controls (446.91,905.62) and (425.49,894.09) .. (423.2,879.37) ;  
\draw  [draw opacity=0][line width=1.5]  (426.43,881.62) .. controls (426.43,881.62) and (426.43,881.62) .. (426.43,881.62) .. controls (426.43,881.62) and (426.43,881.62) .. (426.43,881.62) .. controls (426.43,868.37) and (447.28,857.62) .. (473,857.62) .. controls (498.72,857.62) and (519.57,868.37) .. (519.57,881.62) -- (473,881.62) -- cycle ; \draw  [line width=1.5]  (426.43,881.62) .. controls (426.43,881.62) and (426.43,881.62) .. (426.43,881.62) .. controls (426.43,881.62) and (426.43,881.62) .. (426.43,881.62) .. controls (426.43,868.37) and (447.28,857.62) .. (473,857.62) .. controls (498.72,857.62) and (519.57,868.37) .. (519.57,881.62) ;  

\draw [color={rgb, 255:red, 255; green, 0; blue, 0 }  ,draw opacity=1 ][line width=1.5]    (510.33,827.95) .. controls (495.1,822.99) and (490.7,858.19) .. (485.1,858.19) ;

\end{tikzpicture}}} \; + \; \raisebox{-.95cm}{\resizebox{3cm}{!}{\tikzset{every picture/.style={line width=0.75pt}} 

\begin{tikzpicture}[x=0.75pt,y=0.75pt,yscale=-1,xscale=1]

\draw  [draw opacity=0][fill={rgb, 255:red, 219; green, 219; blue, 219 }  ,fill opacity=1 ][line width=1.5]  (850,797.35) .. controls (920.69,797.35) and (978,838.66) .. (978,889.62) .. controls (978,940.59) and (920.69,981.9) .. (850,981.9) .. controls (779.31,981.9) and (722,940.59) .. (722,889.62) .. controls (722,838.66) and (779.31,797.35) .. (850,797.35) -- cycle (801.73,889.62) .. controls (801.73,902.88) and (823.34,913.62) .. (850,913.62) .. controls (876.66,913.62) and (898.27,902.88) .. (898.27,889.62) .. controls (898.27,876.37) and (876.66,865.62) .. (850,865.62) .. controls (823.34,865.62) and (801.73,876.37) .. (801.73,889.62) -- cycle ;
\draw [color={rgb, 255:red, 255; green, 0; blue, 0 }  ,draw opacity=1 ][line width=1.5]    (820.61,833.94) .. controls (833.35,831.74) and (826.8,798.17) .. (835,797.35) ;
\draw [color={rgb, 255:red, 255; green, 0; blue, 0 }  ,draw opacity=1 ][line width=1.5]  [dash pattern={on 5.63pt off 4.5pt}]  (835,797.35) .. controls (840.47,797.27) and (838.47,865.6) .. (844.97,865.18) ;
\draw  [draw opacity=0][line width=1.5]  (887.33,835.95) .. controls (919.17,845.14) and (941.35,865.71) .. (941.35,889.62) -- (850,889.62) -- cycle ; \draw  [color={rgb, 255:red, 255; green, 0; blue, 0 }  ,draw opacity=1 ][line width=1.5]  (887.33,835.95) .. controls (919.17,845.14) and (941.35,865.71) .. (941.35,889.62) ;  
\draw  [draw opacity=0][line width=1.5]  (820.61,833.94) .. controls (784.57,841.82) and (758.65,863.77) .. (758.65,889.62) -- (850,889.62) -- cycle ; \draw  [color={rgb, 255:red, 255; green, 0; blue, 0 }  ,draw opacity=1 ][line width=1.5]  (820.61,833.94) .. controls (784.57,841.82) and (758.65,863.77) .. (758.65,889.62) ;  
\draw [color={rgb, 255:red, 255; green, 0; blue, 0 }  ,draw opacity=1 ][line width=1.5]    (844.97,865.18) .. controls (850.47,864.93) and (854.13,797.02) .. (859.63,797.52) ;
\draw  [draw opacity=0][line width=1.5]  (796.49,937.28) .. controls (773.56,926.6) and (758.65,909.23) .. (758.65,889.62) -- (850,889.62) -- cycle ; \draw  [color={rgb, 255:red, 255; green, 0; blue, 0 }  ,draw opacity=1 ][line width=1.5]  (796.49,937.28) .. controls (773.56,926.6) and (758.65,909.23) .. (758.65,889.62) ;  
\draw [color={rgb, 255:red, 255; green, 0; blue, 0 }  ,draw opacity=1 ][line width=1.5]  [dash pattern={on 5.63pt off 4.5pt}]  (859.63,797.52) .. controls (867.31,798.24) and (855.77,864.86) .. (862.1,866.19) ;
\draw [color={rgb, 255:red, 255; green, 0; blue, 0 }  ,draw opacity=1 ][line width=1.5]    (887.33,835.95) .. controls (872.1,830.99) and (867.7,866.19) .. (862.1,866.19) ;
\draw [color={rgb, 255:red, 255; green, 0; blue, 0 }  ,draw opacity=1 ][line width=1.5]    (782.65,913.5) .. controls (833.18,973.78) and (937.85,950.12) .. (941.35,889.62) ;
\draw  [draw opacity=0][line width=1.5]  (834.95,841) .. controls (799.72,845.52) and (773.16,865.57) .. (773.16,889.62) .. controls (773.16,898.28) and (776.6,906.42) .. (782.65,913.5) -- (850,889.62) -- cycle ; \draw  [color={rgb, 255:red, 255; green, 0; blue, 0 }  ,draw opacity=1 ][line width=1.5]  (834.95,841) .. controls (799.72,845.52) and (773.16,865.57) .. (773.16,889.62) .. controls (773.16,898.28) and (776.6,906.42) .. (782.65,913.5) ;  
\draw  [draw opacity=0][line width=1.5]  (848.11,840.07) .. controls (846.63,840.09) and (845.16,840.14) .. (843.71,840.22) -- (850,889.62) -- cycle ; \draw  [color={rgb, 255:red, 255; green, 0; blue, 0 }  ,draw opacity=1 ][line width=1.5]  (848.11,840.07) .. controls (846.63,840.09) and (845.16,840.14) .. (843.71,840.22) ;  
\draw  [draw opacity=0][line width=1.5]  (858.02,840.32) .. controls (856.47,840.22) and (854.91,840.14) .. (853.34,840.1) -- (850,889.62) -- cycle ; \draw  [color={rgb, 255:red, 255; green, 0; blue, 0 }  ,draw opacity=1 ][line width=1.5]  (858.02,840.32) .. controls (856.47,840.22) and (854.91,840.14) .. (853.34,840.1) ;  
\draw  [draw opacity=0][line width=1.5]  (869.94,841.74) .. controls (867.92,841.39) and (865.86,841.09) .. (863.77,840.85) -- (850,889.62) -- cycle ; \draw  [color={rgb, 255:red, 255; green, 0; blue, 0 }  ,draw opacity=1 ][line width=1.5]  (869.94,841.74) .. controls (867.92,841.39) and (865.86,841.09) .. (863.77,840.85) ;  
\draw  [draw opacity=0][line width=1.5]  (866.28,938.08) .. controls (900.9,933.26) and (926.84,913.4) .. (926.84,889.62) .. controls (926.84,869.4) and (908.06,852) .. (881.14,844.29) -- (850,889.62) -- cycle ; \draw  [color={rgb, 255:red, 255; green, 0; blue, 0 }  ,draw opacity=1 ][line width=1.5]  (866.28,938.08) .. controls (900.9,933.26) and (926.84,913.4) .. (926.84,889.62) .. controls (926.84,869.4) and (908.06,852) .. (881.14,844.29) ;  
\draw [color={rgb, 255:red, 255; green, 0; blue, 0 }  ,draw opacity=1 ][line width=1.5]    (866.28,938.08) .. controls (855.84,939.52) and (860.78,912.78) .. (855.18,912.78) ;
\draw [color={rgb, 255:red, 255; green, 0; blue, 0 }  ,draw opacity=1 ][line width=1.5]  [dash pattern={on 5.63pt off 4.5pt}]  (855.18,912.78) .. controls (848.18,913.78) and (845.18,981.12) .. (836.85,981.45) ;
\draw [color={rgb, 255:red, 255; green, 0; blue, 0 }  ,draw opacity=1 ][line width=1.5]    (836.85,981.45) .. controls (833.52,981.45) and (830.02,977.75) .. (830.52,964.45) .. controls (831.01,951.15) and (815.52,947.12) .. (796.49,937.28) ;
\draw  [dash pattern={on 1.69pt off 2.76pt}][line width=1.5]  (722,889.62) .. controls (722,838.66) and (779.31,797.35) .. (850,797.35) .. controls (920.69,797.35) and (978,838.66) .. (978,889.62) .. controls (978,940.59) and (920.69,981.9) .. (850,981.9) .. controls (779.31,981.9) and (722,940.59) .. (722,889.62) -- cycle ;
\draw  [draw opacity=0][line width=1.5]  (899.8,887.37) .. controls (897.51,902.09) and (876.09,913.62) .. (850,913.62) .. controls (823.91,913.62) and (802.49,902.09) .. (800.2,887.37) -- (850,884.76) -- cycle ; \draw  [line width=1.5]  (899.8,887.37) .. controls (897.51,902.09) and (876.09,913.62) .. (850,913.62) .. controls (823.91,913.62) and (802.49,902.09) .. (800.2,887.37) ;  
\draw  [draw opacity=0][line width=1.5]  (803.43,889.62) .. controls (803.43,889.62) and (803.43,889.62) .. (803.43,889.62) .. controls (803.43,889.62) and (803.43,889.62) .. (803.43,889.62) .. controls (803.43,876.37) and (824.28,865.62) .. (850,865.62) .. controls (875.72,865.62) and (896.57,876.37) .. (896.57,889.62) -- (850,889.62) -- cycle ; \draw  [line width=1.5]  (803.43,889.62) .. controls (803.43,889.62) and (803.43,889.62) .. (803.43,889.62) .. controls (803.43,889.62) and (803.43,889.62) .. (803.43,889.62) .. controls (803.43,876.37) and (824.28,865.62) .. (850,865.62) .. controls (875.72,865.62) and (896.57,876.37) .. (896.57,889.62) ;

\end{tikzpicture}}} \; + \; \dots
\end{equation}
A similar sum over smooth empty tori \eqref{farey tail state vtft} was considered in \cite{Godet:2024ich}, to describe a wavefunction at~$\mathscr{I}_+$.

This 3d geometric construction projects down to the $\mathbb{C}$LS level. The observer's spectral density can be equivalently computed as the $\mathbb{C}\text{LS} \otimes \mathbb{C}\text{LS}$ annulus amplitude between FZZT states $\ket{\alpha_+} \otimes\ket{\alpha_-}$ and a ``no-boundary'' state in the complex Liouville CFT
\begin{equation} 
    \rho(E,J) = \bigl\langle \alpha_+ ,\alpha_- \bigr| \TT{no-boundary} \bigr\rangle\,.
\end{equation}
This 2d ``no-boundary" BCFT state is (a group average of) the generalized crosscap state:
\begin{equation} \label{cls2 horizon}
    \ket{\TT{no-boundary}} =  \sum_{\mathbb{Z}\backslash \TT{PSL}(2,\mathbb{Z}) } \ket{\TT{C}_{(c,d)}} \otimes \ket{\TT{C}_{(-c,d)}}\,,
\end{equation}
Pictorially, the right-hand side looks like a sum over disk geometries filled in smoothly with all the possible \slz{} crosscaps
\bea \label{farey tail state}
    \ket{\TT{no-boundary}} \is \ \raisebox{-1.05cm}{\resizebox{3cm}{!}{\tikzset{every picture/.style={line width=0.75pt}} 

\begin{tikzpicture}[x=0.75pt,y=0.75pt,yscale=-1,xscale=1]

\draw  [draw opacity=0][fill={rgb, 255:red, 219; green, 219; blue, 219 }  ,fill opacity=1 ] (764,110) -- (1020,110) -- (1020,150) -- (764,150) -- cycle ;
\draw  [draw opacity=0][fill={rgb, 255:red, 219; green, 219; blue, 219 }  ,fill opacity=1 ][line width=1.5]  (1020,142) .. controls (1020,212.69) and (962.69,270) .. (892,270) .. controls (821.3,270) and (764,212.69) .. (764,142) -- (892,142) -- cycle ; \draw  [line width=1.5]  (1020,142) .. controls (1020,212.69) and (962.69,270) .. (892,270) .. controls (821.3,270) and (764,212.69) .. (764,142) ;  
\draw  [draw opacity=0][fill={rgb, 255:red, 219; green, 219; blue, 219 }  ,fill opacity=1 ][line width=1.5]  (764,110.58) .. controls (764,110.58) and (764,110.58) .. (764,110.58) .. controls (764,110.58) and (764,110.58) .. (764,110.58) .. controls (764,93.22) and (821.31,79.15) .. (892,79.15) .. controls (962.69,79.15) and (1020,93.22) .. (1020,110.58) -- (892,110.58) -- cycle ;
\draw  [draw opacity=0][fill={rgb, 255:red, 255; green, 255; blue, 255 }  ,fill opacity=0.31 ][line width=1.5]  (764,111.86) .. controls (764,93.99) and (821.31,79.51) .. (892,79.51) .. controls (962.69,79.51) and (1020,93.99) .. (1020,111.86) .. controls (1020,129.73) and (962.69,144.22) .. (892,144.22) .. controls (821.31,144.22) and (764,129.73) .. (764,111.86) -- cycle ;
\draw [line width=1.5]    (764,111.86) -- (764,142) ;
\draw  [color={rgb, 255:red, 0; green, 0; blue, 0 }  ,draw opacity=1 ][dash pattern={on 1.69pt off 2.76pt}][line width=1.5]  (764,111.86) .. controls (764,93.99) and (821.31,79.51) .. (892,79.51) .. controls (962.69,79.51) and (1020,93.99) .. (1020,111.86) .. controls (1020,129.73) and (962.69,144.22) .. (892,144.22) .. controls (821.31,144.22) and (764,129.73) .. (764,111.86) -- cycle ;
\draw [line width=1.5]    (1020,111.86) -- (1020,142) ;

\end{tikzpicture}}} \; + \; \raisebox{-1cm}{\resizebox{3cm}{!}{\tikzset{every picture/.style={line width=0.75pt}} 

\begin{tikzpicture}[x=0.75pt,y=0.75pt,yscale=-1,xscale=1]

\draw  [draw opacity=0][fill={rgb, 255:red, 219; green, 219; blue, 219 }  ,fill opacity=1 ] (412.02,110.19) -- (668.02,110.19) -- (668.02,232.06) -- (412.02,232.06) -- cycle ;
\draw  [draw opacity=0][fill={rgb, 255:red, 219; green, 219; blue, 219 }  ,fill opacity=1 ][line width=1.5]  (412,110.58) .. controls (412,110.58) and (412,110.58) .. (412,110.58) .. controls (412,110.58) and (412,110.58) .. (412,110.58) .. controls (412,93.22) and (469.31,79.15) .. (540,79.15) .. controls (610.69,79.15) and (668,93.22) .. (668,110.58) -- (540,110.58) -- cycle ;
\draw  [draw opacity=0][fill={rgb, 255:red, 219; green, 219; blue, 219 }  ,fill opacity=1 ][line width=1.5]  (668.02,231.47) .. controls (668.02,231.47) and (668.02,231.47) .. (668.02,231.47) .. controls (668.02,249.95) and (610.71,264.94) .. (540.02,264.94) .. controls (469.33,264.94) and (412.02,249.95) .. (412.02,231.47) -- (540.02,231.47) -- cycle ;
\draw  [draw opacity=0][fill={rgb, 255:red, 255; green, 255; blue, 255 }  ,fill opacity=0.31 ][line width=1.5]  (412.03,111.86) .. controls (412.03,93.99) and (469.34,79.51) .. (540.03,79.51) .. controls (610.72,79.51) and (668.03,93.99) .. (668.03,111.86) .. controls (668.03,129.73) and (610.72,144.22) .. (540.03,144.22) .. controls (469.34,144.22) and (412.03,129.73) .. (412.03,111.86) -- cycle ;
\draw [line width=1.5]    (412.03,111.86) -- (412.03,231.86) ;
\draw [line width=1.5]    (668.03,111.86) -- (668.03,231.86) ;
\draw  [color={rgb, 255:red, 0; green, 0; blue, 0 }  ,draw opacity=1 ][line width=1.5]  (412,231.98) .. controls (412,214.11) and (469.31,199.63) .. (540,199.63) .. controls (610.69,199.63) and (668,214.11) .. (668,231.98) .. controls (668,249.85) and (610.69,264.34) .. (540,264.34) .. controls (469.31,264.34) and (412,249.85) .. (412,231.98) -- cycle ;
\draw  [line width=0.75]  (538.65,231.17) .. controls (539.96,230.98) and (541.63,231.09) .. (542.39,231.42) .. controls (543.14,231.75) and (542.69,232.18) .. (541.39,232.37) .. controls (540.08,232.56) and (538.41,232.45) .. (537.65,232.12) .. controls (536.9,231.79) and (537.35,231.36) .. (538.65,231.17) -- cycle ;
\draw [line width=0.75]  [dash pattern={on 4.5pt off 4.5pt}]  (542.68,231.86) -- (668.03,231.86) ;
\draw  [draw opacity=0][line width=1.5]  (668.02,232.06) .. controls (668.02,249.93) and (610.71,264.41) .. (540.02,264.41) -- (540.02,232.06) -- cycle ; \draw [color={rgb, 255:red, 0; green, 0; blue, 0 }  ,draw opacity=1 ][line width=1.5]    (668.02,232.06) .. controls (668.02,249.93) and (610.71,264.41) .. (540.02,264.41) ; \draw [shift={(540.02,264.41)}, rotate = 357.4] [color={rgb, 255:red, 0; green, 0; blue, 0 }  ,draw opacity=1 ][line width=1.5]    (16.05,-2.99) .. controls (12.43,-1.27) and (9.12,-0.27) .. (6.11,0) .. controls (9.12,0.27) and (12.43,1.27) .. (16.05,2.99)(9.95,-2.99) .. controls (6.32,-1.27) and (3.01,-0.27) .. (0,0) .. controls (3.01,0.27) and (6.32,1.27) .. (9.95,2.99)   ; 
\draw  [draw opacity=0][line width=1.5]  (412.02,232.06) .. controls (412.02,214.19) and (469.33,199.7) .. (540.02,199.7) -- (540.02,232.06) -- cycle ; \draw [color={rgb, 255:red, 0; green, 0; blue, 0 }  ,draw opacity=1 ][line width=1.5]    (412.02,232.06) .. controls (412.02,214.19) and (469.33,199.7) .. (540.02,199.7) ; \draw [shift={(540.02,199.7)}, rotate = 177.04] [color={rgb, 255:red, 0; green, 0; blue, 0 }  ,draw opacity=1 ][line width=1.5]    (16.05,-2.99) .. controls (12.43,-1.27) and (9.12,-0.27) .. (6.11,0) .. controls (9.12,0.27) and (12.43,1.27) .. (16.05,2.99)(9.95,-2.99) .. controls (6.32,-1.27) and (3.01,-0.27) .. (0,0) .. controls (3.01,0.27) and (6.32,1.27) .. (9.95,2.99)   ; 
\draw  [color={rgb, 255:red, 255; green, 0; blue, 0 }  ,draw opacity=1 ][fill={rgb, 255:red, 255; green, 0; blue, 0 }  ,fill opacity=1 ] (664.81,231.86) .. controls (664.81,230.08) and (666.25,228.64) .. (668.03,228.64) .. controls (669.81,228.64) and (671.25,230.08) .. (671.25,231.86) .. controls (671.25,233.64) and (669.81,235.08) .. (668.03,235.08) .. controls (666.25,235.08) and (664.81,233.64) .. (664.81,231.86) -- cycle ;
\draw [line width=0.75]  [dash pattern={on 4.5pt off 4.5pt}]  (413.3,231.46) -- (538.65,231.46) ;
\draw  [color={rgb, 255:red, 255; green, 0; blue, 0 }  ,draw opacity=1 ][fill={rgb, 255:red, 255; green, 0; blue, 0 }  ,fill opacity=1 ] (408.81,231.86) .. controls (408.81,230.08) and (410.25,228.64) .. (412.03,228.64) .. controls (413.81,228.64) and (415.25,230.08) .. (415.25,231.86) .. controls (415.25,233.64) and (413.81,235.08) .. (412.03,235.08) .. controls (410.25,235.08) and (408.81,233.64) .. (408.81,231.86) -- cycle ;
\draw  [color={rgb, 255:red, 0; green, 0; blue, 0 }  ,draw opacity=1 ][dash pattern={on 1.69pt off 2.76pt}][line width=1.5]  (412,112.22) .. controls (412,94.35) and (469.31,79.86) .. (540,79.86) .. controls (610.69,79.86) and (668,94.35) .. (668,112.22) .. controls (668,130.09) and (610.69,144.57) .. (540,144.57) .. controls (469.31,144.57) and (412,130.09) .. (412,112.22) -- cycle ;

    \end{tikzpicture}}} \; + \; \raisebox{-1cm}{\resizebox{3cm}{!}{\input{images/farey_state_general}}} \; + \; \dots\qquad 
\eea
The expression \eqref{cls2 horizon} can be readily generalized for generic 2d CFT, provided an analogue of the generalized crosscaps \eqref{5.23boundary states} can be found. For example, the no-boundary construction could be replicated in the VMS \cite{Collier:2023cyw}, with potential applications to AdS$_3$ gravity.

In 2d gravity and string theory, it seems reasonable to augment this state, by summing over all genera (with one boundary), and including a gas of generalized crosscaps on the worldsheet:
\begin{equation}
    \ket{\TT{no-boundary}} \to\quad  \raisebox{-1.05cm}{\resizebox{3cm}{!}{\input{images/farey_genus_0}}} \; + \; \raisebox{-1.35cm}{\resizebox{3cm}{!}{\input{images/farey_genus_1}}} \; + \; \raisebox{-2.77cm}{\resizebox{3cm}{!}{\input{images/farey_genus_2}}} \; + \; \dots
\end{equation}
The geometries on the right describe the topology of the 2d slice suspended by the observer's worldline. When rotated to Lorentzian signature, 
all these Euclidean geometries lift to a Lorentzian spacetime with the same KdS static patch. It thus seems like a reasonable proposal that the associated $\mathbb{C}\text{LS} \otimes \mathbb{C}\text{LS}$ amplitude with FZZT boundary conditions $\ket{\alpha_+, \alpha_-}$ computes the observer's spectral density, including the contributions from (subleading) off-shell geometries.\footnote{In AdS$_3$ gravity, a gas of Dehn fillings is related with taking into account the (off shell) Seifert manifolds \cite{deBoer:2025rct,Belin:2026pko}. In the near extremal limits, these multiple Dehn fillings are described by a gas of 2d defects \cite{Maxfield:2020ale}.}  It would be interesting to study this further from the perspective of the $\mathbb{C}$LS matrix integral \cite{Collier:2024lys}.


\section*{Acknowledgments}
We thank Dionysios Anninos, Lorenz Eberhardt, Beatrix Muehlmann, Eric Perlmutter, Erez Urbach, and Erik Verlinde for useful discussions. AB is supported by the Leinweber foundation, by the US DOE DE-SC0009988, and the Sivian fund. We acknowledge discussions during the workshop ``Observers, wormholes and complex saddles in cosmology" at the Bernoulli center at EPFL (Lausanne).

\appendix
\section{$\mathbb{C}$LS details}\label{app:stringdetails}
\par In this appendix, we present more details on the \cls{} calculations of section \ref{sect:5cls}. Most steps are \slz{} generalization of the regular crosscap/M\"obius strip case studied in \cite{Blommaert:2025eps}. We summarize our conventions for modular matrices in appendix \ref{app: conventions}. Boundary states involved in the crosscap amplitudes are defined in section \ref{app: states}. The closed and open string channel amplitudes are discussed in appendix \ref{appA.2closed} and \ref{appA.3open} respectively, where we walk through the calculations of the string amplitudes including all the numerical prefactors suppressed in the main text. 

\subsection{Modular matrices} \label{app: conventions}
The worldsheet theory of \cls{} consists of two spacelike Liouville CFTs whose central charge and physical spectrum of Liouville momenta are analytically continued to the regime
\begin{equation}
    c_\pm \in 13 \pm \i \mathbb{R} \,, \quad P_\pm \in e^{\pm \i \pi/4} \mathbb{R}\,.
\end{equation}
It is convenient to parameterize the central charge in terms of a real parameter $\hbar$:
\begin{equation}
    c_\pm = 1 + 6(b_\pm + b_\pm^{-1})^2 = 13+6\i\, (\sfb^2-\sfb^{-2}) \,, \quad  b_\pm = e^{\pm \i \pi/4} \sfb  \,, \quad \hbar=4 \pi \sfb^2 \in \mathbb{R}_+ \, .
\end{equation}
Similarly one parameterizes the conformal weights $h_\pm$ and the Liouville momenta $P_\pm$ as
\begin{equation}
    h_\pm = \frac{c_\pm-1}{24} + P_\pm^2 \,, \qquad   P_\pm = \frac{\pm \i}{4 \pi b_\pm}\alpha = \frac{e^{\pm \i \pi/4}}{2 \sqrt{\pi \hbar}} \alpha\,,\quad \alpha \in \mathbb{R}\,.
\end{equation}
Null stats are
\begin{equation}
\begin{split}
    \alpha_{(n,m)} = \pm 2\pi n + \frac{\i \hbar}{2} m
\end{split}
\end{equation}
The identity representation corresponds to $\mathbb{1} = \alpha_{(1,1)}$. Its null sub-representation $\alpha_{(1,-1)}$ arises at level one. The Virasoro characters for a generic representation reads
\begin{equation} \label{virasoro character definition}
    \chi_\alpha^\pm(\tau) =\Tr_\alpha\LL[e^{2 \pi \i \tau (L_0 - \frac{c_\pm}{24} )} \RR] = \frac{e^{\mp \frac{\alpha^2}{2 \hbar}\tau}}{\eta(\tau)} \, .
\end{equation}
For the identity character one simply subtracts off the contribution of the null states
\begin{equation}
    \chi_{\mathbb{1}}^\pm(\tau) = \chi_{\pm 2 \pi + \i \hbar/2}^\pm(\tau) - \chi_{\pm 2 \pi - \i \hbar/2}^\pm(\tau)
\end{equation}

The Virasoro characters form a representation of $\TT{PSL}(2,\mathbb{Z})$. We focus on the $c_+$ Liouville, dropping the $+$ superscript. The results for the $c_-$ theory are found by $\hbar \to - \hbar$, as evident from \eqref{virasoro character definition}. A matrix
\begin{equation}
    \gamma = \biggl( \begin{matrix} a\! & b \\[-.5mm] c \! &  d \, \end{matrix}\biggr) \; \in \; \text{PSL(2,}\mathbb{Z})\,,
\end{equation}
is represented by the modular matrix $\mathsf{K}^\gamma_{\alpha \beta}$ defined through:
\begin{equation} \label{modular transformation character generic app}
    \chi_{\alpha} \LL(\frac{a \tau + b}{c \tau + d} \RR) = \frac{1}{2} \int \d \beta \, \mathsf{K}^\gamma_{\alpha\beta}  \,\chi_\beta(\tau)\,,
\end{equation}
where the contour of integration is chosen such that the integral converges. $\mathsf{K}^\gamma_{\alpha \beta}$ is given by
\begin{equation}
    \mathsf{K}^\gamma_{\alpha \beta} = \frac{\sqrt{2\i}}{\sqrt{\pi \hbar}} \frac{e^{\i \pi \phi(\gamma)}}{\sqrt{c}} e^{-\frac{a \alpha^2 + d \beta^2}{2 c \hbar}} \cosh\Big(\frac{\alpha \beta}{c \hbar}\Big)\,,\quad \phi(\gamma)=s(d,c) -\frac{a+d}{12 c} \, .
\end{equation}
$s(d,c)$ is the Dedekind sum, and the combination $\phi(c,d)$ arises from the phase picked up by the Dedekind eta function under modular transformation. For the identity representation, one subtracts two terms:
\begin{equation}
    \mathsf{K}^\gamma_{\mathbb{1}\beta} = \ca{N}_{\gamma} \frac{e^{\i \pi \phi(\gamma)}}{\sqrt{c}}e^{-\frac{d}{2c\hbar} \beta^2 + \frac{2\pi}{c \hbar}\beta} \sin\Big(\frac{2\pi a - \beta}{2c}\Big)+ \,(\beta\to-\beta)\,\quad
    \ca{N}_{\gamma} = -\frac{ \sqrt{2\i}}{\sqrt{\pi \hbar}} e^{- \frac{a}{2 c \hbar}\LL(4\pi^2 - \frac{\hbar^2}{4}\RR)}\,. 
\end{equation}
For the generators
\begin{equation}
    \mathsf{S} = \biggl( \begin{matrix} 0\! & -1 \\[-.5mm] 1 \! &  0 \, \end{matrix}\biggr) \,, \quad  \mathsf{T} = \biggl( \begin{matrix} 1\! & 1 \\[-.5mm] 0 \! &  1 \, \end{matrix}\biggr)\,,
\end{equation}
one finds simpler expressions:
\begin{equation}
\label{modularS}
    \mathsf{S}_{\alpha \beta} = \frac{\sqrt{2 \i}}{ \sqrt{\pi \hbar}} \, \cosh\Big(\frac{\alpha \beta}{ \hbar}\Big)  \,, \quad   \mathsf{S}_{\mathbb{1} \beta} = \frac{(2\i)^{3/2}}{\sqrt{\pi \hbar}} \sin\Big(\frac{\beta}{2}\Big) \sinh\Big(\frac{2 \pi \beta}{\hbar}\Big) \,,\quad 
    \mathsf{T}_{\beta} = e^{2 \pi \i \LL(h - \frac{c_+}{24}\RR)} = e^{-\frac{\pi \i}{12}-   \frac{\beta^2}{2\hbar}}  \, .
\end{equation}

\medskip

\subsection{Boundary states} \label{app: states}
We now define the boundary states of the annulus amplitude \eqref{5.31cc}. Starting on the $\varphi_-$ side,  boundary states can be expanded in (crosscap) Ishibashi states as follows:
\begin{equation}
    \ket{\TT{FZZT}(0)} =  \frac{1}{2} \int \d \bar{\beta} \, \frac{\mathsf{S}_{0 \bar{\beta}}}{\sqrt{\mathsf{S}_{\mathbb{1}\bar{\beta}}}} \,\ishiket{\bar{\beta}}\,,\quad e^{-2\pi\i \frac{d}{c} L_0} \ket{\TT{ZZ}} = \frac{1}{2} \int_\mathbb{R} \d \bar{\beta} \, \sqrt{\mathsf{S}_{\mathbb{1}\bar{\beta}}} \,\ishiket{\bar{\beta}}_\TT{C}    \, .
\end{equation}
As for the $\phi_+$ Liouville, the channel duality relation \eqref{5.28chi} fixes the generalized crosscap wavefunction to
\begin{equation} 
    \ket{\text{C}_{\gamma}}  = 
    \frac{1}{2} \int_{\Gamma} \d \beta \, \frac{\mathsf{K}^{\gamma}_{\mathbb{1} \beta} }{ \sqrt{\mathsf{S}_{\mathbb{1} \beta} }} \, {\ishiket{\beta}}_\TT{C}   \, .
\end{equation}
The contour $\Gamma$ is discussed in figure \ref{fig: contour drawing}. The spectral FZZT state is defined as
\begin{align} \label{marked fzzt definition}
    \ket{\TT{FZZT}_\bullet(\alpha_+)} &= -\frac{\sqrt{\i}}{2\pi} \int \d \beta \, \frac{\de_{\alpha_+} (\mathsf{S}_{\beta, \alpha_+ + 2\pi + \i \hbar/2} - \mathsf{S}_{\beta, \alpha_+ + 2\pi - \i \hbar/2})}{\sqrt{\mathsf{S}_{\mathbb{1}\beta}}} \ishiket{\beta}\nonumber\\ &= \frac{1}{\pi \hbar \sqrt{\i}} \int \d \beta  \, \frac{\beta \sin(\beta/2) \mathsf{S}_{\beta, \alpha_++2\pi}}{\sqrt{\mathsf{S}_{\mathbb{1}\beta}}} \ishiket{\beta}  \,.
\end{align}
The derivative $\de_{\alpha_+}$ and the discontinuity arise when defining an amplitude that is interpreted naturally as spectral density \cite{Mertens:2020hbs, Collier:2024mlg_zz, Blommaert:2025eps, Kostov:2002uq}. An FZZT state
\begin{equation}
    \ket{\TT{FZZT}(\alpha)} = \frac{1}{2} \int \d \beta \, \frac{\mathsf{S}_{\beta \alpha}}{\sqrt{\mathsf{S}_{\mathbb{1}\beta}}} \ishiket{\beta}   \, .
\end{equation}
corresponds to fixing the dual boundary cosmological constant;
\begin{equation}
    \tilde{\mu}_{\text{B}}(\alpha) = \cosh\Big(\frac{2 \pi \alpha}{\hbar}\Big)  \, ,\quad I= \tilde{\mu}_{\text{B}} \int \d x \, e^{\varphi_+/b_+^2}\,.
\end{equation}
Denote the corresponding amplitude $\mathcal{Z}(\tilde{\mu}_\text{B})$. Fixed length amplitudes $\mathcal{Z}(\ell)$ are found as \cite{Mertens:2020hbs, Collier:2024mlg_zz, Blommaert:2025eps, Kostov:2002uq}:
\begin{equation} \label{cls fixed length}
    \ca{Z}(\ell) = \frac{1}{2\pi \i} \int_{-\i \infty}^{+\i \infty} \d \tilde{\mu}_{B} \, e^{\tilde{\mu}_{B} \ell} \, \frac{\d}{\d \tilde{\mu}_{B}} \ca{Z}(\tilde{\mu}_{B}) \, .
\end{equation}
The inverse Laplace transform can be performed by contour deformation. Parameterizing the position along the branch using
\begin{equation}
    \tilde{\mu}_{B} = - \cosh\Big(\frac{2 \pi \alpha}{\hbar}\Big) \, ,
\end{equation}
one obtains \cite{Blommaert:2025eps}
\begin{equation} \label{cls fixed length amplitude}
    \ca{Z}(\ell)  = \frac{1}{4\pi \i} \int \d \alpha \, e^{-\ell \cosh(2 \pi \alpha/\hbar)}  \, \frac{\d}{\d \alpha} \Big[ \ca{Z}\Big(\alpha + \frac{\i \hbar}{2}\Big) -  \ca{Z}\Big(\alpha - \frac{\i \hbar}{2}\Big) \Big]= \frac{\sqrt{\i}}{4} \int \d \alpha \, e^{-\ell \cosh(2 \pi \alpha/\hbar)} \ca{Z}_\bullet(\alpha)
\end{equation}
Here $\ca{Z}_\bullet(\alpha)$ deserves the name of spectrum, and is associated with the state \eqref{marked fzzt definition} indeed, after shifting for convenience furthermore $\alpha \to \alpha_+ + 2\pi$.

In the case of the KdS spectral density, we have boundary states $\ket{\TT{FZZT}_\bullet(\alpha_+)}$ and $\ket{\TT{FZZT}_\bullet(\alpha_-)}$, one for each \cls{}. So $\rho_{(c,d)}(\alpha_+,\alpha_-) = \rho_{(c,d)}(\alpha_+) \rho_{(-c,d)}(\alpha_-)$ is a density with respect to the real measure:
\begin{equation}
    \frac{\i}{16} \,  \d \alpha_+ \, \d \alpha_- = \frac{\pi^2}{8}\frac{\d E\,\d J }{\sqrt{E^2 + J^2}} \, .
\end{equation}

\subsection{Closed string channel}\label{appA.2closed}
\begin{figure}[h]
    \centering
\raisebox{2mm}{\resizebox{0.8\linewidth}{!}{\input{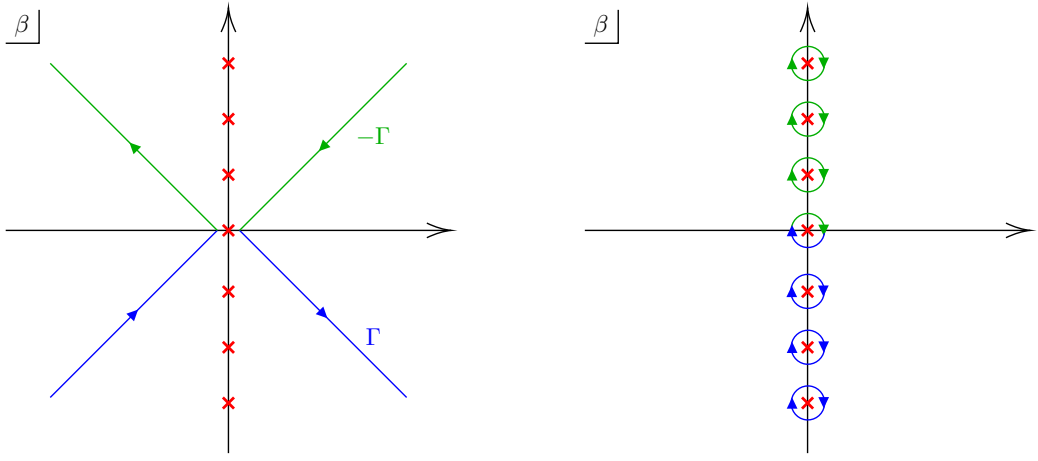}}}
    \caption{The contour $\Gamma \cup -\Gamma$ can be deformed to pick up the poles at momenta $\beta = \i \hbar n/2$, $n \in \mathbb{Z}$ \cite{Blommaert:2025eps}.}
    \label{fig: contour drawing}
\end{figure}

We study the closed string amplitude \eqref{5.31cc}. The contributions due to the $\phi_+$ and $\varphi_-$ Liouvilles read
\begin{equation} 
\begin{split}
    \bra{\TT{FZZT}_\bullet(\alpha_+)} e^{\i \pi \tau H} \ket{\TT{C}_{\gamma}}  = -\frac{1}{2 \sqrt{2 \pi \hbar}}  \int_{\Gamma} \d \beta \; \frac{\beta }{\sinh(2\pi\beta/\hbar)}\,\mathsf{K}^{\gamma}_{\mathbb{1} \beta} \, \mathsf{S}_{\alpha_+ + 2\pi\,\beta}  \, \chi^+_{\beta}\Bigl(\tau - \frac{d}{c} \Bigr)  \,,
\end{split}
\end{equation}
and
\begin{equation}
\begin{split}
    \bra{\TT{FZZT}(0)} e^{\i \pi \tau H} e^{-2\pi \i \frac{d}{c} L_0}\ket{\TT{ZZ}}  & = \frac{1}{\sqrt{2\i \pi \hbar}}  \int \d \bar{\beta}\, \chi^-_{\bar{\beta}}\LL(\tau - \frac{d}{c} \RR)  \, .
\end{split}
\end{equation}
The integral over the modulus $\tau$ is performed before analytically continuing the Liouville momenta \cite{Collier:2024mlg_zz}, leading to the propagator
\begin{equation}
    -\i \int_0^{\i \infty} \d \tau \, \eta\Big(\tau - \frac{d}{c}\Big)^2 \chi^+_\beta\Big(\tau - \frac{d}{c}\Big) \chi^-_{\bar{\beta}}\Big(\tau - \frac{d}{c}\Big) = \frac{2 \i \hbar}{\bar{\beta}^2 - \beta^2} \, .
\end{equation}
The integral over the Ishibashi momentum $\bar{\beta}$ is performed by picking up the pole $\bar{\beta} = \beta$ of the propagator in the lower half plane
\begin{equation}
    \int \d \bar{\beta} \, \frac{2 \i \hbar}{\bar{\beta}^2 - \beta^2} f(\beta,\bar{\beta})   = \frac{2 \pi \hbar}{\beta} f(\beta,\beta) \, .
\end{equation}
This projects on on-shell primary states propagating in the closed channel. We are left with the integral
\begin{equation} 
    \rho^{\mathbb{C}\text{LS}}_\gamma(\alpha_+)  = -\frac{1}{2 \sqrt{\i}} \int_{\Gamma} \d \beta  \, \frac{\mathsf{K}^{\gamma}_{\mathbb{1}\beta}\mathsf{S}_{\alpha_+ + 2\pi\,\beta}}{\sinh(2 \pi \beta/\hbar)}  \, .   
\end{equation}
which can be performed by picking up the poles at $\beta = \i \hbar n/2$ as discussed in figure \ref{fig: contour drawing} 
\begin{equation} 
    \rho^{\mathbb{C}\text{LS}}_\gamma(\alpha_+)  = \frac{\sqrt{\i} \, \hbar}{4}  \sum_{n \in \mathbb{Z}}  \mathsf{K}^{\gamma}_{\mathbb{1}\, \frac{\i \hbar n}{2}} \mathsf{S}_{\alpha_+ \,  \frac{\i \hbar n}{2}} \, .   
\end{equation}
Poisson summation gives:
\begin{equation}
     \frac{\i\hbar}{4} \sum_{n \in \mathbb{Z}} \mathsf{K}^{\gamma}_{\mathbb{1}\,\frac{\i \hbar n}{2}} \mathsf{S}_{\alpha_+\,\frac{\i \hbar n}{2}}  = \frac{1}{2} \sum_{n \in \mathbb{Z}} \int \, \d \beta \, \mathsf{K}^{\gamma}_{\mathbb{1} \beta} \, \mathsf{S}_{\beta\,\alpha_+ + 4 \pi n}  =  \sum_{n \in \mathbb{Z}} \mathsf{K}^{\gamma\cdot S}_{\mathbb{1}\,\alpha_+ + 4 \pi n}  \equiv \sum_{n \in \mathbb{Z}} \mathsf{M}^{\gamma}_{\mathbb{1}, \alpha_+ + 4 \pi n}  \, .
\end{equation}
The gravitational chiral wavefunction is one term in this sum and equals:
\begin{equation}
\begin{split}
     \rho_\gamma(\alpha_+) = \frac{1}{\sqrt{\i}}\mathsf{M}^\gamma_{\mathbb{1}\alpha_+}  = \Big[\frac{\ca{N}_{\gamma \cdot S} }{\sqrt{\i}}\, e^{\i \pi \phi(\gamma \cdot S)}\Big]\, e^{\frac{c}{2 d\hbar}\alpha_+^2 +  \frac{2\pi}{d \hbar}\alpha_+}  \frac{1}{\sqrt{d}} \sin\Big(\frac{2\pi b-\alpha_+}{2 d}\Big) + (\alpha_+ \to - \alpha_+ )  \, .
\end{split}
\end{equation}
Combining the chiral-and anti-chiral wavefunctions, the factors in braces become a real normalization constant, which was dropped in the main text:
\begin{equation}
    \Big[ \frac{\ca{N}_{\gamma_+ \cdot S}}{\sqrt{\i}} \, e^{\i \pi \phi(\gamma_+ \cdot S)}\Big] \,\Big[ \frac{\ca{N}_{\gamma_- \cdot S}}{\sqrt{\i}} \, e^{\i \pi \phi(\gamma_- \cdot S)}\Big] = \frac{2}{\pi \hbar} \, .
\end{equation}

\subsection{Open string channel}\label{appA.3open}
We prove the equivalence of the open trace \eqref{cls open channel integral} with the closed string amplitude \eqref{5.31cc}. In the open channel, the generalized M\"obius strip has length $\tau_\text{open} = -1/c^2 \tau$ and it is twisted by a $2\pi a/c$ rotation; such a twist mixes left and right movers, without affecting the zero-modes. The character of a conformal family propagating on the strip is therefore:
\begin{equation} \label{hatted characters}
    \hat{\chi}^\pm_{\alpha}(\tau_\text{open}) = \Tr_\alpha \Big[e^{\frac{2\pi \i a}{c} (L_0 - h)} e^{2 \pi \i \tau_\text{open} (L_0 - \frac{c_\pm}{24})}\Big] = \frac{e^{\mp \frac{\alpha^2}{2 \hbar} \tau_\text{open}}}{\eta\LL(\tau_\text{open} + \frac{a}{c}\RR)}\,.
\end{equation}
One can then use the inverse of \eqref{modular transformation character generic app}
to replace the closed channel characters in \eqref{cls + amplitude} and \eqref{barbeta} with their open channel counterpart:
\begin{equation}
    \chi_\beta\Big(\tau - \frac{d}{c}\Big) = \frac{e^{\mp \frac{a}{2c\hbar} \beta'^2}}{2} \int \d \beta\, \mathsf{K}^{\gamma^{-1}}_{\beta \beta'} \, \hat{\chi}_{\beta'}\LL(\tau_\text{open}\RR)
\end{equation}
This result in the open/closed relations
\begin{equation}
    \bra{\TT{FZZT}_\bullet(\alpha_+)} e^{\i \pi \tau H} \ket{\TT{C}_{\gamma}} = \frac{\i \sqrt{\hbar} \, c}{2\sqrt{2 \pi}} \, \ca{N}_\gamma \, e^{\i \pi (\phi(\gamma) + \phi(\gamma^{-1}))}  \int \d \beta \, \hat{\rho}^{\, \mathbb{C}\text{LS}}_{\gamma}(\alpha, \beta) \,  \de_\beta \hat{\chi}_\beta(\tau_\text{open})\,,
\end{equation}
with spectrum:
\begin{equation}
    \hat{\rho}^{\, \mathbb{C}\text{LS}}_{\gamma}(\alpha, \beta) =\sum_{n \in \mathbb{Z}} e^{- \i \pi \frac{a}{c}}\, \delta(c (\alpha_+ + 4 \pi n) + 2 \pi + \i \hbar/2) - e^{\i \pi \frac{a}{c}}\, \delta(c (\alpha_+ + 4 \pi n) + 2 \pi - \i\hbar/2) + (\alpha_+ \to - \alpha_+ )\,.
\end{equation}
Furthermore:
\begin{equation}
    \bra{\TT{FZZT}(0)} e^{\i \pi \tau H} e^{-2\pi \i \frac{d}{c} L_0}\ket{\TT{ZZ}}  = - \sqrt{\i} \, e^{\i \pi \phi(\gamma^{-1})} \int \d \bar{\beta} \, e^{\frac{1}{2 c d \hbar}\bar{\beta}^2} \hat{\chi}_{\bar{\beta}}(\tau_\text{open})
\end{equation}
Finally:
\begin{equation}
    -\i \int_0^{\i \infty} \d \tau \, \eta\LL(\tau - \frac{d}{c}\RR)^2 = - \frac{e^{-2\pi \i \phi(\gamma^{-1})}}{c} \int_0^{\i \infty} \frac{\d \tau_\text{open}}{\tau_\text{open}} \, \eta\LL(\tau_\text{open} + \frac{a}{c}\RR)^2   
\end{equation}
This proves the equivalence of the open twisted trace \eqref{cls open channel integral} with the closed crosscap amplitude \eqref{5.31cc}, provided
\begin{equation}
    \hat{\mathsf{K}}^{\gamma}_\TT{clock} = \Big[-\frac{\sqrt{\hbar} \, \ca{N}_\gamma}{2 \sqrt{2\pi \i}}  e^{\i \pi \phi(\gamma)} \Big]  \frac{1}{\sqrt{d}} e^{\frac{2 \pi \i}{c d} (h - \frac{c_- -1}{24})} \, .
\end{equation}
Note that when combining the chiral and anti-chiral gravitational wavefunctions, the prefactors in the braces, which were dropped in equation \eqref{clock twist}, lead to a simple overall constant.

\bibliographystyle{ourbst}
\bibliography{Refs}

@article{Godet:2024ich,
    author = "Godet, Victor",
    title = "{Quantum cosmology as automorphic dynamics}",
    eprint = "2405.09833",
    archivePrefix = "arXiv",
    primaryClass = "hep-th",
    month = "5",
    year = "2024"
}

@article{Okuyama:2025xwx,
    author = "Okuyama, Kazumi",
    title = "{Cap Amplitudes in Random Matrix Models}",
    eprint = "2509.03930",
    archivePrefix = "arXiv",
    primaryClass = "hep-th",
    doi = "10.1093/ptep/ptag030",
    journal = "PTEP",
    volume = "2026",
    number = "3",
    pages = "033B06",
    year = "2026"
}

@article{Marini:2026zjk,
    author = "Marini, Tommaso and Qi, Xiao-Liang and Verlinde, Herman",
    title = "{3D near-de Sitter gravity and the soft mode of DSSYK}",
    eprint = "2604.21014",
    archivePrefix = "arXiv",
    primaryClass = "hep-th",
    month = "4",
    year = "2026"
}

@article{Schouten:2025tvn,
    author = "Schouten, Koen and Isachenkov, Mikhail",
    title = "{The von Neumann algebraic quantum group $\mathrm{SU}_q(1,1)\rtimes \mathbb{Z}_2$ and the DSSYK model}",
    eprint = "2512.10101",
    archivePrefix = "arXiv",
    primaryClass = "math-ph",
    month = "12",
    year = "2025"
}

@article{vanderHeijden:2025zkr,
    author = "van der Heijden, Jeremy and Verlinde, Erik and Xu, Jiuci",
    title = "{Quantum symmetry and geometry in double-scaled SYK}",
    eprint = "2511.08743",
    archivePrefix = "arXiv",
    primaryClass = "hep-th",
    doi = "10.1007/JHEP05(2026)148",
    journal = "JHEP",
    volume = "05",
    pages = "148",
    year = "2026"
}

@article{Belaey:2025ijg,
    author = "Belaey, Andreas and Mertens, Thomas G. and Tappeiner, Thomas",
    title = "{Quantum group origins of edge states in double-scaled SYK}",
    eprint = "2503.20691",
    archivePrefix = "arXiv",
    primaryClass = "hep-th",
    doi = "10.1007/JHEP02(2026)184",
    journal = "JHEP",
    volume = "02",
    pages = "184",
    year = "2026"
}

@article{EllegaardAndersen:2011vps,
    author = "Ellegaard Andersen, J{\o}rgen and Kashaev, Rinat",
    title = {{A TQFT from Quantum Teichm{\"u}ller Theory}},
    eprint = "1109.6295",
    archivePrefix = "arXiv",
    primaryClass = "math.QA",
    doi = "10.1007/s00220-014-2073-2",
    journal = "Commun. Math. Phys.",
    volume = "330",
    pages = "887--934",
    year = "2014"
}

@article{Kashaev:2000ku,
    author = "Kashaev, R. M.",
    title = "{On the spectrum of Dehn twists in quantum Teichmuller theory}",
    eprint = "math/0008148",
    archivePrefix = "arXiv",
    reportNumber = "HIP-2000-44-TH",
    doi = "10.1142/9789812810007_0004",
    month = "8",
    year = "2000"
}

@article{Terashima:2011qi,
    author = "Terashima, Yuji and Yamazaki, Masahito",
    title = "{SL(2,R) Chern-Simons, Liouville, and Gauge Theory on Duality Walls}",
    eprint = "1103.5748",
    archivePrefix = "arXiv",
    primaryClass = "hep-th",
    reportNumber = "PUPT-2368",
    doi = "10.1007/JHEP08(2011)135",
    journal = "JHEP",
    volume = "08",
    pages = "135",
    year = "2011"
}

@article{Mikhaylov:2017ngi,
    author = "Mikhaylov, Victor",
    title = {{Teichm{\"u}ller TQFT vs. Chern-Simons theory}},
    eprint = "1710.04354",
    archivePrefix = "arXiv",
    primaryClass = "hep-th",
    doi = "10.1007/JHEP04(2018)085",
    journal = "JHEP",
    volume = "04",
    pages = "085",
    year = "2018"
}

@article{Teschner:2003em,
    author = "Teschner, J.",
    editor = "Belavin, A. and Corrigan, Edward",
    title = "{On the relation between quantum Liouville theory and the quantized Teichmuller spaces}",
    eprint = "hep-th/0303149",
    archivePrefix = "arXiv",
    doi = "10.1142/S0217751X04020579",
    journal = "Int. J. Mod. Phys. A",
    volume = "19S2",
    pages = "459--477",
    year = "2004"
}

@article{Chekhov:1999tn,
    author = "Chekhov, L. and Fock, V. V.",
    title = "{Quantum Teichmuller space}",
    eprint = "math/9908165",
    archivePrefix = "arXiv",
    doi = "10.1007/BF02557246",
    journal = "Theor. Math. Phys.",
    volume = "120",
    pages = "1245--1259",
    year = "1999"
}

@article{Kashaev:1998fc,
    author = "Kashaev, R. M.",
    title = "{Quantization of Teichmueller spaces and the quantum dilogarithm}",
    doi = "10.1023/A:1007460128279",
    journal = "Lett. Math. Phys.",
    volume = "43",
    pages = "105--115",
    year = "1998"
}

@article{Hikida:2002bt,
    author = "Hikida, Yasuaki",
    title = "{Liouville field theory on a unoriented surface}",
    eprint = "hep-th/0210305",
    archivePrefix = "arXiv",
    reportNumber = "UT-02-57",
    doi = "10.1088/1126-6708/2003/05/002",
    journal = "JHEP",
    volume = "05",
    pages = "002",
    year = "2003"
}

@incollection{blumenhagen2009boundary,
  title={Boundary conformal field theory},
  author={Blumenhagen, Ralph and Plauschinn, Erik},
  booktitle={Introduction to Conformal Field Theory: With Applications to String Theory},
  pages={205--256},
  year={2009},
  publisher={Springer}
}

@article{Kostov:2002uq,
    author = "Kostov, Ivan K.",
    title = "{Boundary correlators in 2-D quantum gravity: Liouville versus discrete approach}",
    eprint = "hep-th/0212194",
    archivePrefix = "arXiv",
    reportNumber = "HUTP-02-A065",
    doi = "10.1016/S0550-3213(03)00147-0",
    journal = "Nucl. Phys. B",
    volume = "658",
    pages = "397--416",
    year = "2003"
}

@article{Collier:2024mlg_zz,
    author = {Collier, Scott and Eberhardt, Lorenz and M\"uhlmann, Beatrix and Rodriguez, Victor A.},
    title = "{The complex Liouville string: worldsheet boundaries and non-perturbative effects}",
    eprint = "2410.09179",
    archivePrefix = "arXiv",
    primaryClass = "hep-th",
    month = "10",
    year = "2024"
}

@article{Strominger:2001pn,
    author = "Strominger, Andrew",
    title = "{The dS / CFT correspondence}",
    eprint = "hep-th/0106113",
    archivePrefix = "arXiv",
    doi = "10.1088/1126-6708/2001/10/034",
    journal = "JHEP",
    volume = "10",
    pages = "034",
    year = "2001"
}

@article{Verlinde:2024zrh,
    author = "Verlinde, Herman and Zhang, Mengyang",
    title = "{SYK Correlators from 2D Liouville-de Sitter Gravity}",
    eprint = "2402.02584",
    archivePrefix = "arXiv",
    primaryClass = "hep-th",
    month = "2",
    year = "2024"
}

@article{Verlinde:1989ua,
    author = "Verlinde, Herman L.",
    title = "{Conformal Field Theory, 2-$D$ Quantum Gravity and Quantization of Teichmuller Space}",
    reportNumber = "PUPT-89/1140",
    doi = "10.1016/0550-3213(90)90510-K",
    journal = "Nucl. Phys. B",
    volume = "337",
    pages = "652--680",
    year = "1990"
}

@article{Gaiotto_Teschner:2024osr,
    author = {Gaiotto, Davide and Teschner, J\"org},
    title = "{Schur Quantization and Complex Chern-Simons theory}",
    eprint = "2406.09171",
    archivePrefix = "arXiv",
    primaryClass = "hep-th",
    month = "6",
    year = "2024"
}

@article{Dimofte:2011py,
    author = "Dimofte, Tudor and Gaiotto, Davide and Gukov, Sergei",
    title = "{3-Manifolds and 3d Indices}",
    eprint = "1112.5179",
    archivePrefix = "arXiv",
    primaryClass = "hep-th",
    doi = "10.4310/ATMP.2013.v17.n5.a3",
    journal = "Adv. Theor. Math. Phys.",
    volume = "17",
    number = "5",
    pages = "975--1076",
    year = "2013"
}

@article{Godet:2025bju,
    author = "Godet, Victor",
    title = {{M{\"o}bius randomness in the Hartle-Hawking state}},
    eprint = "2505.03068",
    archivePrefix = "arXiv",
    primaryClass = "hep-th",
    month = "5",
    year = "2025"
}

@article{Castro:2012gc,
    author = "Castro, Alejandra and Maloney, Alexander",
    title = "{The Wave Function of Quantum de Sitter}",
    eprint = "1209.5757",
    archivePrefix = "arXiv",
    primaryClass = "hep-th",
    reportNumber = "NSF-KITP-12-165",
    doi = "10.1007/JHEP11(2012)096",
    journal = "JHEP",
    volume = "11",
    pages = "096",
    year = "2012"
}

@article{Anninos:2022ujl,
    author = {Anninos, Dionysios and Galante, Dami{\'a}n A. and M{\"u}hlmann, Beatrix},
    title = "{Finite features of quantum de Sitter space}",
    eprint = "2206.14146",
    archivePrefix = "arXiv",
    primaryClass = "hep-th",
    doi = "10.1088/1361-6382/acaba5",
    journal = "Class. Quant. Grav.",
    volume = "40",
    number = "2",
    pages = "025009",
    year = "2023"
}

@article{hayward1990euclidean,
  title={Euclidean action and the thermodynamics of manifolds without boundary},
  author={Hayward, Geoff},
  journal={Physical Review D},
  volume={41},
  number={10},
  pages={3248},
  year={1990},
  publisher={APS}
}

@article{Banihashemi:2022jys,
    author = "Banihashemi, Batoul and Jacobson, Ted",
    title = "{Thermodynamic ensembles with cosmological horizons}",
    eprint = "2204.05324",
    archivePrefix = "arXiv",
    primaryClass = "hep-th",
    doi = "10.1007/JHEP07(2022)042",
    journal = "JHEP",
    volume = "07",
    pages = "042",
    year = "2022"
}

@article{Coleman:2021nor,
    author = "Coleman, Evan and Mazenc, Edward A. and Shyam, Vasudev and Silverstein, Eva and Soni, Ronak M. and Torroba, Gonzalo and Yang, Sungyeon",
    title = "{De Sitter microstates from T$ \overline{T} $ + {\ensuremath{\Lambda}}$_{2}$ and the Hawking-Page transition}",
    eprint = "2110.14670",
    archivePrefix = "arXiv",
    primaryClass = "hep-th",
    doi = "10.1007/JHEP07(2022)140",
    journal = "JHEP",
    volume = "07",
    pages = "140",
    year = "2022"
}

@article{Narovlansky:2025tpb,
    author = "Narovlansky, Vladimir",
    title = "{Towards a microscopic description of de Sitter dynamics}",
    eprint = "2506.02109",
    archivePrefix = "arXiv",
    primaryClass = "hep-th",
    month = "6",
    year = "2025"
}

@article{Anninos:2021ene,
    author = {Anninos, Dionysios and Bautista, Teresa and M{\"u}hlmann, Beatrix},
    title = "{The two-sphere partition function in two-dimensional quantum gravity}",
    eprint = "2106.01665",
    archivePrefix = "arXiv",
    primaryClass = "hep-th",
    doi = "10.1007/JHEP09(2021)116",
    journal = "JHEP",
    volume = "09",
    pages = "116",
    year = "2021"
}

@article{Hikida:2021ese,
    author = "Hikida, Yasuaki and Nishioka, Tatsuma and Takayanagi, Tadashi and Taki, Yusuke",
    title = "{Holography in de Sitter Space via Chern-Simons Gauge Theory}",
    eprint = "2110.03197",
    archivePrefix = "arXiv",
    primaryClass = "hep-th",
    reportNumber = "YITP-21-105; IPMU21-0059",
    doi = "10.1103/PhysRevLett.129.041601",
    journal = "Phys. Rev. Lett.",
    volume = "129",
    number = "4",
    pages = "041601",
    year = "2022"
}

@article{Anninos:2020hfj,
    author = "Anninos, Dionysios and Denef, Frederik and Law, Y. T. Albert and Sun, Zimo",
    title = "{Quantum de Sitter horizon entropy from quasicanonical bulk, edge, sphere and topological string partition functions}",
    eprint = "2009.12464",
    archivePrefix = "arXiv",
    primaryClass = "hep-th",
    doi = "10.1007/JHEP01(2022)088",
    journal = "JHEP",
    volume = "01",
    pages = "088",
    year = "2022"
}

@article{Castro:2020smu,
    author = "Castro, Alejandra and Sabella-Garnier, Philippe and Zukowski, Claire",
    title = "{Gravitational Wilson Lines in 3D de Sitter}",
    eprint = "2001.09998",
    archivePrefix = "arXiv",
    primaryClass = "hep-th",
    doi = "10.1007/JHEP07(2020)202",
    journal = "JHEP",
    volume = "07",
    pages = "202",
    year = "2020"
}

@article{anderson2008boundary,
  title={On boundary value problems for Einstein metrics},
  author={Anderson, Michael T},
  journal={Geometry \& Topology},
  volume={12},
  number={4},
  pages={2009--2045},
  year={2008},
  publisher={Mathematical Sciences Publishers}
}

@article{An:2021fcq,
    author = "An, Zhongshan and Anderson, Michael T.",
    title = "{The initial boundary value problem and quasi-local Hamiltonians in General Relativity}",
    eprint = "2103.15673",
    archivePrefix = "arXiv",
    primaryClass = "gr-qc",
    doi = "10.1088/1361-6382/ac0a86",
    month = "3",
    year = "2021"
}

@article{Anninos:2024wpy,
    author = "Anninos, Dionysios and Galante, Dami{\'a}n A. and Maneerat, Chawakorn",
    title = "{Cosmological observatories}",
    eprint = "2402.04305",
    archivePrefix = "arXiv",
    primaryClass = "hep-th",
    doi = "10.1088/1361-6382/ad5824",
    journal = "Class. Quant. Grav.",
    volume = "41",
    number = "16",
    pages = "165009",
    year = "2024"
}

@article{Chakravarty:2025sbg,
    author = "Chakravarty, Joydeep and Maloney, Alexander and Namjou, Keivan and Ross, Simon F.",
    title = "{The spectrum of pure dS$_{3}$ gravity in the static patch}",
    eprint = "2505.06420",
    archivePrefix = "arXiv",
    primaryClass = "hep-th",
    doi = "10.1007/JHEP10(2025)021",
    journal = "JHEP",
    volume = "10",
    pages = "021",
    year = "2025"
}

@article{Witten:1989ip,
    author = "Witten, Edward",
    title = "{Quantization of {Chern-Simons} Gauge Theory With Complex Gauge Group}",
    reportNumber = "IASSNS-HEP-89/65",
    doi = "10.1007/BF02099116",
    journal = "Commun. Math. Phys.",
    volume = "137",
    pages = "29--66",
    year = "1991"
}

@article{gibbons1977cosmological,
  title={Cosmological event horizons, thermodynamics, and particle creation},
  author={Gibbons, Gary W and Hawking, Stephen W},
  journal={Physical Review D},
  volume={15},
  number={10},
  pages={2738},
  year={1977},
  publisher={APS}
}

@article{Carlip:1994ap,
    author = "Carlip, Steven and Nelson, J. E.",
    title = "{Comparative quantizations of (2+1)-dimensional gravity}",
    eprint = "gr-qc/9411031",
    archivePrefix = "arXiv",
    reportNumber = "UCD-94-37, DFTT-49-94",
    doi = "10.1103/PhysRevD.51.5643",
    journal = "Phys. Rev. D",
    volume = "51",
    pages = "5643--5653",
    year = "1995"
}

@article{Moncrief:1989dx,
    author = "Moncrief, Vincent",
    title = "{Reduction of the Einstein equations in (2+1)-dimensions to a Hamiltonian system over Teichmuller space}",
    doi = "10.1063/1.528274",
    journal = "J. Math. Phys.",
    volume = "30",
    pages = "2907--2914",
    year = "1989"
}

@article{Hosoya:1989sy,
    author = "Hosoya, Akio and Nakao, Ken-ichi",
    title = "{6-dimensional Quantum Town: (2+1)-dimensional Quantum Cosmology}",
    reportNumber = "TIT-HEP-145",
    doi = "10.1143/PTP.82.163",
    journal = "Prog. Theor. Phys.",
    volume = "82",
    pages = "163--174",
    year = "1989"
}

@article{Bousso:2001mw,
    author = "Bousso, Raphael and Maloney, Alexander and Strominger, Andrew",
    title = "{Conformal vacua and entropy in de Sitter space}",
    eprint = "hep-th/0112218",
    archivePrefix = "arXiv",
    doi = "10.1103/PhysRevD.65.104039",
    journal = "Phys. Rev. D",
    volume = "65",
    pages = "104039",
    year = "2002"
}

@article{Balasubramanian:2001nb,
    author = "Balasubramanian, Vijay and de Boer, Jan and Minic, Djordje",
    title = "{Mass, entropy and holography in asymptotically de Sitter spaces}",
    eprint = "hep-th/0110108",
    archivePrefix = "arXiv",
    reportNumber = "VPI-IPPAP-01-01, UPR-964-T",
    doi = "10.1103/PhysRevD.65.123508",
    journal = "Phys. Rev. D",
    volume = "65",
    pages = "123508",
    year = "2002"
}

@article{Brown:1994gs,
    author = "Brown, J. David and Creighton, J. and Mann, Robert B.",
    title = "{Temperature, energy and heat capacity of asymptotically anti-de Sitter black holes}",
    eprint = "gr-qc/9405007",
    archivePrefix = "arXiv",
    reportNumber = "WATPHYS-TH-94-02, CTMP-006-NCSU",
    doi = "10.1103/PhysRevD.50.6394",
    journal = "Phys. Rev. D",
    volume = "50",
    pages = "6394--6403",
    year = "1994"
}

@article{Carlip:2004ba,
    author = "Carlip, Steven",
    title = "{Quantum gravity in 2+1 dimensions}",
    eprint = "gr-qc/0409039",
    archivePrefix = "arXiv",
    doi = "10.1142/9789812702210_0001",
    journal = "World Sci. Ser. 20th Cent. Phys.",
    volume = "33",
    pages = "1--21",
    year = "2004"
}

@article{Castro:2011xb,
    author = "Castro, Alejandra and Lashkari, Nima and Maloney, Alexander",
    title = "{A de Sitter Farey Tail}",
    eprint = "1103.4620",
    archivePrefix = "arXiv",
    primaryClass = "hep-th",
    doi = "10.1103/PhysRevD.83.124027",
    journal = "Phys. Rev. D",
    volume = "83",
    pages = "124027",
    year = "2011"
}

@article{deser1984three,
  title={Three-dimensional cosmological gravity: dynamics of constant curvature},
  author={Deser, Stanley and Jackiw, R},
  journal={Annals of Physics},
  volume={153},
  number={2},
  pages={405--416},
  year={1984},
  publisher={Elsevier}
}

@article{Park:1998qk,
    author = "Park, Mu-In",
    title = "{Statistical entropy of three-dimensional Kerr-de Sitter space}",
    eprint = "hep-th/9806119",
    archivePrefix = "arXiv",
    reportNumber = "MIT-CTP-2750",
    doi = "10.1016/S0370-2693(98)01119-8",
    journal = "Phys. Lett. B",
    volume = "440",
    pages = "275--282",
    year = "1998"
}

@article{Kudler-Flam:2024psh,
    author = "Kudler-Flam, Jonah and Leutheusser, Samuel and Satishchandran, Gautam",
    title = "{Algebraic Observational Cosmology}",
    eprint = "2406.01669",
    archivePrefix = "arXiv",
    primaryClass = "hep-th",
    month = "6",
    year = "2024"
}

@article{Marolf:1996gb,
    author = "Marolf, Donald",
    title = "{Path integrals and instantons in quantum gravity: Minisuperspace models}",
    eprint = "gr-qc/9602019",
    archivePrefix = "arXiv",
    reportNumber = "UCSBTH-96-01",
    doi = "10.1103/PhysRevD.53.6979",
    journal = "Phys. Rev. D",
    volume = "53",
    pages = "6979--6990",
    year = "1996"
}

@article{Blommaert:2025eps,
    author = "Blommaert, Andreas and Tietto, Damiano and Verlinde, Herman",
    title = "{SYK collective field theory as complex Liouville gravity}",
    eprint = "2509.18462",
    archivePrefix = "arXiv",
    primaryClass = "hep-th",
    month = "9",
    year = "2025"
}

@article{Chen:2024rpx,
    author = "Chen, Chang-Han and Penington, Geoff",
    title = "{A clock is just a way to tell the time: gravitational algebras in cosmological spacetimes}",
    eprint = "2406.02116",
    archivePrefix = "arXiv",
    primaryClass = "hep-th",
    month = "6",
    year = "2024"
}

@article{Blommaert:2025bgd,
    author = "Blommaert, Andreas and Kudler-Flam, Jonah and Urbach, Erez Y.",
    title = "{Absolute entropy and the observer{\textquoteright}s no-boundary state}",
    eprint = "2505.14771",
    archivePrefix = "arXiv",
    primaryClass = "hep-th",
    doi = "10.1007/JHEP11(2025)113",
    journal = "JHEP",
    volume = "11",
    pages = "113",
    year = "2025"
}

@article{Cotler:2019nbi,
	archiveprefix = {arXiv},
	author = {Cotler, Jordan and Jensen, Kristan and Maloney, Alexander},
	doi = {10.1007/JHEP06(2020)048},
	eprint = {1905.03780},
	journal = {JHEP},
	pages = {048},
	primaryclass = {hep-th},
	title = {{Low-dimensional de Sitter quantum gravity}},
	volume = {06},
	year = {2020}}

@article{Akers:2025ahe,
    author = "Akers, Chris and Bueller, Gracemarie and DeWolfe, Oliver and Higginbotham, Kenneth and Reinking, Johannes and Rodriguez, Rudolph",
    title = "{On observers in holographic maps}",
    eprint = "2503.09681",
    archivePrefix = "arXiv",
    primaryClass = "hep-th",
    month = "3",
    year = "2025"
}

@article{Shi:2025amq,
    author = "Shi, Xiaoyi and Turiaci, Gustavo J.",
    title = "{The phase of the gravitational path integral}",
    eprint = "2504.00900",
    archivePrefix = "arXiv",
    primaryClass = "hep-th",
    month = "4",
    year = "2025"
}

@article{Svesko:2022txo,
    author = "Svesko, Andrew and Verheijden, Evita and Verlinde, Erik P. and Visser, Manus R.",
    title = "{Quasi-local energy and microcanonical entropy in two-dimensional nearly de Sitter gravity}",
    eprint = "2203.00700",
    archivePrefix = "arXiv",
    primaryClass = "hep-th",
    doi = "10.1007/JHEP08(2022)075",
    journal = "JHEP",
    volume = "08",
    pages = "075",
    year = "2022",
    note = "[Erratum: JHEP 04, 092 (2025)]"
}

@article{Maldacena:2024spf,
    author = "Maldacena, Juan",
    title = "{Real observers solving imaginary problems}",
    eprint = "2412.14014",
    archivePrefix = "arXiv",
    primaryClass = "hep-th",
    month = "12",
    year = "2024"
}

@article{Collier:2024lys,
	archiveprefix = {arXiv},
	author = {Collier, Scott and Eberhardt, Lorenz and M\"uhlmann, Beatrix and Rodriguez, Victor A.},
	eprint = {2410.07345},
	month = {10},
	primaryclass = {hep-th},
	title = {{The complex Liouville string: the matrix integral}},
	year = {2024}}

@article{Ivo:2025yek,
	archiveprefix = {arXiv},
	author = {Ivo, Victor and Maldacena, Juan and Sun, Zimo},
	eprint = {2504.00920},
	month = {4},
	primaryclass = {hep-th},
	title = {{Physical instabilities and the phase of the Euclidean path integral}},
	year = {2025}}

@article{Tietto:2025oxn,
	archiveprefix = {arXiv},
	author = {Tietto, Damiano and Verlinde, Herman},
	eprint = {2502.03869},
	month = {2},
	primaryclass = {hep-th},
	title = {{A microscopic model of de Sitter spacetime with an observer}},
	year = {2025}}

@article{Blommaert:2025avl,
	archiveprefix = {arXiv},
	author = {Blommaert, Andreas and Levine, Adam and Mertens, Thomas G. and Papalini, Jacopo and Parmentier, Klaas},
	eprint = {2501.17091},
	month = {1},
	primaryclass = {hep-th},
	title = {{Wormholes, branes and finite matrices in sine dilaton gravity}},
	year = {2025}}

@article{Maldacena:2004sn,
	archiveprefix = {arXiv},
	author = {Maldacena, Juan Martin and Moore, Gregory W. and Seiberg, Nathan and Shih, David},
	doi = {10.1088/1126-6708/2004/10/020},
	eprint = {hep-th/0408039},
	journal = {JHEP},
	pages = {020},
	reportnumber = {PUPT-2129},
	title = {{Exact vs. semiclassical target space of the minimal string}},
	volume = {10},
	year = {2004}}

@article{Verlinde:2024znh,
	archiveprefix = {arXiv},
	author = {Verlinde, Herman},
	eprint = {2402.00635},
	month = {2},
	primaryclass = {hep-th},
	title = {{Double-scaled SYK, Chords and de Sitter Gravity}},
	year = {2024}}

@article{Gaiotto:2024kze,
	archiveprefix = {arXiv},
	author = {Gaiotto, Davide and Verlinde, Herman},
	eprint = {2409.11551},
	month = {9},
	primaryclass = {hep-th},
	title = {{SYK-Schur duality: Double scaled SYK correlators from $N=2$ supersymmetric gauge theory}},
	year = {2024}}

@article{Collier:2025lux,
	archiveprefix = {arXiv},
	author = {Collier, Scott and Eberhardt, Lorenz and M\"uhlmann, Beatrix},
	eprint = {2501.01486},
	month = {1},
	primaryclass = {hep-th},
	title = {{A microscopic realization of dS$_3$}},
	year = {2025}}

@article{Abdalla:2025gzn,
	archiveprefix = {arXiv},
	author = {Abdalla, Ahmed I. and Antonini, Stefano and Iliesiu, Luca V. and Levine, Adam},
	eprint = {2501.02632},
	month = {1},
	primaryclass = {hep-th},
	title = {{The gravitational path integral from an observer's point of view}},
	year = {2025}}

@article{hartle1983wave,
	author = {Hartle, James B and Hawking, Stephen W},
	journal = {Physical Review D},
	number = {12},
	pages = {2960},
	publisher = {APS},
	title = {Wave function of the universe},
	volume = {28},
	year = {1983}}

@article{Blommaert:2024whf,
	archiveprefix = {arXiv},
	author = {Blommaert, Andreas and Levine, Adam and Mertens, Thomas G. and Papalini, Jacopo and Parmentier, Klaas},
	eprint = {2411.16922},
	month = {11},
	primaryclass = {hep-th},
	title = {{An entropic puzzle in periodic dilaton gravity and DSSYK}},
	year = {2024}}

@article{Witten:2022xxp,
	archiveprefix = {arXiv},
	author = {Witten, Edward},
	doi = {10.4310/ATMP.2023.v27.n1.a6},
	eprint = {2212.08270},
	journal = {Adv. Theor. Math. Phys.},
	number = {1},
	pages = {311--380},
	primaryclass = {hep-th},
	title = {{A note on the canonical formalism for gravity}},
	volume = {27},
	year = {2023}}

@article{Anninos:2011af,
    author = "Anninos, Dionysios and Hartnoll, Sean A. and Hofman, Diego M.",
    title = "{Static Patch Solipsism: Conformal Symmetry of the de Sitter Worldline}",
    eprint = "1109.4942",
    archivePrefix = "arXiv",
    primaryClass = "hep-th",
    doi = "10.1088/0264-9381/29/7/075002",
    journal = "Class. Quant. Grav.",
    volume = "29",
    pages = "075002",
    year = "2012"
}

@article{Held:2024rmg,
	archiveprefix = {arXiv},
	author = {Held, Jesse and Maxfield, Henry},
	eprint = {2410.14824},
	month = {10},
	primaryclass = {hep-th},
	title = {{The Hilbert space of de Sitter JT: a case study for canonical methods in quantum gravity}},
	year = {2024}}

@article{Blommaert:2019wfy,
	archiveprefix = {arXiv},
	author = {Blommaert, Andreas and Mertens, Thomas G. and Verschelde, Henri},
	doi = {10.1007/JHEP02(2021)168},
	eprint = {1911.11603},
	journal = {JHEP},
	pages = {168},
	primaryclass = {hep-th},
	title = {{Eigenbranes in Jackiw-Teitelboim gravity}},
	volume = {02},
	year = {2021}}

@article{Okuyama:2024eyf,
	archiveprefix = {arXiv},
	author = {Okuyama, Kazumi},
	eprint = {2408.03726},
	month = {8},
	primaryclass = {hep-th},
	title = {{Baby universe operators in the ETH matrix model of double-scaled SYK}},
	year = {2024}}

@article{Collier:2024kmo,
	archiveprefix = {arXiv},
	author = {Collier, Scott and Eberhardt, Lorenz and M\"uhlmann, Beatrix and Rodriguez, Victor A.},
	eprint = {2409.17246},
	month = {9},
	primaryclass = {hep-th},
	title = {{The complex Liouville string}},
	year = {2024}}

@article{Aguilar-Gutierrez:2025mxf,
    author = "Aguilar-Gutierrez, Sergio E. and Xu, Jiuci",
    title = "{Geometry of chord intertwiner, multiple shocks and switchback in double-scaled SYK}",
    eprint = "2506.19013",
    archivePrefix = "arXiv",
    primaryClass = "hep-th",
    doi = "10.1007/JHEP02(2026)246",
    journal = "JHEP",
    volume = "02",
    pages = "246",
    year = "2026"
}

@article{Feldbrugge:2017kzv,
    author = "Feldbrugge, Job and Lehners, Jean-Luc and Turok, Neil",
    title = "{Lorentzian Quantum Cosmology}",
    eprint = "1703.02076",
    archivePrefix = "arXiv",
    primaryClass = "hep-th",
    doi = "10.1103/PhysRevD.95.103508",
    journal = "Phys. Rev. D",
    volume = "95",
    number = "10",
    pages = "103508",
    year = "2017"
}

@article{DiUbaldo:2023qli,
    author = "Di Ubaldo, Gabriele and Perlmutter, Eric",
    title = "{AdS$_{3}$/RMT$_{2}$ duality}",
    eprint = "2307.03707",
    archivePrefix = "arXiv",
    primaryClass = "hep-th",
    doi = "10.1007/JHEP12(2023)179",
    journal = "JHEP",
    volume = "12",
    pages = "179",
    year = "2023"
}

@article{Collier:2022emf,
    author = "Collier, Scott and Perlmutter, Eric",
    title = "{Harnessing S-duality in $ \mathcal{N} $ = 4 SYM {\&} supergravity as SL(2, {\ensuremath{\mathbb{Z}}})-averaged strings}",
    eprint = "2201.05093",
    archivePrefix = "arXiv",
    primaryClass = "hep-th",
    doi = "10.1007/JHEP08(2022)195",
    journal = "JHEP",
    volume = "08",
    pages = "195",
    year = "2022"
}

@article{Benjamin:2022pnx,
    author = "Benjamin, Nathan and Chang, Cyuan-Han",
    title = "{Scalar modular bootstrap and zeros of the Riemann zeta function}",
    eprint = "2208.02259",
    archivePrefix = "arXiv",
    primaryClass = "hep-th",
    reportNumber = "CALT-TH 2022-026",
    doi = "10.1007/JHEP11(2022)143",
    journal = "JHEP",
    volume = "11",
    pages = "143",
    year = "2022"
}

@article{Caminiti:2026efx,
    author = "Caminiti, Jacqueline and Herderschee, Aidan",
    title = "{Inner Horizon Saddles and a Spectral KSW Criterion}",
    eprint = "2605.08335",
    archivePrefix = "arXiv",
    primaryClass = "hep-th",
    month = "5",
    year = "2026"
}

@article{Haehl:2023tkr,
    author = "Haehl, Felix M. and Marteau, Charles and Reeves, Wyatt and Rozali, Moshe",
    title = "{Symmetries and spectral statistics in chaotic conformal field theories}",
    eprint = "2302.14482",
    archivePrefix = "arXiv",
    primaryClass = "hep-th",
    doi = "10.1007/JHEP07(2023)196",
    journal = "JHEP",
    volume = "07",
    pages = "196",
    year = "2023"
}

@article{Anninos:2025fer,
    author = "Anninos, Dionysios and Hertog, Thomas and Karlsson, Joel",
    title = "{Quantum Liouville Cosmology}",
    eprint = "2512.15969",
    archivePrefix = "arXiv",
    primaryClass = "hep-th",
    month = "12",
    year = "2025"
}

@article{Dittrich:2024awu,
    author = {Dittrich, Bianca and Jacobson, Ted and Padua-Arg{\"u}elles, Jos{\'e}},
    title = "{de Sitter horizon entropy from a simplicial Lorentzian path integral}",
    eprint = "2403.02119",
    archivePrefix = "arXiv",
    primaryClass = "gr-qc",
    doi = "10.1103/PhysRevD.110.046006",
    journal = "Phys. Rev. D",
    volume = "110",
    number = "4",
    pages = "046006",
    year = "2024"
}

@article{deBoer:2025rct,
    author = "de Boer, Jan and Kames-King, Joshua and Post, Boris",
    title = "{Surgery and statistics in 3d gravity}",
    eprint = "2506.04151",
    archivePrefix = "arXiv",
    primaryClass = "hep-th",
    month = "6",
    year = "2025"
}

@article{Yang:2025lme,
    author = "Yang, Zhenbin and Zhang, Yuzhen and Zheng, Wenwen",
    title = "{Remarks on the de Sitter double cone}",
    eprint = "2505.08647",
    archivePrefix = "arXiv",
    primaryClass = "hep-th",
    reportNumber = "USTC-ICTS/PCFT-25-58",
    doi = "10.1103/jync-pmnz",
    journal = "Phys. Rev. D",
    volume = "113",
    number = "12",
    pages = "126012",
    year = "2026"
}

@article{Yan:2025usw,
    author = "Yan, Cynthia",
    title = "{Puzzles in 3D off-shell geometries via VTQFT}",
    eprint = "2502.16686",
    archivePrefix = "arXiv",
    primaryClass = "hep-th",
    doi = "10.1007/JHEP09(2025)104",
    journal = "JHEP",
    volume = "09",
    pages = "104",
    year = "2025"
}

@article{Benjamin:2019stq,
    author = "Benjamin, Nathan and Ooguri, Hirosi and Shao, Shu-Heng and Wang, Yifan",
    title = "{Light-cone modular bootstrap and pure gravity}",
    eprint = "1906.04184",
    archivePrefix = "arXiv",
    primaryClass = "hep-th",
    reportNumber = "CALT-TH 2019-020, IPMU19-0086, PUPT-2586",
    doi = "10.1103/PhysRevD.100.066029",
    journal = "Phys. Rev. D",
    volume = "100",
    number = "6",
    pages = "066029",
    year = "2019"
}

@article{Bousso:2000nf,
    author = "Bousso, Raphael",
    title = "{Positive vacuum energy and the N bound}",
    eprint = "hep-th/0010252",
    archivePrefix = "arXiv",
    reportNumber = "NSF-ITP-00-113",
    doi = "10.1088/1126-6708/2000/11/038",
    journal = "JHEP",
    volume = "11",
    pages = "038",
    year = "2000"
}

@article{Cui:2025sgy,
    author = "Cui, Chuanxin and Rozali, Moshe",
    title = "{Splitting and gluing in sine-dilaton gravity: matter correlators and the wormhole Hilbert space}",
    eprint = "2509.01680",
    archivePrefix = "arXiv",
    primaryClass = "hep-th",
    doi = "10.1007/JHEP02(2026)160",
    journal = "JHEP",
    volume = "02",
    pages = "160",
    year = "2026"
}

@article{Aguilar-Gutierrez:2026nmd,
    author = "Aguilar-Gutierrez, Sergio E. and Kukolj, Trivko and Seitz, Josef",
    title = "{q-Askey Deformations of Double-Scaled SYK}",
    eprint = "2605.13956",
    archivePrefix = "arXiv",
    primaryClass = "hep-th",
    month = "5",
    year = "2026"
}

@article{Anninos:2024iwf,
	archiveprefix = {arXiv},
	author = {Anninos, Dionysios and Baracco, Chiara and M\"uhlmann, Beatrix},
	eprint = {2406.15271},
	month = {6},
	primaryclass = {hep-th},
	title = {{Remarks on 2D quantum cosmology}},
	year = {2024}}

@article{Kruthoff:2024gxc,
	archiveprefix = {arXiv},
	author = {Kruthoff, Jorrit and Levine, Adam},
	eprint = {2402.10162},
	month = {2},
	primaryclass = {hep-th},
	title = {{Semi-classical dilaton gravity and the very blunt defect expansion}},
	year = {2024}}

@article{Blommaert:2023wad,
	archiveprefix = {arXiv},
	author = {Blommaert, Andreas and Mertens, Thomas G. and Yao, Shunyu},
	doi = {10.1007/JHEP11(2024)054},
	eprint = {2312.00871},
	journal = {JHEP},
	pages = {054},
	primaryclass = {hep-th},
	title = {{The q-Schwarzian and Liouville gravity}},
	volume = {11},
	year = {2024}}

@article{Witten:2023xze,
	archiveprefix = {arXiv},
	author = {Witten, Edward},
	eprint = {2308.03663},
	month = {8},
	primaryclass = {hep-th},
	title = {{A Background Independent Algebra in Quantum Gravity}},
	year = {2023}}

@article{Klemm:2002ir,
	archiveprefix = {arXiv},
	author = {Klemm, Dietmar and Vanzo, Luciano},
	doi = {10.1088/1126-6708/2002/04/030},
	eprint = {hep-th/0203268},
	journal = {JHEP},
	pages = {030},
	reportnumber = {UTF-447, IFUM-710-FT},
	title = {{De Sitter gravity and Liouville theory}},
	volume = {04},
	year = {2002}}

@article{Chen:2025jqm,
    author = "Chen, Yiming and Stanford, Douglas and Tang, Haifeng and Yang, Zhenbin",
    title = "{On the phase of the de Sitter density of states}",
    eprint = "2511.01400",
    archivePrefix = "arXiv",
    primaryClass = "hep-th",
    doi = "10.1007/JHEP05(2026)068",
    journal = "JHEP",
    volume = "05",
    pages = "068",
    year = "2026"
}

@article{Blommaert:2025rgw,
    author = "Blommaert, Andreas and Levine, Adam",
    title = "{Sphere amplitudes and observing the universe's size}",
    eprint = "2505.24633",
    archivePrefix = "arXiv",
    primaryClass = "hep-th",
    month = "5",
    year = "2025"
}

@article{Dijkgraaf:2000fq,
    author = "Dijkgraaf, Robbert and Maldacena, Juan Martin and Moore, Gregory W. and Verlinde, Erik P.",
    title = "{A Black hole Farey tail}",
    eprint = "hep-th/0005003",
    archivePrefix = "arXiv",
    reportNumber = "YCTP-P25-99",
    month = "5",
    year = "2000"
}

@article{Anninos:2026hia,
    author = {Anninos, Dionysios and Baracco, Chiara and Letsios, Vasileios A. and M{\"u}hlmann, Beatrix},
    title = "{dS$^4$ Metamorphosis}",
    eprint = "2602.19812",
    archivePrefix = "arXiv",
    primaryClass = "hep-th",
    month = "2",
    year = "2026"
}

@article{Anninos:2023exn,
    author = {Anninos, Dionysios and Benetti Genolini, Pietro and M{\"u}hlmann, Beatrix},
    title = "{dS$_{2}$ supergravity}",
    eprint = "2309.02480",
    archivePrefix = "arXiv",
    primaryClass = "hep-th",
    doi = "10.1007/JHEP11(2023)145",
    journal = "JHEP",
    volume = "11",
    pages = "145",
    year = "2023"
}

@article{Cotler:2025gui,
    author = "Cotler, Jordan and Jensen, Kristan",
    title = "{Norm of the no-boundary state}",
    eprint = "2506.20547",
    archivePrefix = "arXiv",
    primaryClass = "hep-th",
    doi = "10.1007/JHEP03(2026)180",
    journal = "JHEP",
    volume = "03",
    pages = "180",
    year = "2026"
}

@article{Anninos:2011ui,
    author = "Anninos, Dionysios and Hartman, Thomas and Strominger, Andrew",
    title = "{Higher Spin Realization of the dS/CFT Correspondence}",
    eprint = "1108.5735",
    archivePrefix = "arXiv",
    primaryClass = "hep-th",
    doi = "10.1088/1361-6382/34/1/015009",
    journal = "Class. Quant. Grav.",
    volume = "34",
    number = "1",
    pages = "015009",
    year = "2017"
}

@inproceedings{Witten:2001kn,
    author = "Witten, Edward",
    title = "{Quantum gravity in de Sitter space}",
    booktitle = "{Strings 2001: International Conference}",
    eprint = "hep-th/0106109",
    archivePrefix = "arXiv",
    month = "6",
    year = "2001"
}

@article{Keller:2014xba,
    author = "Keller, Christoph A. and Maloney, Alexander",
    title = "{Poincare Series, 3D Gravity and CFT Spectroscopy}",
    eprint = "1407.6008",
    archivePrefix = "arXiv",
    primaryClass = "hep-th",
    reportNumber = "RUNHETC-2014-13",
    doi = "10.1007/JHEP02(2015)080",
    journal = "JHEP",
    volume = "02",
    pages = "080",
    year = "2015"
}

@article{Benjamin:2020mfz,
    author = "Benjamin, Nathan and Collier, Scott and Maloney, Alexander",
    title = "{Pure Gravity and Conical Defects}",
    eprint = "2004.14428",
    archivePrefix = "arXiv",
    primaryClass = "hep-th",
    doi = "10.1007/JHEP09(2020)034",
    journal = "JHEP",
    volume = "09",
    pages = "034",
    year = "2020"
}

@article{Goto:2026ipq,
    author = "Goto, Kanato and Milekhin, Alexey and Verlinde, Herman and Xu, Jiuci",
    title = "{Generalized Free Fields in de Sitter from 1D CFT}",
    eprint = "2605.03037",
    archivePrefix = "arXiv",
    primaryClass = "hep-th",
    reportNumber = "OU-HET-1309, RIKEN-iTHEMS-Report-26",
    month = "5",
    year = "2026"
}

@article{Teschner:2000md,
    author = "Teschner, J.",
    editor = "Bernard, Denis and Bonora, Loriano and Mussardo, Giuseppe and Corrigan, Edward and Gomez, Cesar and Nahm, Werner and Julia, Bernard",
    title = "{Remarks on Liouville theory with boundary}",
    eprint = "hep-th/0009138",
    archivePrefix = "arXiv",
    doi = "10.22323/1.006.0041",
    journal = "PoS",
    volume = "tmr2000",
    pages = "041",
    year = "2000"
}

@article{Ammon:2013hba,
    author = "Ammon, Martin and Castro, Alejandra and Iqbal, Nabil",
    title = "{Wilson Lines and Entanglement Entropy in Higher Spin Gravity}",
    eprint = "1306.4338",
    archivePrefix = "arXiv",
    primaryClass = "hep-th",
    reportNumber = "NSF-KITP-13-112",
    doi = "10.1007/JHEP10(2013)110",
    journal = "JHEP",
    volume = "10",
    pages = "110",
    year = "2013"
}

@article{Castro:2018srf,
    author = "Castro, Alejandra and Iqbal, Nabil and Llabr{\'e}s, Eva",
    title = "{Wilson lines and Ishibashi states in AdS$_{3}$/CFT$_{2}$}",
    eprint = "1805.05398",
    archivePrefix = "arXiv",
    primaryClass = "hep-th",
    doi = "10.1007/JHEP09(2018)066",
    journal = "JHEP",
    volume = "09",
    pages = "066",
    year = "2018"
}

@article{DiazDorronsoro:2017hti,
    author = "Diaz Dorronsoro, Juan and Halliwell, Jonathan J. and Hartle, James B. and Hertog, Thomas and Janssen, Oliver",
    title = "{Real no-boundary wave function in Lorentzian quantum cosmology}",
    eprint = "1705.05340",
    archivePrefix = "arXiv",
    primaryClass = "gr-qc",
    doi = "10.1103/PhysRevD.96.043505",
    journal = "Phys. Rev. D",
    volume = "96",
    number = "4",
    pages = "043505",
    year = "2017"
}

@article{Gaiotto:2024osr,
    author = {Gaiotto, Davide and Teschner, J{\"o}rg},
    title = "{Schur Quantization and Complex Chern-Simons theory}",
    eprint = "2406.09171",
    archivePrefix = "arXiv",
    primaryClass = "hep-th",
    month = "6",
    year = "2024"
}

@article{Maldacena:1997re,
    author = "Maldacena, Juan Martin",
    title = "{The Large $N$ limit of superconformal field theories and supergravity}",
    eprint = "hep-th/9711200",
    archivePrefix = "arXiv",
    reportNumber = "HUTP-97-A097, HUTP-98-A097",
    doi = "10.4310/ATMP.1998.v2.n2.a1",
    journal = "Adv. Theor. Math. Phys.",
    volume = "2",
    pages = "231--252",
    year = "1998"
}

@article{Gubser:1998bc,
    author = "Gubser, S. S. and Klebanov, Igor R. and Polyakov, Alexander M.",
    title = "{Gauge theory correlators from noncritical string theory}",
    eprint = "hep-th/9802109",
    archivePrefix = "arXiv",
    reportNumber = "PUPT-1767",
    doi = "10.1016/S0370-2693(98)00377-3",
    journal = "Phys. Lett. B",
    volume = "428",
    pages = "105--114",
    year = "1998"
}

@article{Witten:1998qj,
    author = "Witten, Edward",
    title = "{Anti de Sitter space and holography}",
    eprint = "hep-th/9802150",
    archivePrefix = "arXiv",
    reportNumber = "IASSNS-HEP-98-15",
    doi = "10.4310/ATMP.1998.v2.n2.a2",
    journal = "Adv. Theor. Math. Phys.",
    volume = "2",
    pages = "253--291",
    year = "1998"
}

@article{Banihashemi:2024aal,
    author = "Banihashemi, Batoul and Jacobson, Ted",
    title = "{On the lapse contour in the gravitational path integral}",
    eprint = "2405.10307",
    archivePrefix = "arXiv",
    primaryClass = "hep-th",
    doi = "10.1103/PhysRevD.111.066014",
    journal = "Phys. Rev. D",
    volume = "111",
    number = "6",
    pages = "066014",
    year = "2025"
}

@article{halliwell1988derivation,
  title={Derivation of the Wheeler-DeWitt equation from a path integral for minisuperspace models},
  author={Halliwell, Jonathan J},
  journal={Physical Review D},
  volume={38},
  number={8},
  pages={2468},
  year={1988},
  publisher={APS}
}

@article{Maloney:2007ud,
    author = "Maloney, Alexander and Witten, Edward",
    title = "{Quantum Gravity Partition Functions in Three Dimensions}",
    eprint = "0712.0155",
    archivePrefix = "arXiv",
    primaryClass = "hep-th",
    doi = "10.1007/JHEP02(2010)029",
    journal = "JHEP",
    volume = "02",
    pages = "029",
    year = "2010"
}

@article{Belin:2026pko,
    author = "Belin, Alexandre and Collier, Scott and Eberhardt, Lorenz and Liska, Diego and Post, Boris",
    title = "{A universal sum over topologies in 3d gravity}",
    eprint = "2601.07906",
    archivePrefix = "arXiv",
    primaryClass = "hep-th",
    month = "1",
    year = "2026"
}

@article{berkooz2019towards,
	author = {Berkooz, Micha and Isachenkov, Mikhail and Narovlansky, Vladimir and Torrents, Genis},
	journal = {Journal of High Energy Physics},
	number = {3},
	pages = {1--72},
	publisher = {Springer},
	title = {Towards a full solution of the large N double-scaled SYK model},
	volume = {2019},
	year = {2019}}

@article{Maxfield:2020ale,
	archiveprefix = {arXiv},
	author = {Maxfield, Henry and Turiaci, Gustavo J.},
	doi = {10.1007/JHEP01(2021)118},
	eprint = {2006.11317},
	journal = {JHEP},
	pages = {118},
	primaryclass = {hep-th},
	title = {{The path integral of 3D gravity near extremality; or, JT gravity with defects as a matrix integral}},
	volume = {01},
	year = {2021}}

@article{Collier:2023cyw,
	archiveprefix = {arXiv},
	author = {Collier, Scott and Eberhardt, Lorenz and M\"uhlmann, Beatrix and Rodriguez, Victor A.},
	eprint = {2309.10846},
	month = {9},
	primaryclass = {hep-th},
	title = {{The Virasoro Minimal String}},
	year = {2023}}

@article{Belaey:2023jtr,
	archiveprefix = {arXiv},
	author = {Belaey, Andreas and Mariani, Francesca and Mertens, Thomas G.},
	eprint = {2310.04245},
	month = {10},
	primaryclass = {hep-th},
	title = {{Branes in JT (super)gravity from group theory}},
	year = {2023}}

@article{Lin:2023trc,
	archiveprefix = {arXiv},
	author = {Lin, Henry W. and Stanford, Douglas},
	eprint = {2307.15725},
	month = {7},
	primaryclass = {hep-th},
	title = {{A symmetry algebra in double-scaled SYK}},
	year = {2023}}

@article{Berkooz:2018jqr,
	archiveprefix = {arXiv},
	author = {Berkooz, Micha and Isachenkov, Mikhail and Narovlansky, Vladimir and Torrents, Genis},
	doi = {10.1007/JHEP03(2019)079},
	eprint = {1811.02584},
	journal = {JHEP},
	pages = {079},
	primaryclass = {hep-th},
	title = {{Towards a full solution of the large N double-scaled SYK model}},
	volume = {03},
	year = {2019}}

@article{Saad:2019lba,
	archiveprefix = {arXiv},
	author = {Saad, Phil and Shenker, Stephen H. and Stanford, Douglas},
	eprint = {1903.11115},
	month = {3},
	primaryclass = {hep-th},
	title = {{JT gravity as a matrix integral}},
	year = {2019}}

@article{Blommaert:2023opb,
	archiveprefix = {arXiv},
	author = {Blommaert, Andreas and Mertens, Thomas G. and Yao, Shunyu},
	doi = {10.1007/JHEP02(2024)067},
	eprint = {2306.00941},
	journal = {JHEP},
	pages = {067},
	primaryclass = {hep-th},
	title = {{Dynamical actions and q-representation theory for double-scaled SYK}},
	volume = {02},
	year = {2024}}

@article{Witten:1988hc,
	author = {Witten, Edward},
	doi = {10.1016/0550-3213(88)90143-5},
	journal = {Nucl. Phys. B},
	pages = {46},
	reportnumber = {IASSNS-HEP-88-32},
	title = {{(2+1)-Dimensional Gravity as an Exactly Soluble System}},
	volume = {311},
	year = {1988}}

@article{Harlow:2018tqv,
	archiveprefix = {arXiv},
	author = {Harlow, Daniel and Jafferis, Daniel},
	doi = {10.1007/JHEP02(2020)177},
	eprint = {1804.01081},
	journal = {JHEP},
	pages = {177},
	primaryclass = {hep-th},
	title = {{The Factorization Problem in Jackiw-Teitelboim Gravity}},
	volume = {02},
	year = {2020}}

@article{Iliesiu:2019xuh,
	archiveprefix = {arXiv},
	author = {Iliesiu, Luca V. and Pufu, Silviu S. and Verlinde, Herman and Wang, Yifan},
	doi = {10.1007/JHEP11(2019)091},
	eprint = {1905.02726},
	journal = {JHEP},
	pages = {091},
	primaryclass = {hep-th},
	reportnumber = {PUPT-2584},
	title = {{An exact quantization of Jackiw-Teitelboim gravity}},
	volume = {11},
	year = {2019}}

@article{witten1989quantum,
	author = {Witten, Edward},
	journal = {Communications in Mathematical Physics},
	number = {3},
	pages = {351--399},
	publisher = {Springer},
	title = {Quantum field theory and the Jones polynomial},
	volume = {121},
	year = {1989}}

@article{elitzur1989remarks,
	author = {Elitzur, Shmuel and Moore, Gregory and Schwimmer, Adam and Seiberg, Nathan},
	journal = {Nuclear Physics B},
	number = {1},
	pages = {108--134},
	publisher = {Elsevier},
	title = {Remarks on the canonical quantization of the Chern-Simons-Witten theory},
	volume = {326},
	year = {1989}}

@article{Cotler:2020ugk,
	archiveprefix = {arXiv},
	author = {Cotler, Jordan and Jensen, Kristan},
	doi = {10.1007/JHEP04(2021)033},
	eprint = {2006.08648},
	journal = {JHEP},
	pages = {033},
	primaryclass = {hep-th},
	title = {{AdS$_{3}$ gravity and random CFT}},
	volume = {04},
	year = {2021}}

@article{Lin:2022nss,
	archiveprefix = {arXiv},
	author = {Lin, Henry and Susskind, Leonard},
	eprint = {2206.01083},
	month = {6},
	primaryclass = {hep-th},
	title = {{Infinite Temperature's Not So Hot}},
	year = {2022}}

@article{Susskind:2021esx,
	archiveprefix = {arXiv},
	author = {Susskind, Leonard},
	doi = {10.22128/jhap.2021.455.1005},
	eprint = {2109.14104},
	journal = {JHAP},
	number = {1},
	pages = {1--22},
	primaryclass = {hep-th},
	title = {{Entanglement and Chaos in De Sitter Space Holography: An SYK Example}},
	volume = {1},
	year = {2021}}

@article{Lin:2022rbf,
	archiveprefix = {arXiv},
	author = {Lin, Henry W.},
	doi = {10.1007/JHEP11(2022)060},
	eprint = {2208.07032},
	journal = {JHEP},
	pages = {060},
	primaryclass = {hep-th},
	title = {{The bulk Hilbert space of double scaled SYK}},
	volume = {11},
	year = {2022}}

@article{Collier:2023fwi,
	archiveprefix = {arXiv},
	author = {Collier, Scott and Eberhardt, Lorenz and Zhang, Mengyang},
	eprint = {2304.13650},
	month = {4},
	primaryclass = {hep-th},
	title = {{Solving 3d Gravity with Virasoro TQFT}},
	year = {2023}}

@article{kitaev2015simple,
	author = {Kitaev, Alexei},
	journal = {Entanglement in Strongly-Correlated Quantum Matter},
	pages = {38},
	title = {A simple model of quantum holography (part 2)},
	year = {2015}}

@article{Maldacena:2016hyu,
	archiveprefix = {arXiv},
	author = {Maldacena, Juan and Stanford, Douglas},
	doi = {10.1103/PhysRevD.94.106002},
	eprint = {1604.07818},
	journal = {Phys. Rev. D},
	number = {10},
	pages = {106002},
	primaryclass = {hep-th},
	title = {{Remarks on the Sachdev-Ye-Kitaev model}},
	volume = {94},
	year = {2016}}

@article{Berkooz:2018qkz,
	archiveprefix = {arXiv},
	author = {Berkooz, Micha and Narayan, Prithvi and Simon, Joan},
	doi = {10.1007/JHEP08(2018)192},
	eprint = {1806.04380},
	journal = {JHEP},
	pages = {192},
	primaryclass = {hep-th},
	title = {{Chord diagrams, exact correlators in spin glasses and black hole bulk reconstruction}},
	volume = {08},
	year = {2018}}

@article{Narovlansky:2023lfz,
	archiveprefix = {arXiv},
	author = {Narovlansky, Vladimir and Verlinde, Herman},
	eprint = {2310.16994},
	month = {10},
	primaryclass = {hep-th},
	title = {{Double-scaled SYK and de Sitter Holography}},
	year = {2023}}

@article{Mertens:2020hbs,
	archiveprefix = {arXiv},
	author = {Mertens, Thomas G. and Turiaci, Gustavo J.},
	doi = {10.1007/JHEP01(2021)073},
	eprint = {2006.07072},
	journal = {JHEP},
	pages = {073},
	primaryclass = {hep-th},
	title = {{Liouville quantum gravity -- holography, JT and matrices}},
	volume = {01},
	year = {2021}}

@article{Zamolodchikov:2001ah,
	archiveprefix = {arXiv},
	author = {Zamolodchikov, Alexander B. and Zamolodchikov, Alexei B.},
	editor = {Ivanov, Evgeny A. and Zupnik, Boris M.},
	eprint = {hep-th/0101152},
	month = {1},
	pages = {280--299},
	reportnumber = {RUNHETC-2001-02, LPM-01-01},
	title = {{Liouville field theory on a pseudosphere}},
	year = {2001}}

@article{Fateev:2000ik,
	archiveprefix = {arXiv},
	author = {Fateev, V. and Zamolodchikov, Alexander B. and Zamolodchikov, Alexei B.},
	eprint = {hep-th/0001012},
	month = {1},
	reportnumber = {RUNHETC-2000-01},
	title = {{Boundary Liouville field theory. 1. Boundary state and boundary two point function}},
	year = {2000}}

@article{Cotler:2016fpe,
	archiveprefix = {arXiv},
	author = {Cotler, Jordan S. and Gur-Ari, Guy and Hanada, Masanori and Polchinski, Joseph and Saad, Phil and Shenker, Stephen H. and Stanford, Douglas and Streicher, Alexandre and Tezuka, Masaki},
	doi = {10.1007/JHEP05(2017)118},
	eprint = {1611.04650},
	journal = {JHEP},
	note = {[Erratum: JHEP 09, 002 (2018)]},
	pages = {118},
	primaryclass = {hep-th},
	reportnumber = {SU-ITP-16-19, SU-ITP-16/19, YITP-16-124},
	title = {{Black Holes and Random Matrices}},
	volume = {05},
	year = {2017}}

@article{Heydeman:2020hhw,
	archiveprefix = {arXiv},
	author = {Heydeman, Matthew and Iliesiu, Luca V. and Turiaci, Gustavo J. and Zhao, Wenli},
	doi = {10.1088/1751-8121/ac3be9},
	eprint = {2011.01953},
	journal = {J. Phys. A},
	number = {1},
	pages = {014004},
	primaryclass = {hep-th},
	reportnumber = {PUPT-2621},
	title = {{The statistical mechanics of near-BPS black holes}},
	volume = {55},
	year = {2022}}

@article{Almheiri:2024xtw,
	archiveprefix = {arXiv},
	author = {Almheiri, Ahmed and Goel, Akash and Hu, Xu-Yao},
	eprint = {2403.18333},
	month = {3},
	primaryclass = {hep-th},
	title = {{Quantum gravity of the Heisenberg algebra}},
	year = {2024}}

@article{Chandrasekaran:2022cip,
	archiveprefix = {arXiv},
	author = {Chandrasekaran, Venkatesa and Longo, Roberto and Penington, Geoff and Witten, Edward},
	doi = {10.1007/JHEP02(2023)082},
	eprint = {2206.10780},
	journal = {JHEP},
	pages = {082},
	primaryclass = {hep-th},
	title = {{An algebra of observables for de Sitter space}},
	volume = {02},
	year = {2023}}

@article{jafferis2022jt,
	author = {Jafferis, Daniel Louis and Kolchmeyer, David K and Mukhametzhanov, Baur and Sonner, Julian},
	journal = {arXiv preprint arXiv:2209.02131},
	title = {JT gravity with matter, generalized ETH, and Random Matrices},
	year = {2022}}

@article{Blommaert:2024ydx,
	archiveprefix = {arXiv},
	author = {Blommaert, Andreas and Mertens, Thomas G. and Papalini, Jacopo},
	eprint = {2404.03535},
	month = {4},
	primaryclass = {hep-th},
	title = {{The dilaton gravity hologram of double-scaled SYK}},
	year = {2024}}

@article{Berkooz:2024lgq,
	archiveprefix = {arXiv},
	author = {Berkooz, Micha and Mamroud, Ohad},
	eprint = {2407.09396},
	month = {7},
	primaryclass = {hep-th},
	title = {{A Cordial Introduction to Double Scaled SYK}},
	year = {2024}}

@article{Okuyama:2023kdo,
	archiveprefix = {arXiv},
	author = {Okuyama, Kazumi},
	doi = {10.1007/JHEP09(2023)133},
	eprint = {2306.15981},
	journal = {JHEP},
	pages = {133},
	primaryclass = {hep-th},
	title = {{Discrete analogue of the Weil-Petersson volume in double scaled SYK}},
	volume = {09},
	year = {2023}}

@article{Louko:1995jw,
	archiveprefix = {arXiv},
	author = {Louko, Jorma and Sorkin, Rafael D.},
	doi = {10.1088/0264-9381/14/1/018},
	eprint = {gr-qc/9511023},
	journal = {Class. Quant. Grav.},
	pages = {179--204},
	reportnumber = {SU-GP-95-5-1, WISC-MILW-95-TH-16, MDDP-PP-96-40},
	title = {{Complex actions in two-dimensional topology change}},
	volume = {14},
	year = {1997}}

@article{Harlow:2025pvj,
	archiveprefix = {arXiv},
	author = {Harlow, Daniel and Usatyuk, Mykhaylo and Zhao, Ying},
	eprint = {2501.02359},
	month = {1},
	primaryclass = {hep-th},
	reportnumber = {MIT-CTP/5824},
	title = {{Quantum mechanics and observers for gravity in a closed universe}},
	year = {2025}}

@article{Collier:2024kwt,
	archiveprefix = {arXiv},
	author = {Collier, Scott and Eberhardt, Lorenz and M\"uhlmann, Beatrix and Rodriguez, Victor A.},
	eprint = {2409.18759},
	month = {9},
	primaryclass = {hep-th},
	title = {{The complex Liouville string: the worldsheet}},
	year = {2024}}

@article{Heller:2024ldz,
	archiveprefix = {arXiv},
	author = {Heller, Michal P. and Papalini, Jacopo and Schuhmann, Tim},
	eprint = {2412.17785},
	month = {12},
	primaryclass = {hep-th},
	title = {{Krylov spread complexity as holographic complexity beyond JT gravity}},
	year = {2024}}

\end{document}